\pdfoutput=1

\documentclass[%
 reprint,
 superscriptaddress,
 nobibnotes,
 amsmath,amssymb,
 aps,
floatfix,
 showkeys,
 longbibliography,
]{revtex4-2}

\usepackage{graphicx}
\usepackage{dcolumn}
\usepackage{bm}
\usepackage{chemformula}
\usepackage[separate-uncertainty=true,separate-uncertainty,multi-part-units=single]{siunitx}
\usepackage{hyperref}
\usepackage{float} 
\usepackage{multirow}
\usepackage{subfigure}
\hypersetup{colorlinks=true,allcolors=magenta}

\renewcommand{\thetable}{\arabic{table}}
\renewcommand{\figurename}{Figure}
\renewcommand{\tablename}{Table}

\begin{document}

\title{Evaluating Blended Refrigerants for Thermochemical Energy Storage and Circular Refrigerant Recovery using Activated Carbons}

\author{H. Lucassen}
    \affiliation{Energy Technology, Department of Mechanical Engineering, Eindhoven University of Technology, P. O. Box 513, 5600 MB Eindhoven, The Netherlands}
\author{A. Luna-Triguero}
    \affiliation{Energy Technology, Department of Mechanical Engineering, Eindhoven University of Technology, P. O. Box 513, 5600 MB Eindhoven, The Netherlands}
    \affiliation{Eindhoven Institute for Renewable Energy Systems (EIRES), Eindhoven University of Technology, Eindhoven 5600 MB, The Netherlands}
\author{J. M. Vicent-Luna}
    \email[Corresponding author: ]{j.vicent.luna@tue.nl}
    \affiliation{Materials Simulation \& Modelling, Department of Applied Physics and Science Education, Eindhoven University of Technology, 5600 MB, Eindhoven, The Netherlands}
    \affiliation{Eindhoven Institute for Renewable Energy Systems (EIRES), Eindhoven University of Technology, Eindhoven 5600 MB, The Netherlands}

\date{\today}

\begin{abstract}

The climate crisis demands a rapid shift to sustainable energy technologies and higher efficiency in existing energy systems. Adsorption-based thermochemical energy storage is a promising alternative due to its high energy density and compatibility with renewable heat sources. In this work, we investigate the adsorption behavior of pure refrigerants (R32, R125, R134a, and R600) and their commercial blends (R410A, R407F, R417A, and R417C) in six activated carbons for thermochemical energy storage and circular refrigerant recovery.
A multiscale computational workflow combining Monte Carlo simulations, thermodynamic modeling, and breakthrough simulations is developed to predict adsorption, storage, and separation behavior from pure-component adsorption data. The methodology integrates adsorption potential theory (APT), ideal adsorbed solution theory (IAST), and models for the isosteric heat of adsorption. In addition, an in-house computational framework is developed to calculate heats of adsorption and energy storage densities for both pure refrigerants and multicomponent mixtures.
Although developed using molecular simulations as a benchmark, the methodology is directly applicable to experimental studies, since it only requires adsorption isotherms of the pure components as input to evaluate the performance of refrigerant blends. The results show that refrigerant blends can achieve higher storage densities than their pure counterparts due to cooperative adsorption and more efficient molecular packing. Furthermore, the activated carbons selectively separate key refrigerant components, highlighting their potential for sustainable refrigerant recovery. Overall, this work provides a general framework for the rational design and screening of next-generation refrigerant blends for adsorption-driven energy storage and separation applications.

\end{abstract}

\keywords{Refrigerant blends, Storage density, Multiscale modelling, Hydrofluorocarbon regeneration}

\maketitle


\section{Introduction}
\label{sec:intro}

Climate change is one of the most significant challenges facing the 21st century. Concerns regarding global warming emerged during the mid-20th century, culminating in the 2015 Paris Agreement signed by 195 countries \cite{UnitedNationsTheAgreement}. The primary objective of the agreement is to limit global warming to below 2 $^\circ$C above pre-industrial levels while pursuing efforts to restrict it to 1.5 $^\circ$C. Achieving these goals requires substantial reductions in greenhouse gas emissions, including carbon dioxide, methane, and fluorinated gases (F-gases), together with the transition toward sustainable energy technologies and improved energy efficiency in existing industrial processes.

Refrigerants constitute one of the most widely used classes of F-gases. They are employed as working fluids in heating and cooling systems ranging from domestic to industrial scales in conventional vapor-compression technologies \cite{Vuppaladadiyam2022ProgressConsequences}. According to EERA (2022), the implementation of Thermal Energy Storage (TES) technologies in industrial processes could reduce emissions in Europe by up to 513 Mt CO$_2$ per year by decreasing the environmental impact of industrial heating and cooling, which accounts for nearly 80\% of industrial energy consumption \cite{EERA2023PolicyIntegration}. TES also provides an effective strategy for balancing energy supply and demand associated with intermittent renewable energy generation \cite{TNOEnergyStorage}. Among TES technologies, adsorption-based systems, including adsorption heat pumps (AHPs) and adsorption cooling systems (ACSs), are increasingly considered sustainable alternatives to conventional heating and cooling technologies because they operate at lower thermal requirements \cite{Islam2024HarnessingSystems, Madero-Castro2023Alcohol-basedFrameworks, DeLange2015Adsorption-DrivenFrameworks, Yang2024ExperimentalSystems, Xia2020AdsorptionCOF-5}. Refrigerants are attractive working fluids for these systems due to their favorable thermodynamic properties, high energy densities, and strong adsorption compatibility with porous materials \cite{ Yagnamurthy2021AdsorptionCharacterization, Peng2010ComparisonSilicalite-1, Askalany2012ExperimentalPair, Vicent-Luna2024AdsorptionFrameworks, Saha2008IsothermsCarbon}. Furthermore, adsorption systems operate with significantly smaller refrigerant inventories than conventional vapor-compression systems, reducing their overall environmental impact.

Currently, hydrofluorocarbons (HFCs) remain among the most widely used refrigerants despite exhibiting global warming potentials (GWPs) up to 23,000 times larger than that of CO$_2$ \cite{Sosa2023ExploringPotential}. Consequently, the European Union established the objective of reducing F-gas emissions by two-thirds by 2030 \cite{Sosa2023ExploringPotential}. Nevertheless, approximately 20,000 tons of HFCs are still produced annually \cite{Yasaka2023Life-CycleDestruction}, corresponding to 35--45 Mt of CO$_2$-equivalent emissions if released into the atmosphere. To address both environmental and regulatory constraints, industry is transitioning toward fourth-generation refrigerants, primarily consisting of blends of HFCs and hydrofluoroolefins (HFOs). These blends exhibit lower GWPs and improved thermodynamic performance \cite{Sosa2023ExploringPotential}. In addition, they facilitate circular refrigerant management through the recovery and reuse of existing HFCs. Yasaka et al. (2023) reported that refrigerant reclamation produces significantly lower emissions than refrigerant production or destruction, highlighting the importance of refrigerant recovery strategies \cite{Yasaka2023Life-CycleDestruction}. However, most refrigerants are still inadequately recovered at end-of-life and are instead released into the atmosphere \cite{Chen2025SustainableAction}.

The separation of modern refrigerant blends remains challenging because many of these mixtures are azeotropic or near-azeotropic. Conventional recovery methods, such as cryogenic separation, often fail to achieve the purity levels required for industrial reuse \cite{Wanigarathna2020MetalReview}. In this context, adsorption-based separation using nanoporous materials has emerged as a promising alternative. This work focuses on two complementary applications of refrigerant adsorption systems: thermochemical energy storage, which requires understanding adsorption thermodynamics and heats of adsorption, and the separation of refrigerant blends for circular recovery, which requires accurate prediction of mixture adsorption and separation dynamics.

Among nanoporous adsorbents, metal-organic frameworks (MOFs) \cite{Xia2020AdsorptionCOF-5, DeLange2015Adsorption-DrivenFrameworks}, zeolites \cite{Lehmann2017AssessmentApplications, Ristic2018ImprovedStorage, Madero-Castro2023OnTransfer}, and activated carbons (ACs) \cite{Critoph1995HeatGases, Xiao2010SimulationStorage, Madero-Castro2022AdsorptionApplications} are the most commonly investigated materials for adsorption-based energy storage applications. While MOFs often exhibit extremely high surface areas, activated carbons remain attractive due to their stability, scalability, and broad pore-size distributions. Most studies on refrigerant adsorption focus on MOFs and rely primarily on experimental characterization \cite{Vicent-Luna2024AdsorptionFrameworks}. In contrast, studies addressing refrigerant adsorption in activated carbons, particularly for refrigerant mixtures, remain relatively limited.

Adsorption equilibrium and thermophysical properties have been investigated experimentally for several pure refrigerants in activated carbons, including R32 \cite{Yagnamurthy2021AdsorptionCharacterization}, R125 \cite{Peng2010ComparisonSilicalite-1}, R134a \cite{Askalany2012ExperimentalPair}, and R600 \cite{Saha2008IsothermsCarbon}. Existing studies on refrigerant blends mainly focus either on separation performance for mixtures containing R32, R125, and R134a \cite{Ribeiro2023VacuumBlend, Sosa2020AdsorptionSeparation, Sosa2023SupportingPotential, Sosa2024ExploringPotential} or on energetic and thermodynamic properties of adsorption \cite{El-Sharkawy2016AdsorptionCarbon, Askalany2014AdsorptionCarbons, Askalany2016HighlyDifluoromethane}. However, these studies typically investigate only a limited number of adsorbent/adsorbate systems and rarely combine adsorption thermodynamics, storage performance, and separation behavior within a unified framework. Sosa et al. (2020) presented the first systematic experimental study of pure refrigerant components on different activated carbons for the separation of commercial refrigerant blends \cite{Sosa2020AdsorptionSeparation}. They investigated the separation of R410A and R407F on activated carbons using Ideal Adsorbed Solution Theory (IAST) to predict mixture adsorption. Subsequent works extended these analyses to MOFs, zeolitic imidazolate frameworks (ZIFs), and biomass-derived activated carbons \cite{Sosa2023ExploringPotential, Sosa2024ExploringPotential}. These studies demonstrated the potential of porous materials for refrigerant separation and highlighted the superior adsorption capacities of some micro-mesoporous materials under high-pressure conditions. Nevertheless, they remain largely limited to experimental selectivity analyses and do not address adsorption-based energy storage or breakthrough separation dynamics.

In parallel, molecular simulation methods have become increasingly important for studying gas adsorption in porous materials. Grand Canonical Monte Carlo (GCMC) simulations are widely regarded as the standard approach for calculating adsorption isotherms in nanoporous systems \cite{Dubbeldam2013MONTECodes}. In GCMC simulations, temperature, volume, and chemical potential are fixed while the number of adsorbed molecules fluctuates to satisfy thermodynamic equilibrium. Although numerous numerical studies have investigated refrigerant adsorption in MOFs \cite{Garcia2021SystematicRefrigerants, Cai2020MolecularNanoparticles, Hu2025MolecularIRMOF-1}, comparatively few studies focus on activated carbons because of the challenges associated with representing their heterogeneous pore structures \cite{Beltran-Larrotta2025NewCarbon}. Moreover, numerical studies on refrigerant mixtures remain scarce and are generally restricted to separation analyses.\cite{Elhussien2025ExploringSimulations}.

To date, no study has systematically combined adsorption thermodynamics, energy storage performance, and separation behavior of refrigerant blends within a unified predictive framework. Furthermore, the adsorption behavior of refrigerant mixtures remains poorly understood, particularly regarding how the composition of the adsorbed phase differs from the bulk mixture and how individual refrigerant components contribute to the overall storage and separation performance.

To address these gaps, this work introduces a multiscale workflow for predicting adsorption, energy storage, and separation properties of refrigerant blends from pure-component adsorption data. The methodology combines adsorption potential theory (APT) \cite{Stavarache2024AdaptedMaterials}, Ideal Adsorbed Solution Theory (IAST), and models for the isosteric heat of adsorption. In addition, a Python-based software package was developed for calculating heats of adsorption and energy storage densities for both pure refrigerants and multicomponent mixtures.\cite{lucassen_2026_20134100} While individual tools exist for estimating adsorption thermodynamics \cite{Iacomi2019PyGAPS:Characterisation, Hassan2026AIM:Simulation, Dubbeldam2020RASPAMaterials}, no available framework combines adsorption equilibrium, heats of adsorption, storage density, and mixture thermodynamics into a unified workflow applicable to both simulations and experimental measurements.

The proposed methodology enables efficient screening of adsorbent/adsorbate working pairs and provides new insight into the adsorption thermodynamics of refrigerant blends. In particular, it allows the relationship between bulk and adsorbed-phase compositions to be analyzed systematically, opening new possibilities for the rational design of refrigerant blends tailored to specific porous materials and operating conditions. The workflow is applied to four commercial refrigerant mixtures, R410A, R407F, R417A, and R417C, composed of different fractions of R32, R125, R134a, and R600, adsorbed in six activated carbons: Bhatia-01, Bhatia-02, Bhatia-03, CS400, CS1000, and CS1000a. 

\section{Methodology}
\label{sec:methods}

The workflow developed in this work combines atomistic simulations, thermodynamic modeling, and fixed-bed breakthrough simulations to evaluate refrigerant adsorption systems for thermochemical energy storage and circular refrigerant recovery. First, pure-component adsorption isotherms are obtained from Grand Canonical Monte Carlo (GCMC) simulations using RASPA and fitted with analytical adsorption models using RUPTURA \cite{Dubbeldam2020RASPAMaterials, Sharma2023RUPTURA:Models}. These pure-component isotherms are then used as input for three complementary tasks: (i) the generation of adsorption isotherms at different temperatures using adsorption potential theory (APT), (ii) the calculation of isosteric heats of adsorption and gravimetric storage densities, and (iii) the prediction of multicomponent adsorption equilibria using ideal adsorbed solution theory (IAST). Finally, separation-relevant properties are evaluated through equilibrium selectivity, working capacity, separation factors, and breakthrough simulations.

A Python-based software package was developed to calculate isosteric heats of adsorption and storage densities for both pure-component and multicomponent adsorption systems.\cite{lucassen_2026_20134100} The implementation is provided in the Supplementary Material \ref{Github repository}. The framework is designed to use either simulation-derived or experimentally measured adsorption isotherms as input, enabling efficient screening of adsorbent/adsorbate working pairs.

This study analyzes the adsorption behavior of four commercial refrigerant blends: R407F (30 wt\% R32, 30 wt\% R125, and 40 wt\% R134a), R410A (50 wt\% R32 and 50 wt\% R125), R417A (3.4 wt\% R600, 46.6 wt\% R125, and 50 wt\% R134a), and R417C (1.7 wt\% R600, 19.5 wt\% R125, and 78.8 wt\% R134a). These blends are evaluated in six activated carbons: Bhatia-01, Bhatia-02, Bhatia-03, CS1000, CS400, and CS1000a.

Accurate prediction of blend properties requires prior characterization of the corresponding pure components. The saturation pressure ($P_s$) at 303 K, critical volume ($V_c$), critical temperature ($T_c$), and global warming potential (GWP) of the individual refrigerants are summarized in \autoref{tab:gas_properties}. R600 has a significantly lower GWP than the other refrigerants and has been reported to exhibit a high coefficient of performance (COP) \cite{Ibrahim2025HydrocarbonR600a}. However, its higher flammability requires additional safety considerations when used as a pure refrigerant. Its use at low concentrations in blends may therefore improve performance while remaining within established safety limits. The lower and upper explosive limits of R600 are 1.86--8.41 vol\% \cite{TheEngineeringToolboxGasesLimits}. Based on the individual GWP values, the refrigerant blends R407F, R410A, R417A, and R417C have GWPs of 2077, 1790, 2288, and 1745 CO$_2$-eq, respectively.

\begin{table}[t]
\centering
\scriptsize
\setlength{\tabcolsep}{2pt}

\begin{tabular}{l c c c c c}
\hline
\textbf{Molecule} & \textbf{Structure} &
\begin{tabular}{c}
\textbf{$P_s$} \\
(MPa)
\end{tabular} &
\begin{tabular}{c}
\textbf{$V_c$} \\
(cm$^3$/mol)
\end{tabular} &
\begin{tabular}{c}
\textbf{$T_c$} \\
(K)
\end{tabular} &
\begin{tabular}{c}
\textbf{$GWP$} \\
(CO$_2$-eq)
\end{tabular} \\
\hline

\begin{tabular}[c]{@{}l@{}}
R32 \\
(Difluoro- \\
methane)
\end{tabular}
&
\includegraphics[width=0.075\columnwidth]{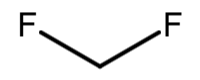}
& 1.93 & 120.8 & 351.26 & 675 \\

\begin{tabular}[c]{@{}l@{}}
R125 \\
(Pentafluoro- \\
ethane)
\end{tabular}
&
\includegraphics[width=0.075\columnwidth]{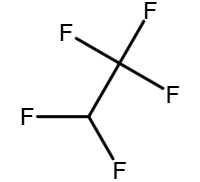}
& 1.57 & 211.3 & 339.7 & 3500 \\

\begin{tabular}[c]{@{}l@{}}
R134a \\
(1,1,1,2-Tetra- \\
fluoroethane)
\end{tabular}
&
\includegraphics[width=0.075\columnwidth]{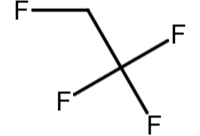}
& 0.77 & 198.8 & 374.21 & 1430 \\

\begin{tabular}[c]{@{}l@{}}
R600 \\
(Butane)
\end{tabular}
&
\includegraphics[width=0.075\columnwidth]{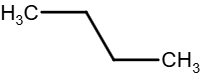}
& 3.80 & 255.0 & 425.13 & 3 \\

\hline
\end{tabular}

\caption{Properties of pure refrigerants \cite{NISTChemistry69}.}
\label{tab:gas_properties}

\end{table}

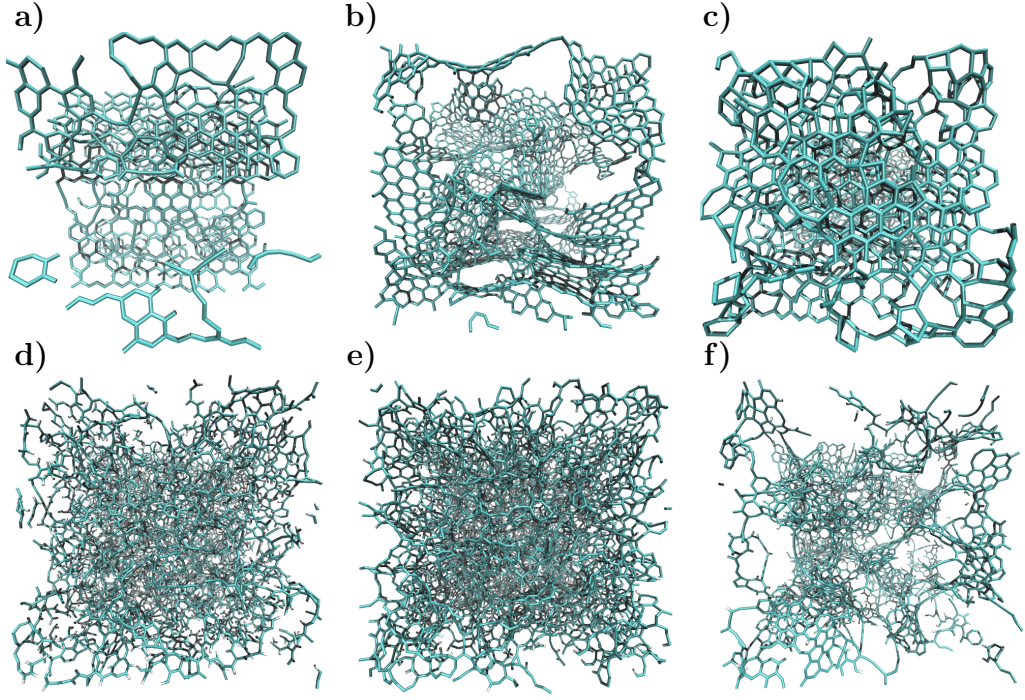
\begin{figure*}[t]
    \centering
    \begin{tikzpicture}
        \node[anchor=south west, inner sep=0] (image) at (0,0)
            {\includegraphics[width=0.75\linewidth]{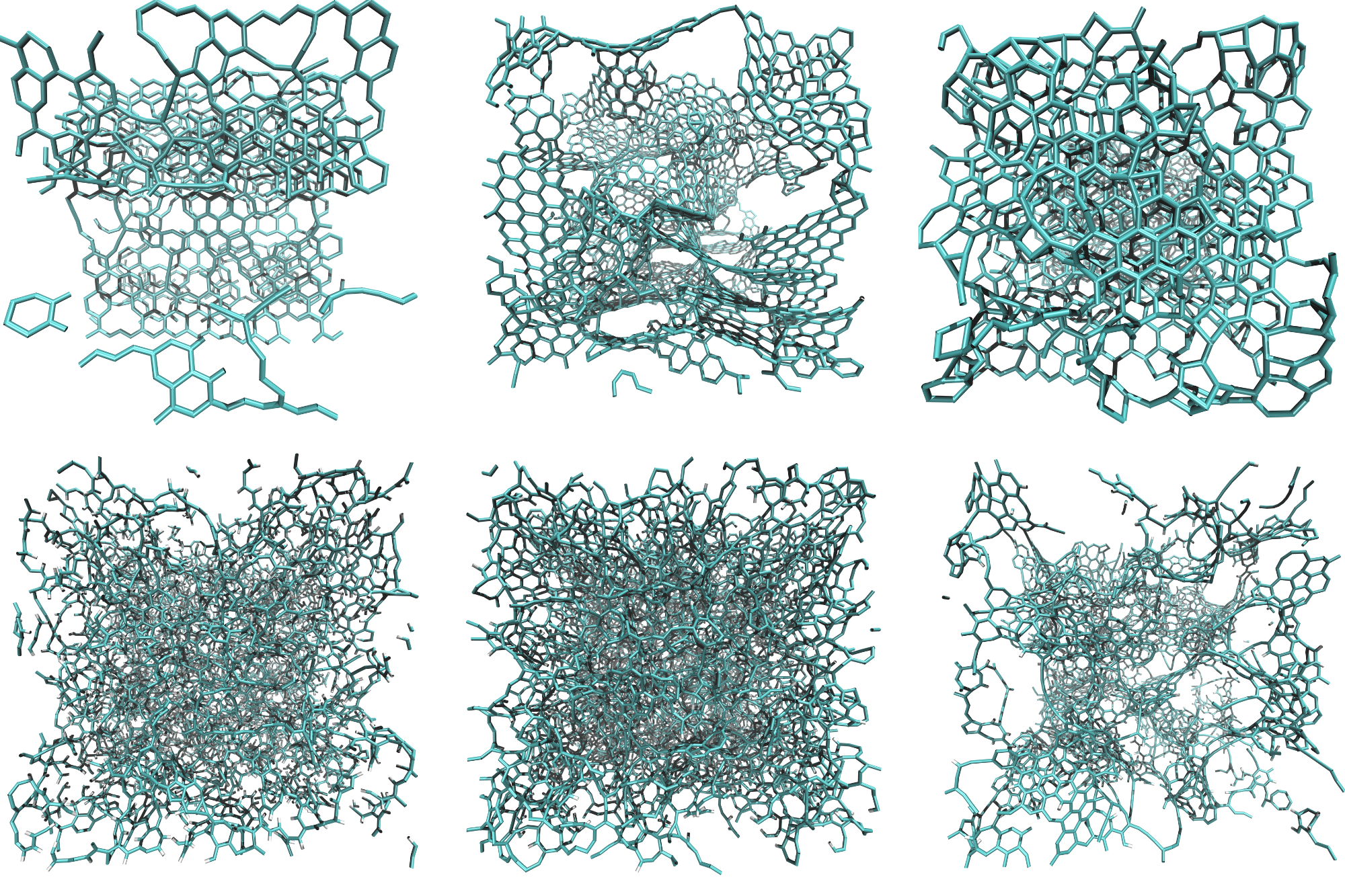}};
        \begin{scope}[x={(image.south east)}, y={(image.north west)}]

            \node[font=\large\bfseries, anchor=west] at (0.00,1.02) {a)};
            \node[font=\large\bfseries]             at (0.35,1.02) {b)};
            \node[font=\large\bfseries]             at (0.70,1.02) {c)};

            \node[font=\large\bfseries, anchor=west] at (0.00,0.52) {d)};
            \node[font=\large\bfseries]             at (0.35,0.52) {e)};
            \node[font=\large\bfseries]             at (0.70,0.52) {f)};

        \end{scope}
    \end{tikzpicture}

    \caption{Snapshot of Bhatia-01 (a), Bhatia-02 (b), Bhatia-03 (c), CS400 (d), CS1000 (e) and CS1000a (f).}
    \label{fig:AtomisticVisualization}
\end{figure*}

Bhatia-01, Bhatia-02, and Bhatia-03 are activated carbon models based on surface-based representations,\cite{Bathia-Langmuir-2008, Bathia-Carbon-2016, Bathia-JPCC_2017} whereas CS400, CS1000, and CS1000a are matrix-based activated carbon models.\cite{Jain-Langmuir-2006, JAIN20062445} The six structures are shown in \autoref{fig:AtomisticVisualization}, and their textural properties are summarized in \autoref{tab:ac_properties}. These properties are consistent with previous data reported in the literature \cite{Peng2018UnderstandingCarbons, Thyagarajan2020AMaterials, Peng2020SeparationSimulations}. Bhatia-01 and Bhatia-02 exhibit lower densities and larger surface areas than Bhatia-03, suggesting higher adsorption capacities. CS400 and CS1000 display dense structures with limited accessible porosity, whereas CS1000a combines larger accessible pores with a high surface area, suggesting a higher adsorption capacity than the other matrix-based activated carbons.

\begin{table}[H]
\centering
\scriptsize
\setlength{\tabcolsep}{3pt}
\renewcommand{\arraystretch}{0.98}

\begin{tabular}{l c c c c c}
\hline
\textbf{Material} &
\begin{tabular}{c}
\textbf{SA} \\
(m$^2$g$^{-1}$)
\end{tabular} &
\begin{tabular}{c}
\textbf{PV} \\
(cm$^3$g$^{-1}$)
\end{tabular} &
\begin{tabular}{c}
\textbf{PLD} \\
(\AA)
\end{tabular} &
\begin{tabular}{c}
\textbf{LCD} \\
(\AA)
\end{tabular} &
\begin{tabular}{c}
\textbf{Density} \\
(g/cm$^3$)
\end{tabular} \\
\hline

Bhatia-01 & 1895.9 & 0.690 & 7.94  & 11.06 & 876  \\
Bhatia-02 & 1622.6 & 0.634 & 6.93  & 13.16 & 951  \\
Bhatia-03 & 1173.3 & 0.503 & 11.39 & 15.14 & 1023 \\
CS400     & 305.3  & 0.186 & 2.56  & 6.48  & 1094 \\
CS1000    & 134.6  & 0.107 & 2.02  & 6.60  & 1497 \\
CS1000a   & 2678.0 & 0.846 & 7.39  & 11.94 & 727  \\

\hline
\end{tabular}

\caption{Structural properties of Bhatia-01, Bhatia-02, Bhatia-03, CS400, CS1000, and CS1000a.}
\label{tab:ac_properties}

\end{table}

\subsection{Pure-component adsorption simulations and isotherm fitting}

GCMC simulations were performed with RASPA to compute adsorption equilibrium points \cite{Dubbeldam2020RASPAMaterials}. Simulations were carried out from 283 to 353 K, with temperature intervals between 10 and 30 K, and pressures between 1 and $3\cdot10^{6}$ Pa distributed logarithmically. Further details on the simulation parameters are provided in the Supplementary Material \ref{subsec: Settings Raspa}.

Pure-component adsorption isotherms were fitted into an isotherm equation model using the RUPTURA software \cite{Sharma2023RUPTURA:Models}. Two adsorption models were considered: the multi-site Sips model and the multi-site Langmuir-Freundlich model.

The Sips model is expressed as
\begin{equation}\label{Eq: Sips}
    q(p)=\sum_i q_i^{sat}\frac{(bp)^{1/\nu}}{1+(bp)^{1/\nu}},
\end{equation}
where $q(p)$ $[mol/kg]$ is the loading as a function of pressure, $q_i^{sat}$ $[mol/kg]$ is the saturation loading of site $i$, $b$ $[Pa^{-1}]$ is the adsorption affinity, and $\nu$ $[-]$ is the heterogeneity parameter \cite{Sharma2023RUPTURA:Models}.

The multi-site Langmuir-Freundlich model is mathematically similar and is given by
\begin{equation}\label{eq: Langmuir-Freundlich model}
      q(p)=\sum_i q_i^{sat}\frac{bp^{\nu}}{1+bp^{\nu}}.
\end{equation}

The quality of the isotherm fits was evaluated using the coefficient of determination ($R^2$), the residual sum of squares (RSS), and the root-mean-square error (RMSE) \cite{ApXMachineLearningRegressionMetrics}.

\subsection{Adsorption potential theory}

The adsorption characteristic curve of an adsorbate/adsorbent pair relates the adsorption potential to the filling volume \cite{Stavarache2024AdaptedMaterials}. According to adsorption potential theory (APT), this characteristic curve is independent of temperature and pressure, enabling adsorption isotherms at different temperatures to be reconstructed from a reference isotherm. The adsorption potential is defined as
\begin{equation}\label{eq: adsorption potential}
    A=RT \ln\left(\frac{P_{sat}}{P}\right),
\end{equation}
where $R$ $[J/(mol\cdot K)]$ is the universal gas constant, $T$ $[K]$ is the temperature, $P_{sat}$ $[Pa]$ is the saturation pressure, and $P$ $[Pa]$ is the pressure of the adsorbate. The adsorption potential corresponds to the opposite value of the molar Gibbs free energy change.

The filling volume is calculated using the Dubinin expression
\begin{equation}\label{eq: Dubinin formula}
    W=\frac{q}{\rho_{ads}},
\end{equation}
where $q$ $[mol/kg]$ is the adsorbed amount and $\rho_{ads}$ $[kg/m^3]$ is the adsorbed-phase density.

The method of Stavarache et al. requires a reference isotherm and fluid properties of the adsorbate to construct the characteristic curve \cite{Stavarache2024AdaptedMaterials}. Fluid properties were obtained from the NIST Chemistry WebBook and include bulk densities, saturation pressures, critical isochores, and critical constants \cite{NISTChemistry69}. Since the studied conditions lie below the critical regime in both pressure and density, extrapolation was used to determine saturation pressures at temperatures not directly available in the database.

The adsorbed-phase density was estimated using Hauer's method,\cite{Hauer2010} which assumes a linear temperature dependence:
\begin{equation}\label{eq: Hauer density}
    \rho_{ads}=\rho_r(1-\alpha(T-T_r)),
\end{equation}
where $T$ $[K]$ is the temperature, $T_r$ $[K]$ is the reference temperature, $\rho_r$ $[kg/m^3]$ is the density at the reference temperature, and $\alpha$ $[K^{-1}]$ is the thermal expansion coefficient.

APT has previously been validated for simple gases, such as CO$_2$ and CH$_4$, adsorbed in MOFs \cite{Stavarache2024AdaptedMaterials}. In the present work, APT is applied to refrigerant/activated-carbon systems, which are more complex due to the polarity, polarizability, and molecular size of the refrigerants, as well as the heterogeneous pore structures of activated carbons. The APT-derived isotherms are subsequently evaluated not only for loading prediction, but also as input for isosteric heat, storage density, and IAST-based mixture calculations.

\subsection{Isosteric heat of adsorption}

The isosteric heat of adsorption, $Q_{st}$ $[kJ/mol]$, corresponds to the positive value of the adsorption enthalpy. In RASPA, $Q_{st}$ can be obtained directly from energy and loading fluctuations in the GCMC ensemble \cite{Dubbeldam2013MONTECodes, Dubbeldam2020RASPAMaterials}:
\begin{equation}\label{eq: fluctuation heat}
\begin{aligned}
    -Q_{st} &= \Delta H \\
       &= \frac{\left\langle U N \right\rangle_\mu-\left\langle U \right\rangle_\mu \left\langle N \right\rangle_\mu}{\left\langle N^2 \right\rangle_\mu-\left\langle N \right\rangle_\mu^2}
       -\left\langle U_g \right\rangle - RT ,
\end{aligned}
\end{equation}
where $\langle \rangle_{\mu}$ denotes the ensemble average in the GCMC ensemble, $U$ $[kJ/mol]$ is the total internal energy of the simulation box, $U_g$ $[kJ/mol]$ is the internal energy associated with the guest molecules, $N$ is the number of guest molecules, and $Q_{st}$ is the heat released upon adsorption. Although the fluctuation method provides direct access to adsorption enthalpies, it does not yield a smooth continuous curve and may suffer from larger uncertainties at high loading \cite{Torres-Knoop2017BehaviorConditions}.

To obtain continuous heat-of-adsorption profiles from adsorption isotherms, two analytical methods were implemented: the Clausius--Clapeyron approach and the Virial method. The Clausius--Clapeyron equation is written as \cite{Islam2024HarnessingSystems}
\begin{equation}\label{eq: Clausius Claperyon}
    Q_{st}=-RT^2\left(\frac{\partial \ln(P)}{\partial T}\right)_q ,
\end{equation}
where $P$ $[Pa]$ is the pressure and $q$ $[mol/kg]$ is the loading. The derivative $\left(\partial \ln(P)/\partial T\right)_q$ was computed using two approaches: a fitting-based method and a hybrid interpolation method combining Piecewise Cubic Hermite Interpolating Polynomial (PCHIP) with linear interpolation \cite{MATLABPiecewisePCHIP}. At least three temperatures were required for each loading point, and only fits with $R^2 > 0.95$ were accepted. Recent work by Heinselman et al. (2026) highlighted methodological limitations and uncertainties associated with Clausius-Clapeyron-derived heats of adsorption, particularly regarding the sensitivity to the adsorption isotherms quality and experimental reproducibility \cite{Heinselman2026AnApproach}. In the present work, these uncertainties are partially reduced by using simulated adsorption isotherms, which, for these types of molecules, provide highly equilibrated and internally consistent adsorption data under controlled thermodynamic conditions. Nevertheless, care must still be taken when comparing the calculated heats of adsorption with values reported in the literature.

The Virial equation expresses the adsorption isotherm as
\begin{equation}\label{eq: Virial}
    \ln\left(\frac{P}{n}\right)=\frac{1}{T}\sum_{j=0}^l a_jn^j+\sum_{j=0}^m b_jn^j ,
\end{equation}
where $n$ $[mol/kg]$ is the adsorbed amount, and $a_j$ and $b_j$ are polynomial coefficients. The Virial formulation assumes
\begin{equation}\label{eq:VirialAssumptions}
\begin{aligned}
    \left[ \frac{\partial \ln P}{\partial (1/T)} \right]_n &= g(n), \\
    \left[ \frac{\partial g(n)}{\partial T} \right]_n &= 0.
\end{aligned}
\end{equation}
The isosteric heat is then obtained as
\begin{equation}\label{eq: virial isosteric enthalpy}
    Q_{st}(n)=-R \sum_{j=0}^l a_jn^j .
\end{equation}

The polynomial coefficients were obtained through a global multi-temperature fit combined with a BFGS optimizer \cite{Brownlee2021AAlgorithm}. Since no general rule exists for selecting the polynomial orders, candidate polynomial combinations were evaluated using their goodness of fit, and the $R^2$ value was used to guide the selection. Following recommendations by Nuhnen et al. \cite{Nuhnen2020AMOFs}, particular care was taken in low-pressure regions where Virial fits can become unstable. For consistency with the Clausius--Clapeyron analysis, at least three temperatures per loading point were required.

The implementation of the heat-of-adsorption calculation was validated in the Supplementary Material  \ref{appendix: Validation code} by benchmarking against experimental and simulation data.\cite{Gooijer2025TAMOF-1Components, Queen2014ComprehensiveZn}

\subsection{Energy storage density}

The energy storage density is defined as the total energy stored between adsorption and desorption states per unit mass of adsorbent. It is calculated as the integral of the adsorption enthalpy over the working capacity \cite{Vicent-Luna2024AdsorptionFrameworks}:
\begin{equation}\label{eq: Storage density}
    SD = \int_{q_{des}}^{q_{ads}}\Delta h(q)\,dq .
\end{equation}

Three operating modes that mirror pressure-swing adsorption (PSA), temperature-swing adsorption (TSA), and combined pressure-temperature-swing adsorption (PTSA) conditions were considered. In PSA, adsorption and desorption occur at the same temperature but different pressures. In TSA, adsorption and desorption occur at the same pressure but different temperatures. In PTSA, both pressure and temperature are varied.

For thermochemical energy storage, the adsorption state satisfies $P_{ads} > P_{des}$ and $T_{ads} < T_{des}$. In this work, the maximum storage density was evaluated for different adsorption pressures. The adsorption temperature was set to $T_{ads}=283$ K, and adsorption pressures between 100 and 500 kPa were considered. The minimum desorption pressure was set to $P_{des}=1$ Pa to estimate the maximum theoretical working capacity, defined as the difference in loading between adsorption and desorption conditions.

\subsection{Multicomponent adsorption and mixture heats of adsorption}

Mixture adsorption is defined by the equilibrium uptake of each component when multiple species compete for adsorption sites. Multicomponent adsorption was predicted using IAST with RUPTURA \cite{Sharma2023RUPTURA:Models} and benchmarked against explicit mixture GCMC simulations performed with RASPA \cite{Dubbeldam2020RASPAMaterials}. IAST requires only fitted pure-component adsorption isotherms as input and predicts mixture loadings and adsorbed-phase compositions. However, it does not directly provide mixture enthalpies. Therefore, two approaches were implemented to estimate the isosteric heat of adsorption of mixtures from IAST-derived adsorption data.

The first approach is based on a Clausius--Clapeyron expression for mixture adsorption:
\begin{equation}\label{eq: clausius clapeyron mixture}
    Q_{st,i}^{mix}=RT^2\left(\frac{\partial \ln(P y_i)}{\partial T}\right)_{n_i},
\end{equation}
where $i$ denotes the component in the mixture, and $y_i$ is the adsorbed-phase mole fraction of component $i$ at pressure $P$ \cite{Hamid2023EstimationIsotherm}. For the overall mixture heat of adsorption, $y_i=1$. The same filtering procedure used for pure-component Clausius--Clapeyron calculations was applied, including the requirement of at least three temperatures and an acceptance criterion of $R^2 > 0.95$.

The second approach uses a linear mixing rule, where the mixture heat of adsorption is obtained by weighting the pure-component isosteric heats by their adsorbed-phase mole fractions:
\begin{equation}\label{eq: linear mixing rule}
    Q_{st}^{mix}=\sum_i y_iQ_{st,i}^{0},
\end{equation}
where $Q_{st,i}^{0}$ is the isosteric heat of adsorption of pure component $i$. This approach assumes ideal adsorbed-phase behavior and neglects cross-interactions between adsorbed components. By contrast, the Clausius--Clapeyron-based mixture approach is not explicitly restricted to ideal mixtures but depends on the accuracy of the underlying mixture adsorption data.

The agreement between the IAST-derived heat-of-adsorption estimates and the GCMC reference values was evaluated using the average relative deviation (ARD\%) \cite{Hamid2023EstimationIsotherm}:
\begin{equation}\label{eq: ARD}
    ARD\%=\frac{100}{N}\sum_{l=1}^N\left|\frac{Q_{st,l}^{IAST}-Q_{st,l}^{GCMC}}{Q_{st,l}^{GCMC}}\right|,
\end{equation}
where $N$ is the number of points considered.

\subsection{Separation performance and breakthrough modeling}

The adsorption-based separation performance of each mixture was evaluated using working capacity, productivity, adsorption selectivity, separation factor, and breakthrough curves. The working capacity of component $i$ is defined as \cite{Gonzalez-Galan2024UnderstandingMixtures}
\begin{equation}\label{Eq: working capacity}
    \Delta q_i=q_{ads,i}-q_{des,i}=q_i(p_{ads},T_{ads})-q_i(p_{des},T_{des}) .
\end{equation}
The fractional contribution of component $i$ to the total working capacity is then given by
\begin{equation}\label{Eq: total working capacity}
    P_i=\frac{\Delta q_i}{\sum_i \Delta q_i}.
\end{equation}

For a binary mixture of components $i$ and $j$, the adsorption selectivity is calculated as \cite{Gonzalez-Galan2024UnderstandingMixtures}
\begin{equation}\label{Eq: adsorption selectivity}
    S_{ads}(i/j)=\frac{y_i/y_j}{x_i/x_j},
\end{equation}
where $y_i$ and $y_j$ are adsorbed-phase mole fractions, and $x_i$ and $x_j$ are bulk-phase mole fractions. For ternary and quaternary mixtures, the adsorption selectivity is calculated as \cite{Bessa2023AnProducts}
\begin{equation}\label{eq: multicomponent selectivity}
    S_{ads}=\frac{q_i(1-y_i)}{y_i\sum_{j=1,j\neq i}^n q_j}.
\end{equation}

To combine adsorption selectivity with working capacity, the separation factor is defined as \cite{Gonzalez-Galan2024UnderstandingMixtures}
\begin{equation}\label{Eq: separation factor}
    \alpha(i/j)=\ln(S_{ads}(i,j))\Delta q_i .
\end{equation}
This factor reduces the influence of extreme selectivity values that can arise when one component is only weakly adsorbed.

The dynamic separation behavior was evaluated through fixed-bed breakthrough simulations using a Linear Driving Force (LDF) model implemented in RUPTURA \cite{Sharma2023RUPTURA:Models}. In this approach, a gas mixture flows through a cylindrical column packed with adsorbent particles, and the outlet composition evolves with time as the bed approaches saturation. The resulting breakthrough curves provide information on competitive adsorption, mass-transfer effects, and the temporal separation of mixture components. For the breakthrough simulations, the column void fraction was set to 0.40, the operating pressure to $100$ kPa, the pressure gradient to 0, the inlet velocity to $0.10$ m$\cdot$s$^{-1}$, and the column length to 0.10 m.

\begin{figure}[t]
    \centering
    \begin{tikzpicture}
        \node[anchor=south west, inner sep=0] (image) at (0,0)
            {\includegraphics[width=\linewidth]{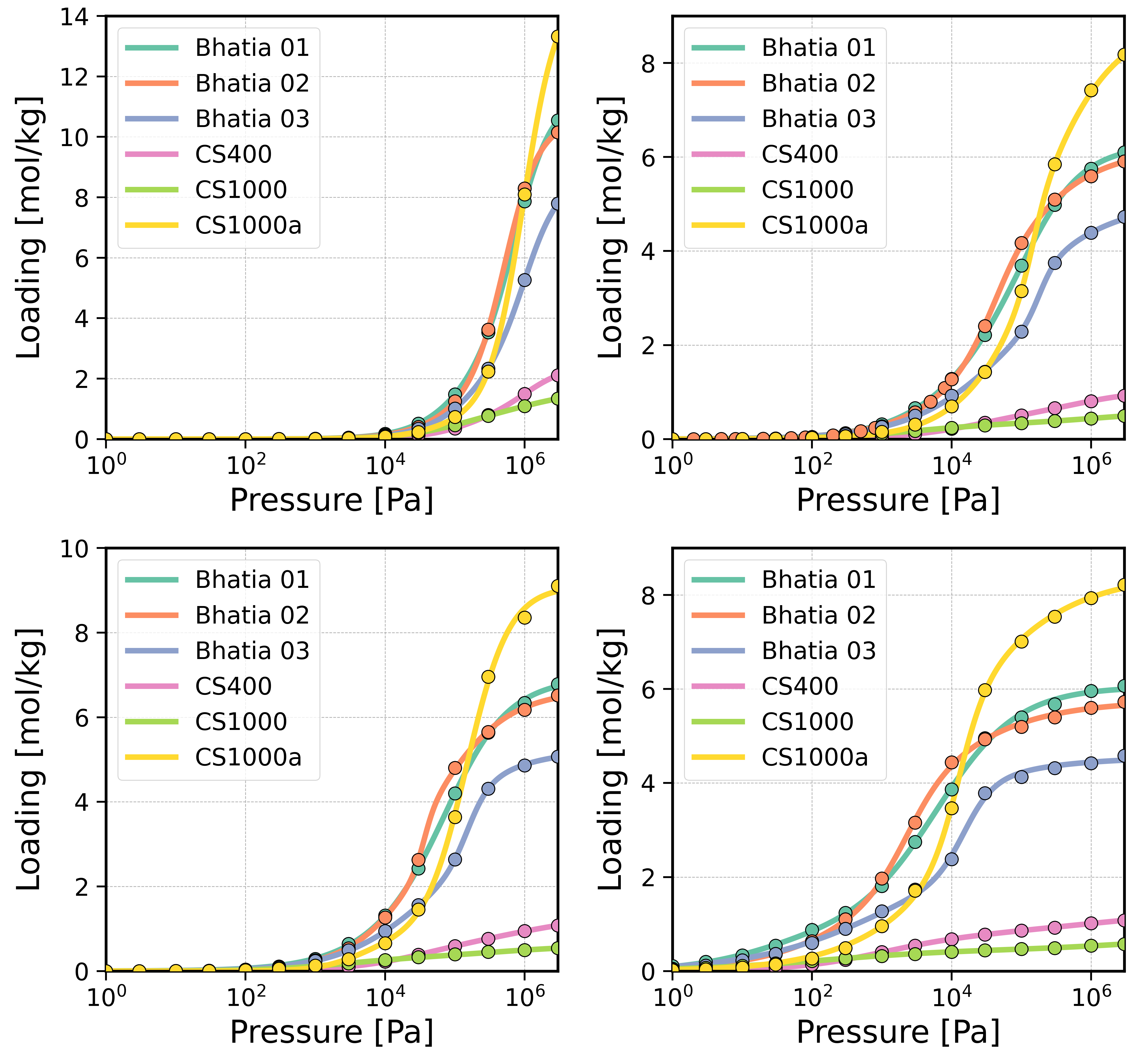}};
            
        \begin{scope}[x={(image.south east)}, y={(image.north west)}]
            \node[font=\bfseries, anchor=west] at (0.00,1.02) {a)};
            \node[font=\bfseries]             at (0.55,1.02) {b)};

            \node[font=\bfseries, anchor=west] at (0.00,0.52) {c)};
            \node[font=\bfseries]             at (0.55,0.52) {d)};
        \end{scope}
    \end{tikzpicture}

    \caption{Adsorption isotherms of R32 (a), R125 (b), R134a (c), and R600 (d) at $303$ K, comparing GCMC data ($\bullet$), and dual-site Sips model (---).}
    
    \label{fig:Adsorption_isotherms_of_all_adsorbents_adsorbates}
\end{figure}
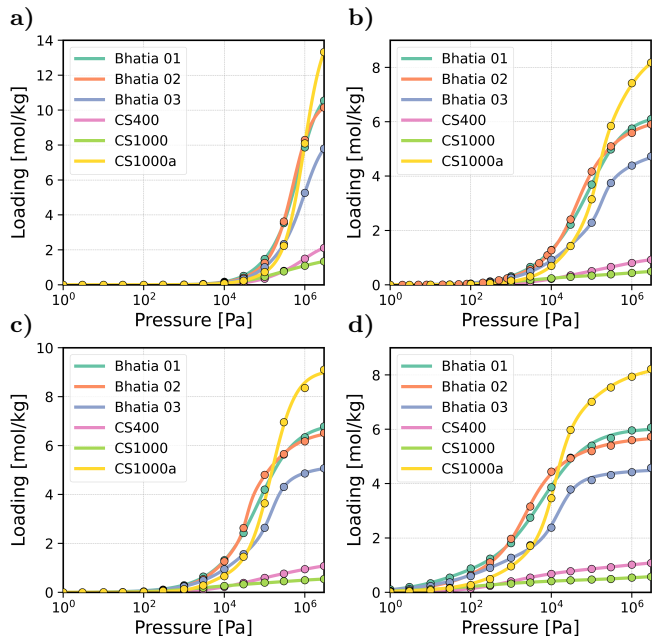

\section{Results and Discussion}
\label{sec:results}


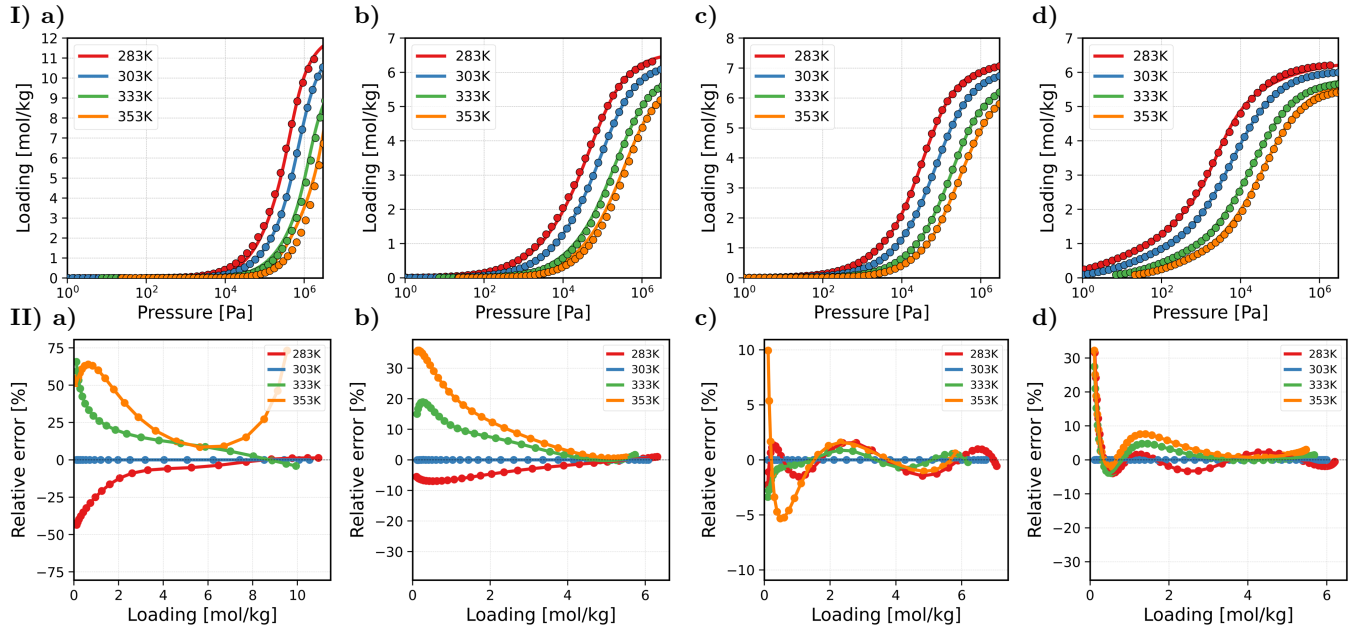
\begin{figure*}[t]
    \centering
    \begin{tikzpicture}
        \node[anchor=south west, inner sep=0] (image) at (0,0)
            {\includegraphics[width=\linewidth]{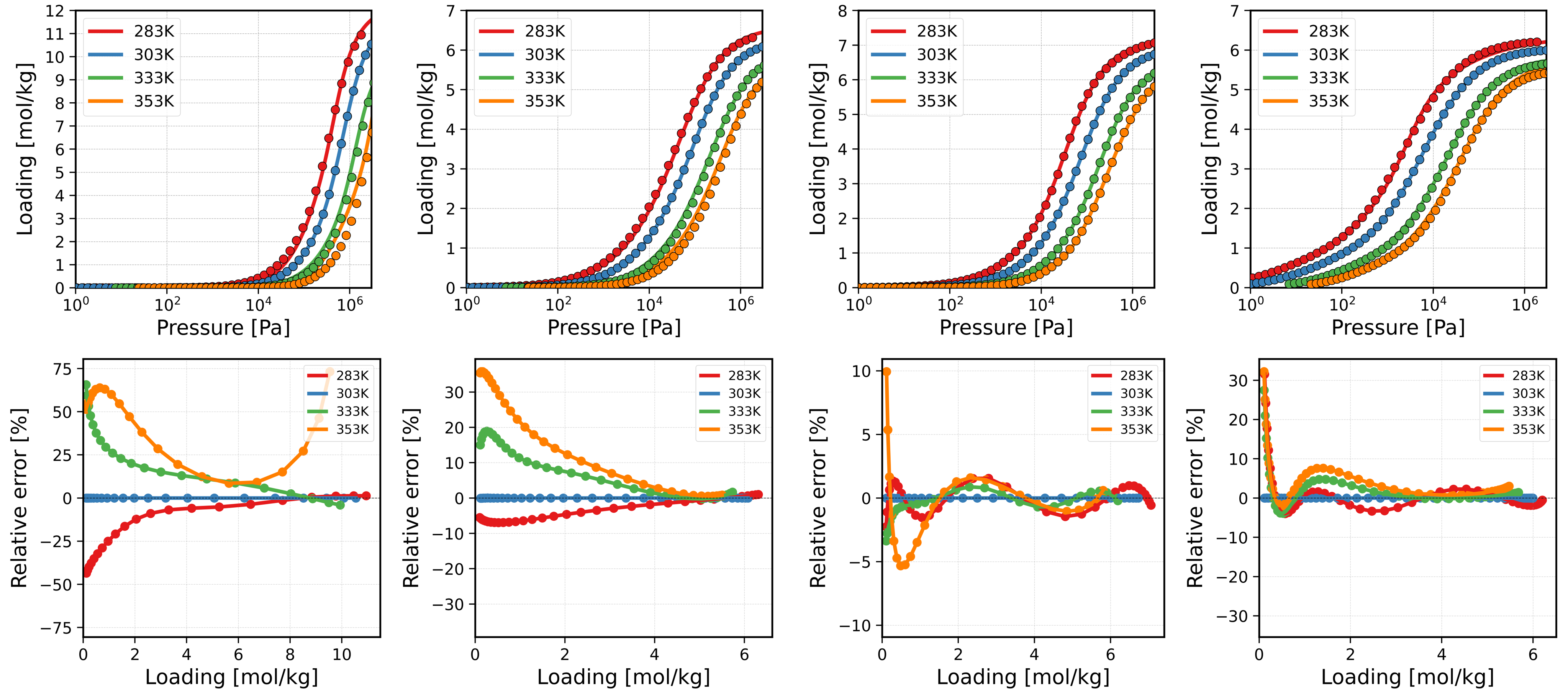}};
            
        \begin{scope}[x={(image.south east)}, y={(image.north west)}]
            \node[font=\bfseries, anchor=west] at (0.00,1.02) {I) a)};
            \node[font=\bfseries]             at (0.27,1.02) {b)};
            \node[font=\bfseries]             at (0.52,1.02) {c)};
            \node[font=\bfseries]             at (0.77,1.02) {d)};

            \node[font=\bfseries, anchor=west] at (0.00,0.52) {II) a)};
            \node[font=\bfseries]             at (0.27,0.52) {b)};
            \node[font=\bfseries]             at (0.52,0.52) {c)};
            \node[font=\bfseries]             at (0.77,0.52) {d)};
        \end{scope}
    \end{tikzpicture}

    \caption{Comparison of adsorption isotherms (I) dual-site Sips fitting of GCMC data(---) and APT prediction ($\bullet$) and (II) Relative error for Bhatia-01 at $283$-$353$ K for R32 (a), R125 (b), R134a (c), and R600 (d). Lines in (II) are visual guides.}

    \label{fig:Bathia01_APT_validation}
\end{figure*}

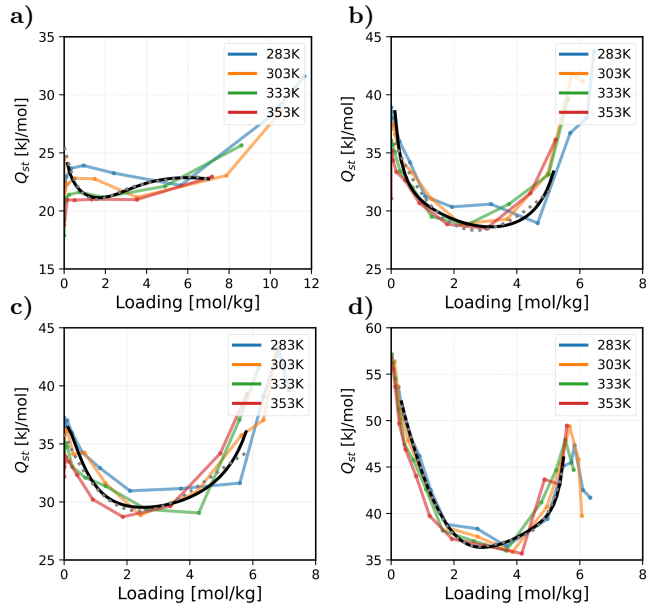
\begin{figure}[t]
    \centering

    \begin{tikzpicture}
        \node[anchor=south west, inner sep=0] (image) at (0,0)
            {\includegraphics[width=\linewidth]{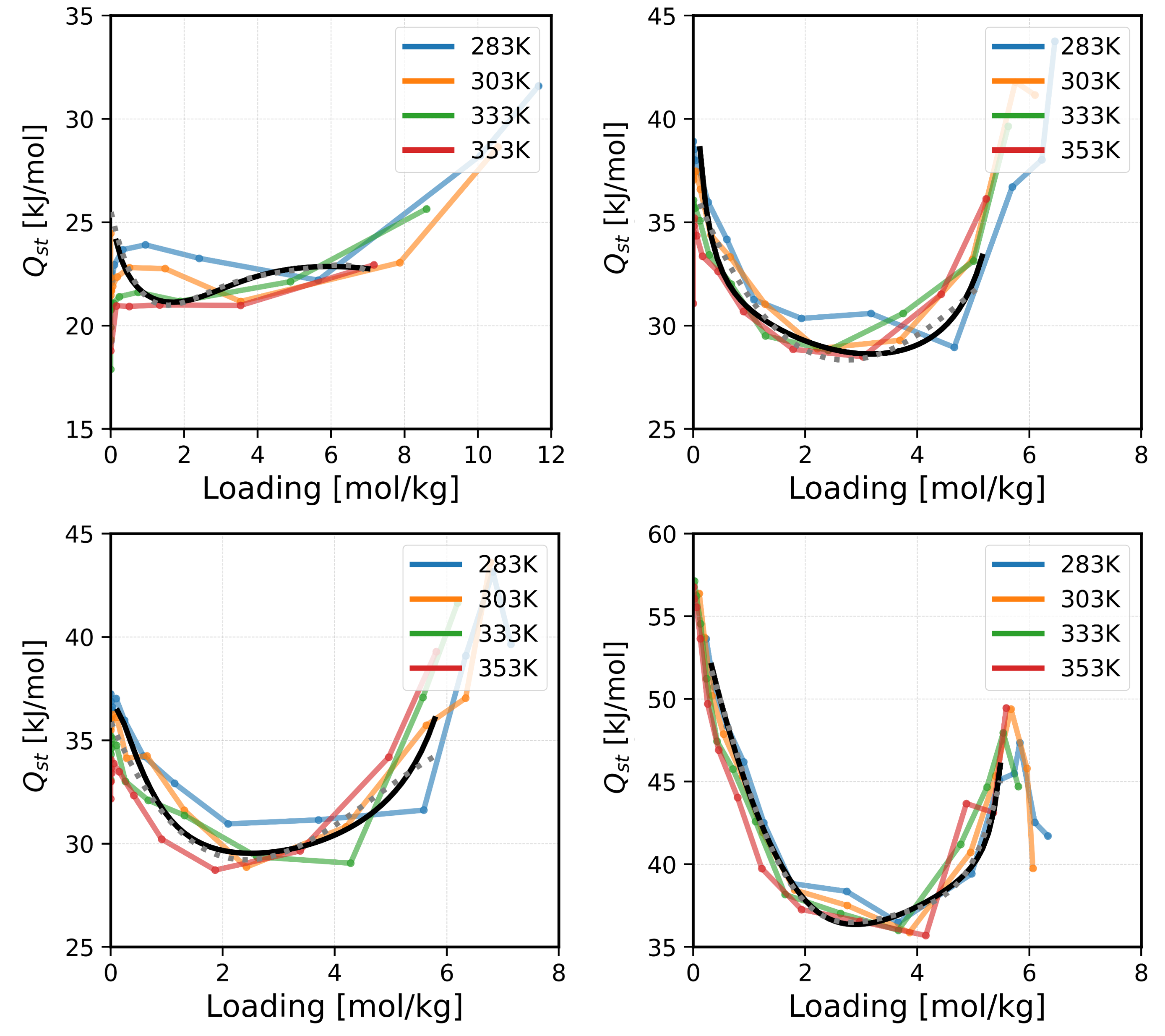}};
            
        \begin{scope}[x={(image.south east)}, y={(image.north west)}]
            \node[font=\bfseries, anchor=west] at (0.00,1.02) {a)};
            \node[font=\bfseries]             at (0.55,1.02) {b)};

            \node[font=\bfseries, anchor=west] at (0.00,0.52) {c)};
            \node[font=\bfseries]             at (0.55,0.52) {d)};
        \end{scope}
    \end{tikzpicture}

    \caption{Isosteric heat of adsorption for R32 (a), R125 (b), R134a (c), and R600 (d) in Bhatia-01 using the Clausius-Clapeyron equation (---), the Virial equation (-- --), and GCMC data ($-\bullet$).}

    \label{fig:validation_heats_adsorbate_adsorbent}
\end{figure}

We first evaluate the adsorption behavior of the pure refrigerants in the six activated carbons. \autoref{fig:Adsorption_isotherms_of_all_adsorbents_adsorbates} shows the simulated adsorption isotherms of R32, R125, R134a, and R600 at $303$ K for all adsorbents, together with the corresponding fits to the dual-site Sips isotherm model. The dual-site Langmuir-Freundlich model produced comparable but slightly poorer fits, as reported in the Supplementary Material \ref{tab: Fit quality metrics for the Dual-site Sips and Dual-site Langmuir-Freundlich models.}. Therefore, the dual-site Sips model is used throughout the remainder of this work. The corresponding fitting parameters are provided in the Supplementary Material \ref{sec: Dual-site Sips parameters}. All refrigerants exhibit similar adsorption trends within each activated carbon. CS400 and CS1000 show low loadings over the entire pressure range, indicating limited adsorption capacity, in agreement with their restricted accessible porosity. In contrast, CS1000a exhibits the highest uptake among the investigated materials. Bhatia-01 and Bhatia-02 show comparable adsorption capacities, whereas Bhatia-03 exhibits lower uptake. These trends are consistent with the textural properties of the activated carbons discussed above. Adsorption isotherms at the remaining temperatures are provided in the Supplementary Material \ref{supplementary fig: Adsorption isotherms pure components other structures}.

The differences between refrigerants can be rationalized from their molecular size and adsorption affinity. R32, the smallest molecule, starts adsorbing at higher pressures than the other refrigerants but reaches the highest adsorption capacity because it can access smaller pores. Its higher adsorption pressure is consistent with weaker surface interactions resulting from its smaller contact area with the carbon surface. In CS1000a, the uptake of R32 increases across the full pressure range, suggesting that saturation has not yet been reached. R125 and R134a exhibit similar capacities and adsorption pressures, consistent with their comparable molecular sizes and shapes. R600 starts adsorbing at lower pressures but reaches saturation at lower capacities because it is the largest molecule among the four refrigerants.

To compute heats of adsorption from adsorption data, isotherms at different temperatures are required. In this context, adsorption potential theory (APT) is evaluated as a strategy to predict adsorption isotherms at multiple temperatures from a single reference isotherm and its characteristic curve. In \autoref{fig:Bathia01_APT_validation}, the adsorption isotherms of R32, R125, R134a, and R600 in Bhatia-01 at $283$, $333$, and $353$ K are predicted using the adsorption isotherm at $303$ K as reference. The predicted isotherms agree closely with the GCMC simulations. Similar behavior is observed for the remaining working pairs, as shown in the Supplementary Material \ref{Supplementary: APT Isotherms}.

Despite this good agreement in absolute loading, numerical deviations between the APT predictions and GCMC reference data can affect subsequent thermodynamic post-processing. \autoref{fig:Bathia01_APT_validation} also shows the relative error as a function of adsorption capacity. The relative error is largest at low loading, particularly in the Henry regime, where small absolute deviations translate into large relative errors. Small inconsistencies between characteristic curves and increasing temperature differences from the reference isotherm also contribute to these deviations. This behavior is relevant because derivative-based thermodynamic methods, such as the Clausius--Clapeyron and Virial equations, are sensitive to small differences between isotherms. The corresponding relative errors for the remaining working pairs are provided in the Supplementary Material \ref{Supplementary: APT Isotherms}.

Overall, APT provides accurate adsorption loadings over a broad range of conditions, particularly at medium-to-high pressures and for temperature differences up to approximately $30$ K from the reference isotherm. R134a and R600 show the closest agreement with the GCMC values, followed by R125 and R32. The relative errors also depend on the adsorbent, with larger deviations generally observed for systems with higher adsorption capacities. These results show that APT is suitable for predicting adsorption capacities and for comparative screening, but caution is required when using APT-derived isotherms as input for heat-of-adsorption calculations.

The isosteric heat of adsorption is then calculated using the Clausius-Clapeyron equation (\autoref{eq: Clausius Claperyon}) and the Virial equation (\autoref{eq: Virial}). These calculations are performed with an open source Python-based software package developed in this work,\cite{lucassen_2026_20134100} which is validated against experimental and simulation data available in the literature (see the Supplementary Material  \ref{appendix: Validation code}). To assess the applicability of these methods to the refrigerant/activated-carbon working pairs, the heats of adsorption obtained from the fluctuation method in RASPA are used as reference. The fluctuation method provides direct molecular-scale information on the interaction strength between adsorbates and adsorbents at different loadings. However, the resulting data may contain statistical noise and larger uncertainties, especially at high pressures, which limits its use for smooth post-processing. The Clausius-Clapeyron and Virial methods are therefore used to obtain continuous heat-of-adsorption profiles. \autoref{fig:validation_heats_adsorbate_adsorbent} compares the fluctuation, Clausius-Clapeyron, and Virial method results for each pure refrigerant in Bhatia-01. The three approaches show good agreement. For the Virial method, the software suggests the polynomial orders $(a_j,b_j)$ used in the fit. The suggested orders are used for R125 $(3,2)$, R134a $(4,3)$, and R600 $(3,2)$. For R32, the suggested order $(1,0)$ was replaced by $(5,5)$ to improve the fit quality.

Both the Clausius-Clapeyron and Virial equations show sensitivity to the input adsorption isotherms. Although the isosteric heat of adsorption should be independent of the specific temperature combination used, different selections of temperature-dependent isotherms lead to variations in the calculated heat profiles; see Figure \ref{subsection: Sensitivity Temperature Selection simulations} in the Supplementary Material. Consequently, the small relative deviations introduced by APT can be amplified when APT-derived isotherms are used for the heat of adsorption calculations. The use of the dual-site Langmuir-Freundlich model instead of the dual-site Sips model produces negligible differences because both models fit the isotherms with comparable accuracy.

The pure-component heats of adsorption follow consistent trends across all activated carbons. R600 exhibits the highest heat of adsorption, R125 and R134a show comparable values, and R32 has the lowest heat of adsorption. This trend is consistent with the molecular sizes of the refrigerants. Larger molecules have greater contact areas with the carbon surface and therefore experience stronger dispersion interactions, resulting in higher isosteric heats of adsorption.

\begin{figure*}[t]
    \centering

    \begin{tikzpicture}
        \node[anchor=south west, inner sep=0] (image) at (0,0)
            {\includegraphics[width=0.9\linewidth]{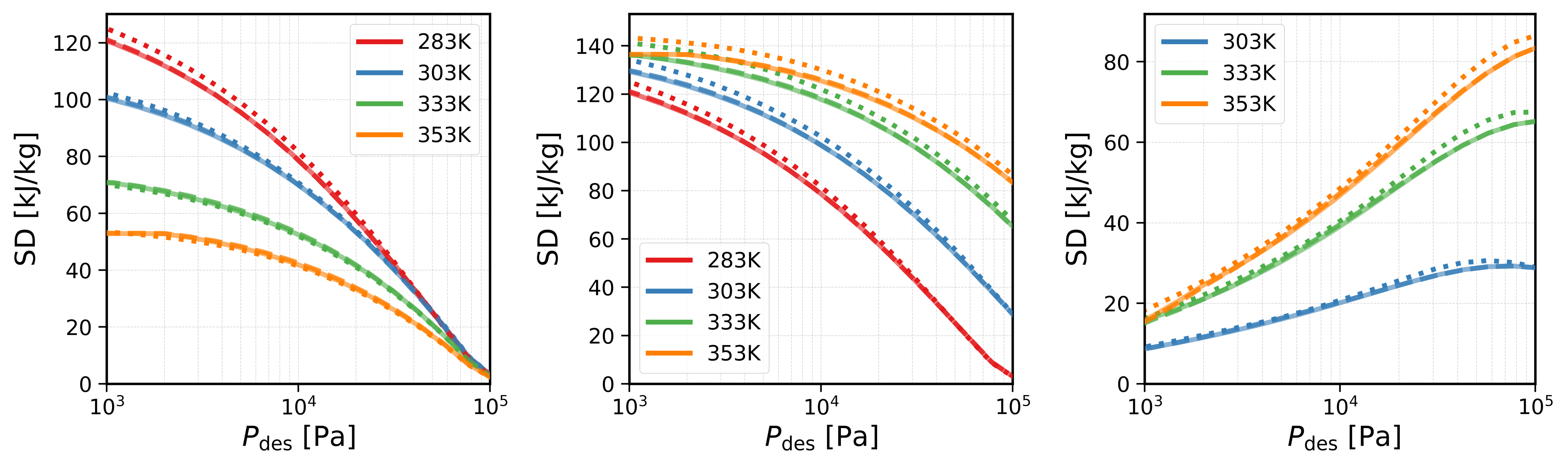}};
            
        \begin{scope}[x={(image.south east)}, y={(image.north west)}]
            \node[font=\bfseries, anchor=west] at (0.00,1.02) {a)};
            \node[font=\bfseries]             at (0.35,1.02) {b)};
            \node[font=\bfseries]             at (0.70,1.02) {c)};
        \end{scope}
    \end{tikzpicture}

    \caption{Storage density validation for Bhatia-01 with R125 at $283$--$353$ K for PSA ($P_{ads}=100$ kPa) (a), PTSA ($P_{ads}=100$ kPa, $T_{ads}=283$ K) (b), and TSA ($T_{ads}=283$ K) (c), using Clausius-Clapeyron (---), Virial (-- --), and fluctuation method (:).}

    \label{fig:Storage_density_validation_Bhatia01_R125}
\end{figure*}
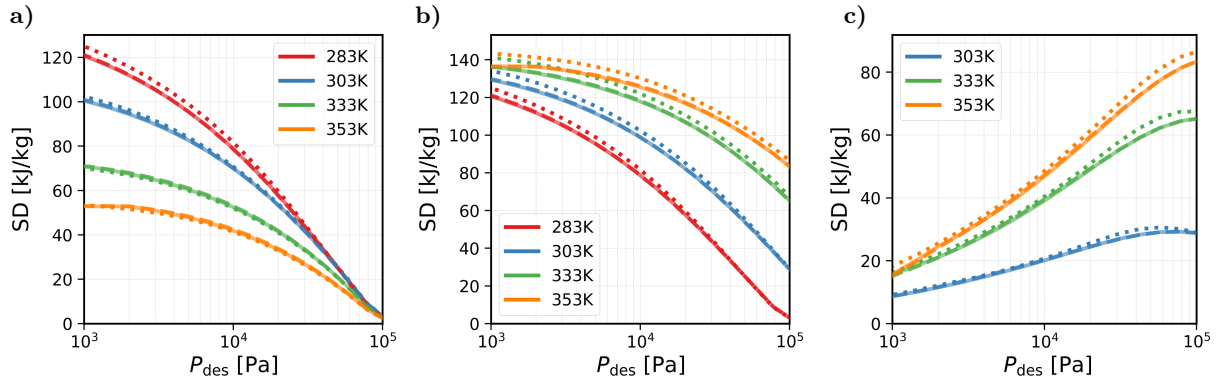

\begin{figure}[H]
    \centering

    \begin{tikzpicture}
        \node[anchor=south west, inner sep=0] (image) at (0,0)
            {\includegraphics[width=0.95\linewidth]{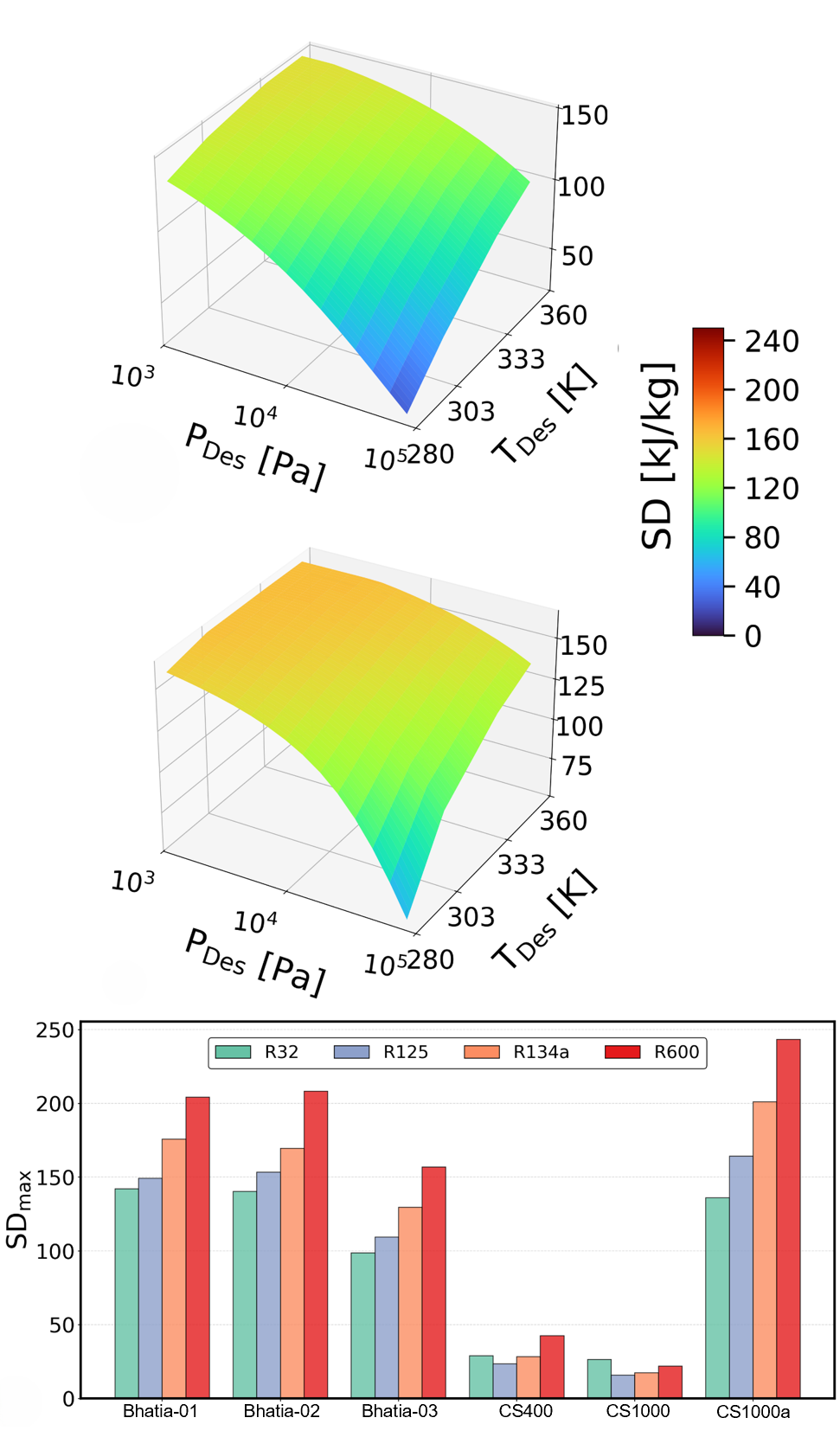}};
            
        \begin{scope}[x={(image.south east)}, y={(image.north west)}]
            \node[font=\bfseries, anchor=west] at (0.02,1.02) {a)};
            \node[font=\bfseries, anchor=west] at (0.02,0.70) {b)};
            \node[font=\bfseries, anchor=west] at (0.02,0.35) {c)};
        \end{scope}
    \end{tikzpicture}

    \caption{Storage density 3D plots of R125 in Bhatia-01 (a) and CS1000a (b) for PTSA ($P_{ads}=100$ kPa, $T_{ads}=283$ K), and maximum storage density per working pair (c).}

    \label{fig:storage_density_3D_plots}
\end{figure}
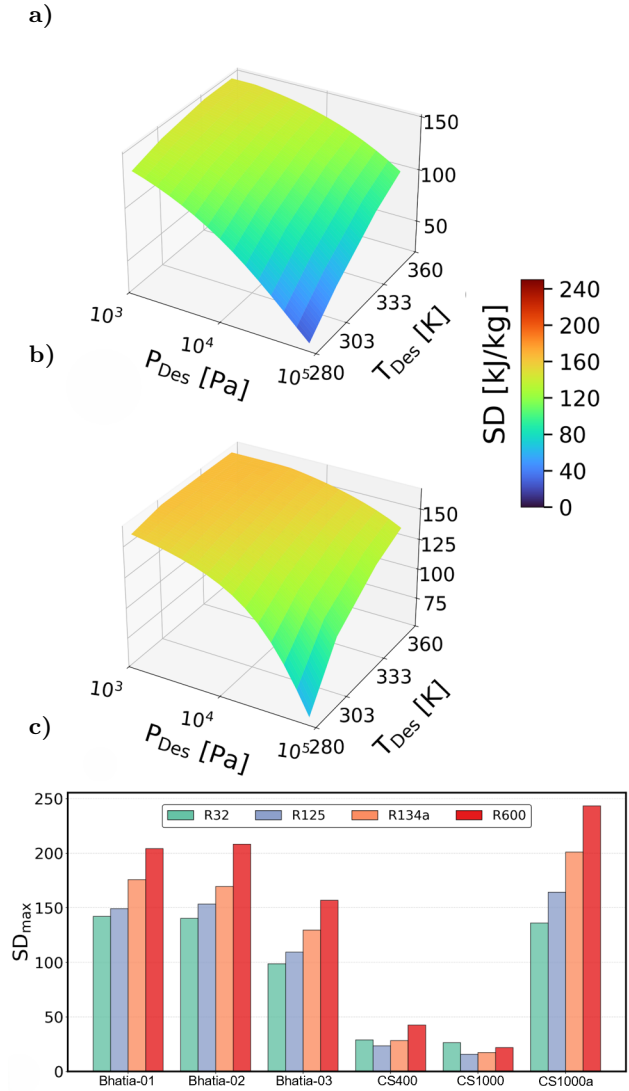

The energy storage density is calculated by integrating the heat of adsorption between the desorption and adsorption states. Three representative operating modes are considered: pressure-swing adsorption (PSA), temperature-swing adsorption (TSA), and pressure-temperature-swing adsorption (PTSA). The storage density calculation is validated by comparing the values obtained from the Clausius-Clapeyron and Virial methods with those obtained from the fluctuation method.

\autoref{fig:Storage_density_validation_Bhatia01_R125} compares the three approaches for Bhatia-01/R125 under different operating modes. \autoref{fig:Storage_density_validation_Bhatia01_R125}a) corresponds to PSA, with fixed $P_{ads}=100$ kPa and $T_{ads}=T_{des}$. \autoref{fig:Storage_density_validation_Bhatia01_R125}b) corresponds to TSA, with $P_{ads}=P_{des}$ and $T_{ads}=283$ K. \autoref{fig:Storage_density_validation_Bhatia01_R125}c) corresponds to PTSA, with fixed $P_{ads}=100$ kPa and $T_{ads}=283$ K.

Two conclusions can be drawn from this comparison. First, the storage densities obtained from the different heat-of-adsorption methods are in good agreement. For PTSA at high temperatures, the analytical approaches deviate slightly from the fluctuation-based values due to limitations in the loading range over which the heat of adsorption can be reliably calculated. Second, PTSA provides the highest storage density, followed by PSA, whereas TSA gives the lowest values under the conditions considered.

Based on this result, the remaining storage-density analysis focuses on PTSA using the Clausius--Clapeyron method. \autoref{fig:storage_density_3D_plots}a) and \autoref{fig:storage_density_3D_plots}b) show the storage density of R125 as a function of pressure and temperature for Bhatia-01 and CS1000a, respectively. CS1000a exhibits a steeper increase in storage density and reaches a plateau at lower pressures, whereas Bhatia-01 increases more gradually and does not reach a clear plateau within the investigated range. This behavior is consistent with the adsorption isotherms: CS1000a starts adsorbing at relatively high pressures but reaches a larger saturation capacity, producing a sharp increase in storage density once adsorption becomes significant. Results for the remaining working pairs are provided in the Supplementary Material \ref{sec:Pure component storage density}.

When comparing the maximum storage density at $P_{ads}=500$ kPa and $P_{des,min}=1$ Pa in \autoref{fig:storage_density_3D_plots}c), it should be noted that the maximum adsorption pressure is not always reached within the valid range of the heat-of-adsorption data. For several adsorbent/adsorbate combinations, the maximum pressure derived from the available isosteric heat data is lower than $500$ kPa. In these cases, the integration uses the highest valid pressure instead of the nominal $P_{ads}$; (see the Supplementary Material \ref{Storage density maximum}). Similarly, when $P_{des,min}=1$ Pa lies outside the valid pressure range, the integration terminates at the lowest accessible pressure, leading to a plateau in storage density. Thus, the reported values correspond to the maximum storage densities that can be derived from the available isotherm data. Since most isotherms are already close to saturation, extending the pressure range is not expected to substantially increase the storage density for most systems.

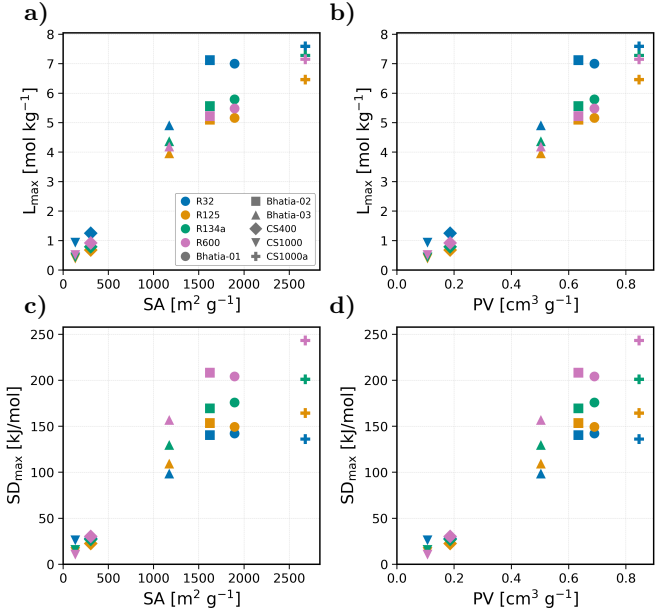
\begin{figure}[t!]
    \centering

    \begin{tikzpicture}
        \node[anchor=south west, inner sep=0] (image) at (0,0)
            {\includegraphics[width=\linewidth]{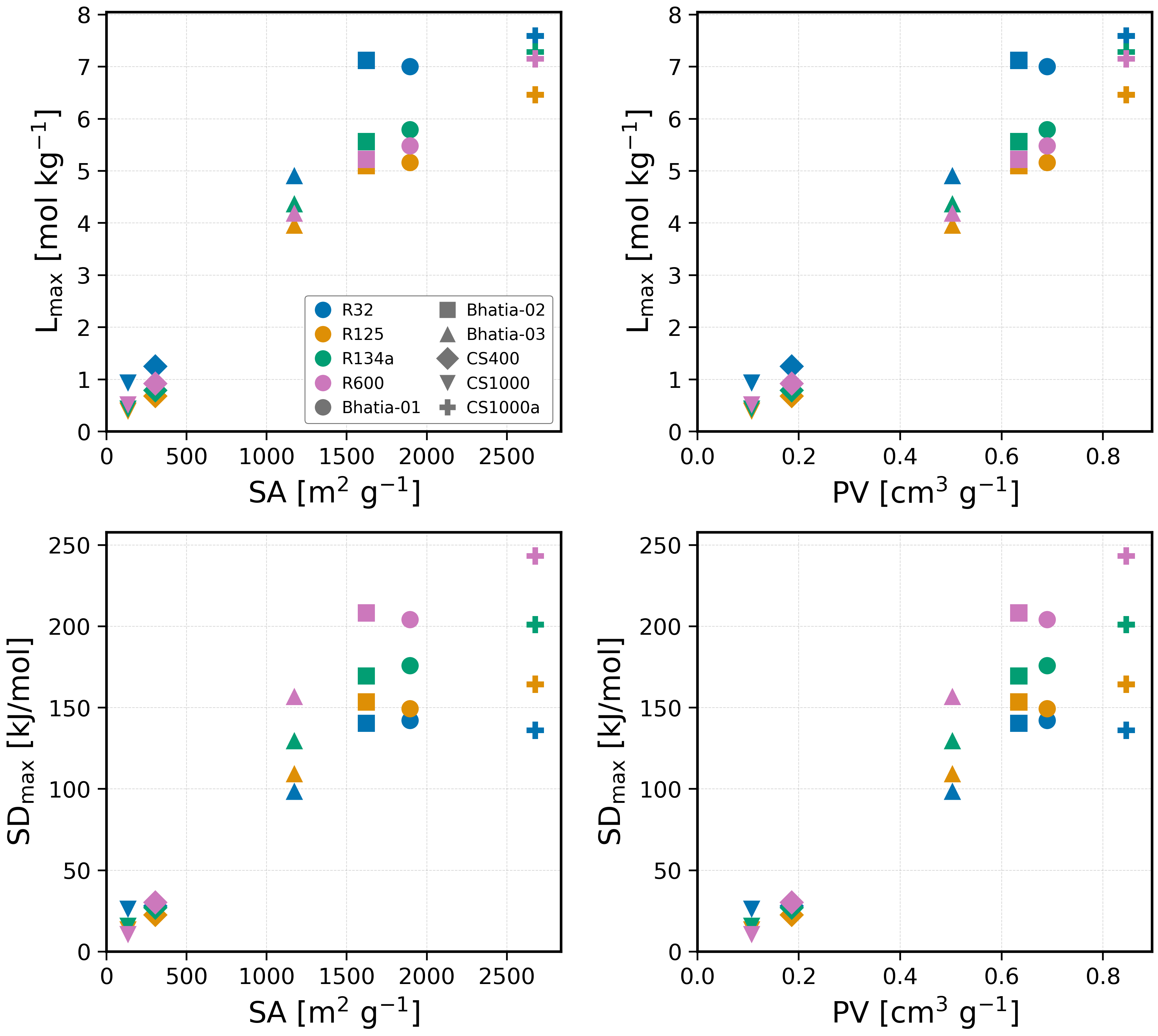}};
            
        \begin{scope}[x={(image.south east)}, y={(image.north west)}]
            \node[font=\bfseries, anchor=west] at (0.02,1.02) {a)};
            \node[font=\bfseries]             at (0.52,1.02) {b)};
            \node[font=\bfseries, anchor=west] at (0.02,0.52) {c)};
            \node[font=\bfseries]             at (0.52,0.52) {d)};
        \end{scope}
    \end{tikzpicture}

    \caption{Correlation between maximum achievable loading ($L_{\max}$) and energy storage density ($SD_{\max}$) with $SA$ and $PV$.}

    \label{fig:correlation_textural_properties}
\end{figure}

The relationship between maximum achievable loading, maximum storage density, and the textural properties of the activated carbons is shown in \autoref{fig:correlation_textural_properties}. The general performance follows the order CS1000 $<$ CS400 $<$ Bhatia-03 $<$ Bhatia-02 $\approx$ Bhatia-01 $<$ CS1000a. In general, higher surface area and pore volume lead to larger maximum loadings and storage densities. R32 in CS1000a is an exception because saturation is not reached within the investigated pressure range, and therefore, its maximum loading and storage density are not fully attained. R32 exhibits the lowest pure-component storage density, whereas R600 shows the highest. Among the activated carbons, Bhatia-01 and Bhatia-02 perform similarly and outperform Bhatia-03. CS400 and CS1000 provide very limited usable storage density, whereas CS1000a achieves the highest values for all pure components except R32, for which Bhatia-02 performs best within the investigated pressure range.

We next evaluate the adsorption and energy storage performance of the refrigerant blends. First, IAST is validated for predicting multicomponent adsorption isotherms and for providing the mixture data required to calculate heats of adsorption. The total mixture adsorption isotherms for R407F (30 wt\% R32, 30 wt\% R125, and 40 wt\% R134a), R410A (50 wt\% R32 and 50 wt\% R125), R417A (3.4 wt\% R600, 46.6 wt\% R125, and 50 wt\% R134a), and R417C (1.7 wt\% R600, 19.5 wt\% R125, and 78.8 wt\% R134a) are provided in Figure \ref{total_adsorption_isotherm_mixtures} in the Supplementary Material. \autoref{fig:summarry_mixtures_heat_adsorption_R407F_R417A}a) shows the component and total adsorption isotherms of R407F and R417A in Bhatia-01 at $303$ K, comparing IAST predictions with GCMC reference data. The total adsorption isotherm of a refrigerant blend is given by the sum of the equilibrium adsorption loadings of all components of the mixture in the adsorbed phase. For R407F, the IAST-predicted adsorption isotherms agree well with GCMC. The uptake of R32 is slightly underestimated, resulting in a small underestimation of the total mixture loading. For R417A, IAST overestimates the uptake of R600 and underestimates that of R125, although the total mixture adsorption remains in good agreement with GCMC. This behavior indicates the presence of competitive adsorption. The effect is more pronounced at low temperatures, where the more strongly adsorbing component can displace weaker components from the adsorption sites. As temperature increases, the adsorption strength of the components decreases, and their differences become less pronounced, reducing the degree of competitive adsorption. Consequently, the deviations in the individual component loadings become smaller at higher temperatures, and the corresponding isotherms may converge or even intersect. Such intersections are not observed in the GCMC data and indicate that the component-wise IAST decomposition can introduce nonphysical behavior under strong competitive adsorption. These results show that IAST captures the total mixture adsorption with sufficient accuracy over the investigated temperature range. However, the predicted individual component loadings should be interpreted with caution, especially for mixtures in which one component exhibits significantly stronger interactions with the adsorbent than the others.

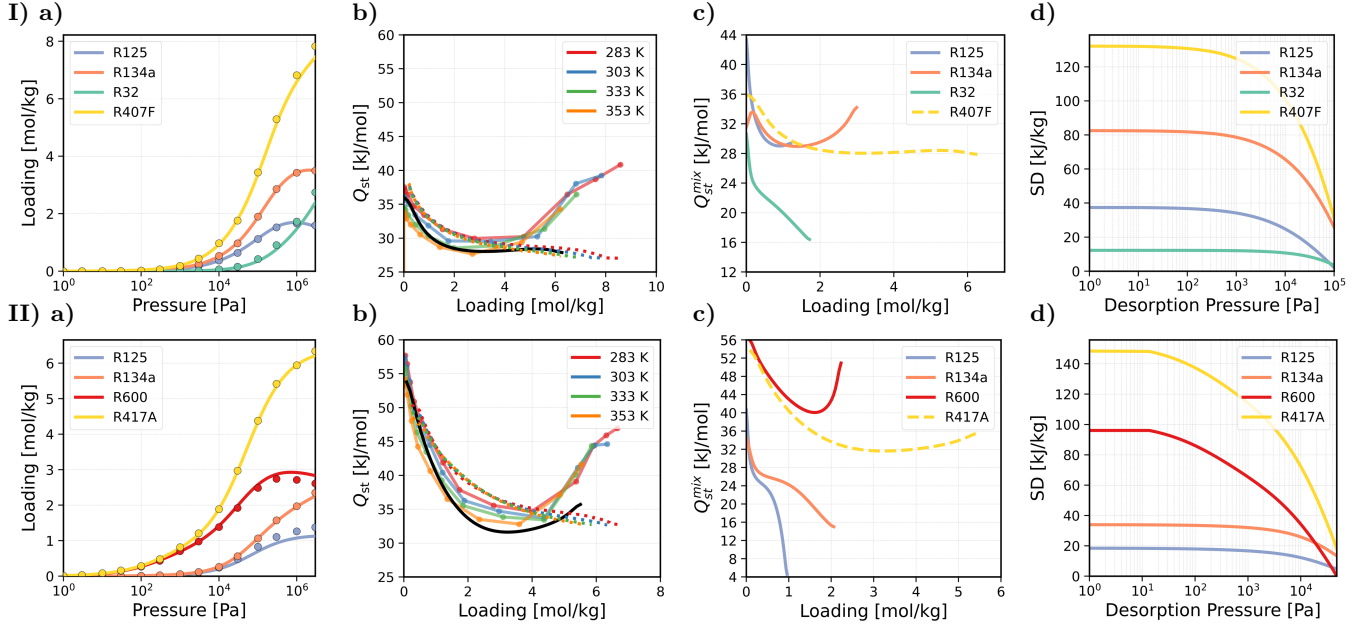
\begin{figure*}[t]
    \centering

    \begin{tikzpicture}
        \node[anchor=south west, inner sep=0] (image) at (0,0)
            {\includegraphics[width=\linewidth]{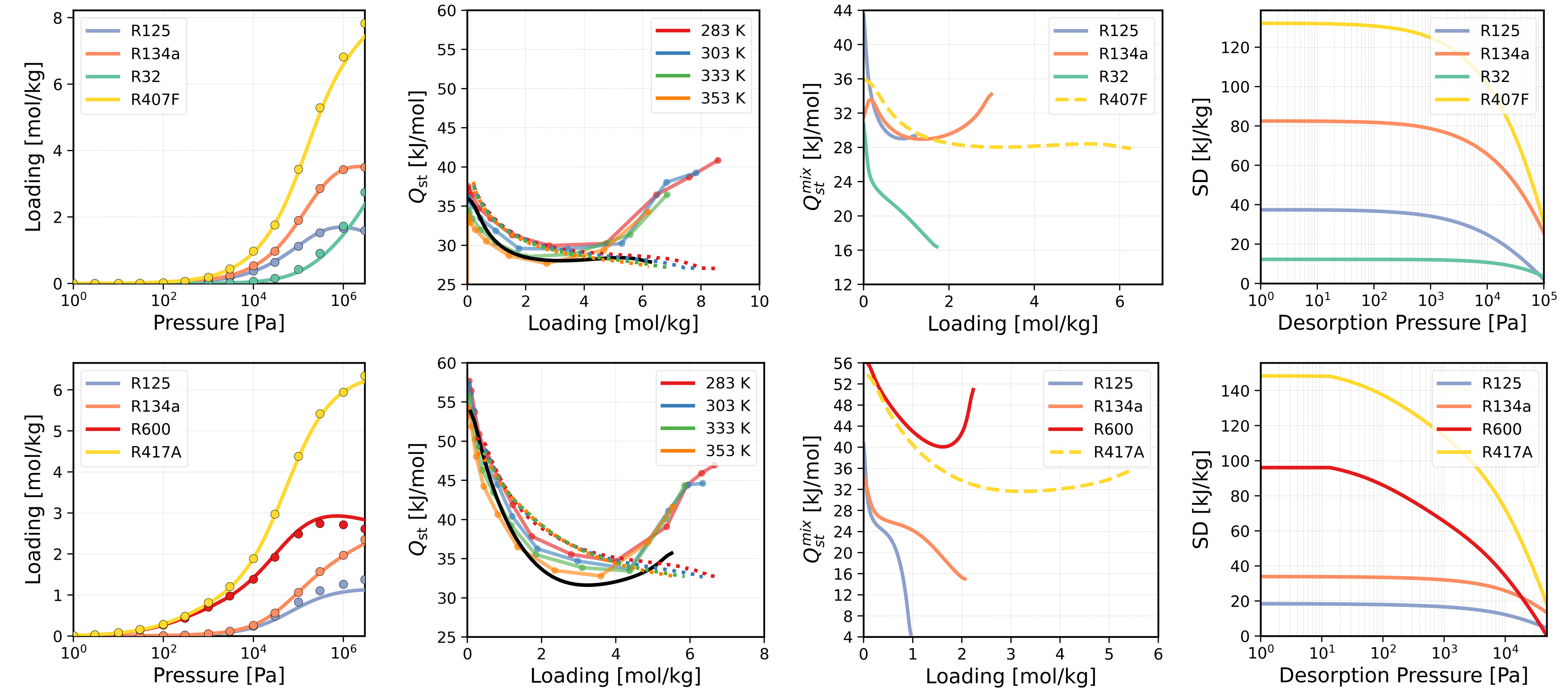}};
            
        \begin{scope}[x={(image.south east)}, y={(image.north west)}]
            \node[font=\bfseries, anchor=west] at (0.00,1.02) {I) a)};
            \node[font=\bfseries]             at (0.27,1.02) {b)};
            \node[font=\bfseries]             at (0.52,1.02) {c)};
            \node[font=\bfseries]             at (0.77,1.02) {d)};

            \node[font=\bfseries, anchor=west] at (0.00,0.52) {II) a)};
            \node[font=\bfseries]             at (0.27,0.52) {b)};
            \node[font=\bfseries]             at (0.52,0.52) {c)};
            \node[font=\bfseries]             at (0.77,0.52) {d)};
        \end{scope}
    \end{tikzpicture}

    \caption{Adsorption isotherm (a), isosteric heat of adsorption methods (b), isosteric heat of adsorption per component (c), and storage density per component (d) for R407F (I) and R417A (II), using GCMC data ($\bullet$), IAST (---), Clausius-Clapeyron total (---), linear mixing equation (-- --), and fluctuation method ($-\bullet$).}

    \label{fig:summarry_mixtures_heat_adsorption_R407F_R417A}
\end{figure*}

\autoref{fig:summarry_mixtures_heat_adsorption_R407F_R417A}b) shows the mixture heats of adsorption obtained from the Clausius-Clapeyron mixture approach (\autoref{eq: clausius clapeyron mixture}) and from the linear mixing rule (\autoref{eq: linear mixing rule}), using the pure-component Clausius-Clapeyron heats as input. \autoref{tab:ard_heat_adsorption} compares the ARD\% between both IAST-based approaches and the GCMC fluctuation reference. To ensure a consistent comparison, the ARD\% is calculated over a common loading range of $0.5$-$5.5$ mol/kg using interpolation. Both approaches show good agreement with the fluctuation method. A complete overview of the mixture heats of adsorption is provided in the Supplementary Material  ~\ref{sup: HoA linear and CC}.

\begin{table}[H]
\centering
\setlength{\tabcolsep}{4pt}
\renewcommand{\arraystretch}{1.0}

\begin{tabular}{l c c c c}
\hline
& \textbf{R407F} & \textbf{R410A} & \textbf{R417A} & \textbf{R417C} \\
\hline

$\mathbf{Q_{st,mix}}$ &
\begin{tabular}[c]{@{}c@{}}
5.26 -- \\
9.37\%
\end{tabular}
&
\begin{tabular}[c]{@{}c@{}}
6.79 -- \\
12.57\%
\end{tabular}
&
\begin{tabular}[c]{@{}c@{}}
5.83 -- \\
9.37\%
\end{tabular}
&
\begin{tabular}[c]{@{}c@{}}
7.61 -- \\
11.56\%
\end{tabular}
\\

$\mathbf{Q_{st,cc}}$ &
\begin{tabular}[c]{@{}c@{}}
5.51 -- \\
8.63\%
\end{tabular}
&
\begin{tabular}[c]{@{}c@{}}
5.61 -- \\
7.18\%
\end{tabular}
&
\begin{tabular}[c]{@{}c@{}}
4.81 -- \\
11.18\%
\end{tabular}
&
\begin{tabular}[c]{@{}c@{}}
6.23 -- \\
10.68\%
\end{tabular}
\\

\hline
\end{tabular}

\caption{ARD\% of the fluctuation method compared to the Clausius-Clapeyron mixture and linear mixture equations.}
\label{tab:ard_heat_adsorption}

\end{table}

The isosteric heats of adsorption obtained from the Clausius-Clapeyron mixture approach remain reliable because the total mixture isotherm is less affected by component-wise competitive adsorption errors. However, the convergence or intersection of individual component isotherms affects the calculation of component-specific heats of adsorption. When the Clausius-Clapeyron analysis is applied to IAST-generated component isotherms, nonmonotonic or inconsistent temperature dependencies can appear. As a result, the component-wise heats of adsorption (\autoref{fig:summarry_mixtures_heat_adsorption_R407F_R417A}c) and the corresponding component storage densities (\autoref{fig:summarry_mixtures_heat_adsorption_R407F_R417A}d) are not physically meaningful for R417A. Similar behavior is observed for R417C; see Figure~\ref{sup: seperation R410A R417C} in the Supplementary Material. The linear mixing rule is also affected by the over- and underestimation of individual component loadings in R417A and R417C. In contrast, this limitation is less severe for R407F and R410A, where the component isotherms remain sufficiently separated, and the resulting heats of adsorption show consistent trends; see Figure~\ref{sup: seperation R410A R417C} in the Supplementary Material.

Overall, the Clausius-Clapeyron mixture approach provides mixture heats of adsorption that are in good agreement with the GCMC fluctuation method and generally closer to the reference than the linear mixing rule. Component-wise Clausius-Clapeyron heats are reliable only when the mixture behaves approximately ideally and when component isotherms remain well separated. For the investigated blends, R417A exhibits a slightly higher mixture heat of adsorption than R417C due to its higher R600 content, whereas R410A outperforms R407F due to the absence of R134a and the higher fraction of R125. In general, the R600-containing mixtures show higher heats of adsorption than the R32-containing mixtures.

\begin{figure*}[t]
    \centering

    \begin{tikzpicture}
        \node[anchor=south west, inner sep=0] (image) at (0,0)
            {\includegraphics[width=\linewidth]{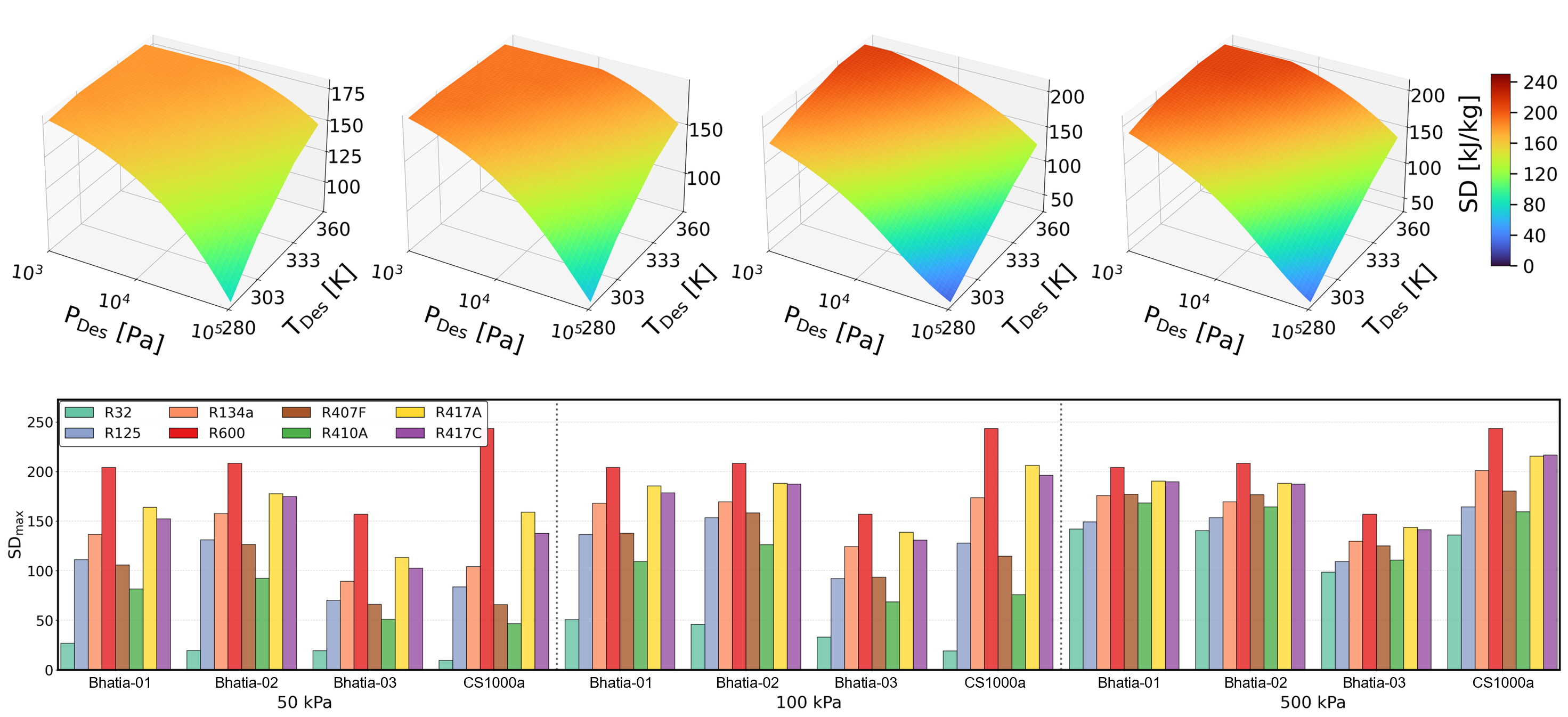}};
            
        \begin{scope}[x={(image.south east)}, y={(image.north west)}]
            \node[font=\bfseries, anchor=west] at (-0.02,1.02) {I) a)};
            \node[font=\bfseries]             at (0.25,1.02) {b)};
            \node[font=\bfseries]             at (0.50,1.02) {II) a)};
            \node[font=\bfseries]             at (0.72,1.02) {b)};

            \node[font=\bfseries, anchor=west] at (-0.02,0.52) {III)};
        \end{scope}
    \end{tikzpicture}

    \caption{Storage density 3D plots of R407F (a) and R417A (b) in Bhatia-01 (I) and CS1000a (II) for PTSA ($P_{ads}=100$ kPa, $T_{ads}=283$ K), and the theoretical maximum storage density per working pair (III).}

    \label{fig:Summary_3D_maximum_storage_densities}
\end{figure*}
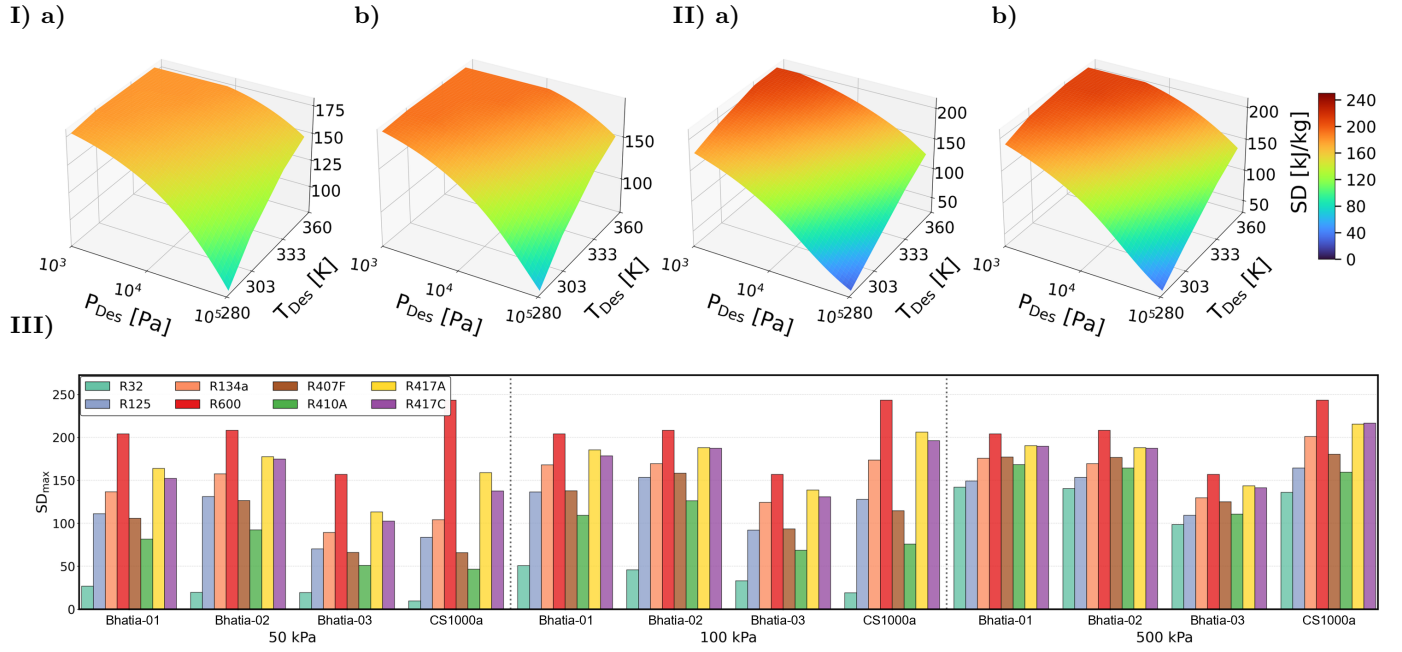

The Clausius-Clapeyron mixture approach and the linear mixing rule are further validated against the GCMC fluctuation method for the storage density calculations in the Supplementary Material \ref{sec: control sd mixture}. As for the pure components, three-dimensional storage-density surfaces provide insight into the influence of operating pressure and temperature. \autoref{fig:Summary_3D_maximum_storage_densities} shows the storage-density surfaces of R407F and R417A in Bhatia-01 and CS1000a. Under the reference adsorption conditions of $283$ K and $100$ kPa, R417A generally achieves a higher storage density than R407F. For Bhatia-01, the storage density increases steadily with pressure and temperature, without reaching a clear plateau, indicating that additional refrigerant can still be adsorbed under a broader range of desorption conditions. In contrast, CS1000a exhibits a more defined plateau at lower pressures and higher temperatures, suggesting that its adsorption capacity is reached more rapidly. The remaining three-dimensional storage density plots are provided in the Supplementary Material  ~\ref{3D plots mix}.

To compare the maximum achievable storage densities, adsorption pressures of $50$, $100$, and $500$ kPa are considered in \autoref{fig:Summary_3D_maximum_storage_densities}. As discussed for the pure components, the integration is limited by the thermodynamically valid pressure range of the heat of adsorption data. Therefore, the imposed $P_{ads}$ and $P_{des,min}$ are not always reached. The highest valid pressures and corresponding maximum storage densities are reported in the Supplementary Material \ref{Maximum Storage Density}. At low adsorption pressure, the R600-containing blends, R417A and R417C, exhibit the highest maximum storage densities. Their similar performance indicates that reducing the R600 content has only a moderate effect on the total storage density, despite R600 having the highest pure-component storage density. At higher adsorption pressures, the R32-containing blends, R407F and R410A, show a more pronounced increase in storage density, reflecting the stronger pressure sensitivity of R32 adsorption. In some activated carbons, these blends exceed the maximum storage density of the corresponding pure components. This enhancement is due to cooperative adsorption and more efficient molecular packing within the pore space, which reduces void volume and increases total uptake.

Although pure R600 has very low GWP and high storage density, its high flammability limits its practical use as a pure working fluid. Blends containing R600 can partially retain its favorable adsorption properties while remaining within safer and more practical composition ranges. Nevertheless, R600 reaches saturation at relatively low pressure in the R600-containing mixtures. At higher pressures, the additional uptake is dominated by the other components, decreasing the relative contribution of R600 to the overall storage density. To compare environmental impact and storage performance simultaneously, the ratio between GWP and storage density is evaluated. In Bhatia-01, R417C gives the lowest value, $8.05$ GWP/kJ, which is approximately $24\%$ lower than R417A ($10.6$ GWP/kJ), $28\%$ lower than R410A ($11.2$ GWP/kJ), and $30\%$ lower than R407F ($11.5$ GWP/kJ). This comparison shows that the best-performing blend in terms of absolute storage density is not necessarily the most favorable when environmental impact is included.

The relative performance of the activated carbons follows the same trends observed for the pure components. Bhatia-01 and Bhatia-02 exhibit similar behavior, Bhatia-03 shows lower performance, and CS1000a achieves the highest storage densities. This is consistent with the broader pore size distribution and larger accessible pore volume of CS1000a. These results demonstrate that blend composition and adsorbent textural properties jointly determine the final storage performance. Overall, the storage density trends provide guidelines for the rational design of refrigerant blends for adsorption-based thermochemical energy storage. Because the adsorbed-phase composition can differ substantially from the bulk composition, the optimal blend composition depends on the porous material and operating pressure. For mixtures without severe competitive adsorption artifacts, the component-wise contributions indicate which refrigerants most strongly enhance storage density. This opens the possibility of designing new refrigerant blends by tuning the bulk composition to maximize the adsorbed phase contribution of the most beneficial components for each adsorbent.

\begin{figure*}[t]
    \centering

    \begin{tikzpicture}
        \node[anchor=south west, inner sep=0] (image) at (0,0)
            {\includegraphics[width=\linewidth]{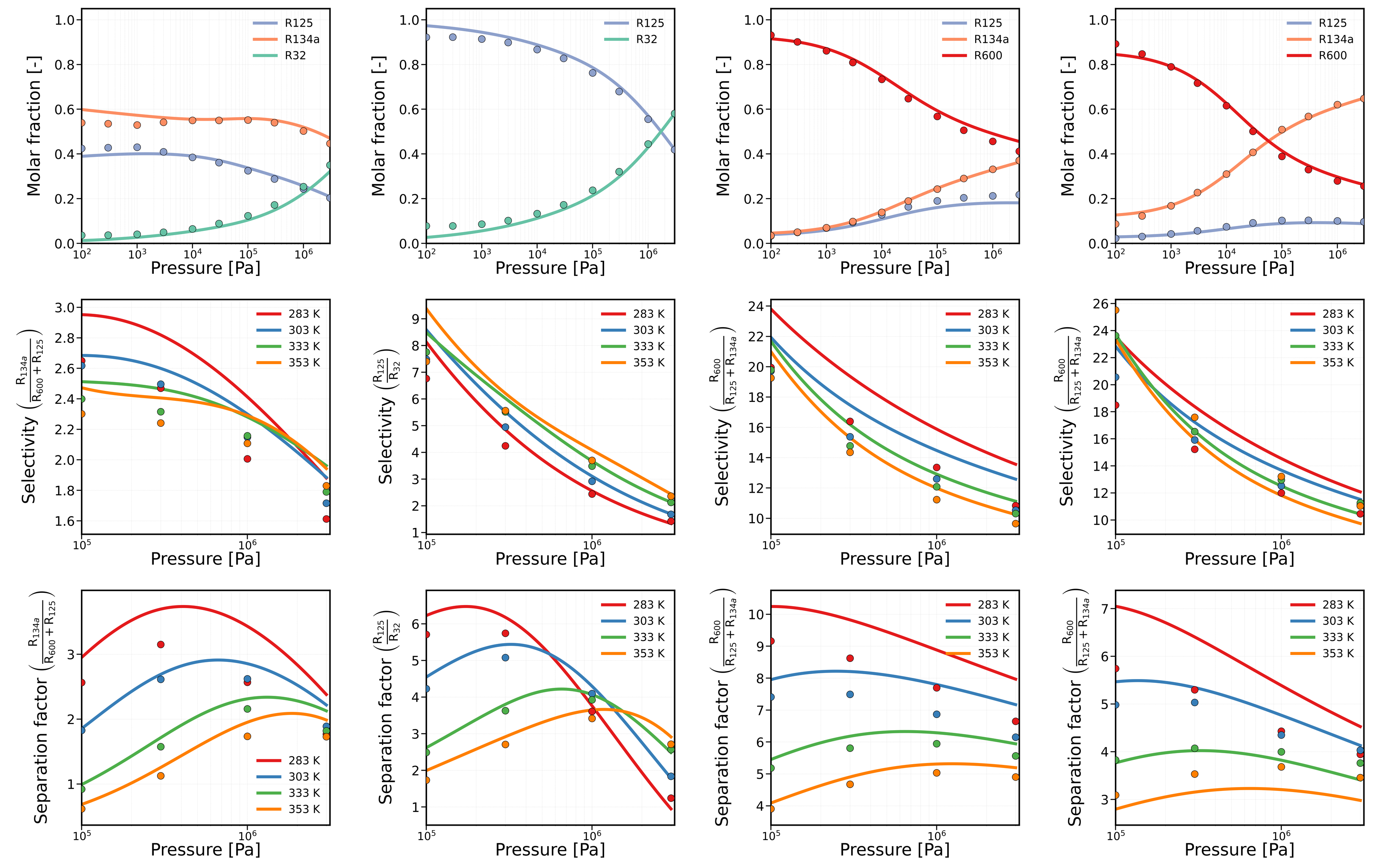}};
            
        \begin{scope}[x={(image.south east)}, y={(image.north west)}]
            \node[font=\bfseries, anchor=west] at (0.00,1.02) {I) a)};
            \node[font=\bfseries]             at (0.28,1.02) {b)};
            \node[font=\bfseries]             at (0.52,1.02) {c)};
            \node[font=\bfseries]             at (0.78,1.02) {d)};

            \node[font=\bfseries, anchor=west] at (0.00,0.68) {II) a)};
            \node[font=\bfseries]             at (0.28,0.68) {b)};
            \node[font=\bfseries]             at (0.52,0.68) {c)};
            \node[font=\bfseries]             at (0.78,0.68) {d)};

            \node[font=\bfseries, anchor=west] at (0.00,0.34) {III) a)};
            \node[font=\bfseries]             at (0.28,0.34) {b)};
            \node[font=\bfseries]             at (0.52,0.34) {c)};
            \node[font=\bfseries]             at (0.78,0.34) {d)};
        \end{scope}
    \end{tikzpicture}

    \caption{IAST mixture mole fractions at $303\,\mathrm{K}$ (I), selectivity (II), and separation factors (III) for R407F (a), R410A (b), R417A (c), and R417C (d), using GCMC data ($\bullet$) and IAST ($-$).}

    \label{fig:IAST_3x4_summary}
\end{figure*}
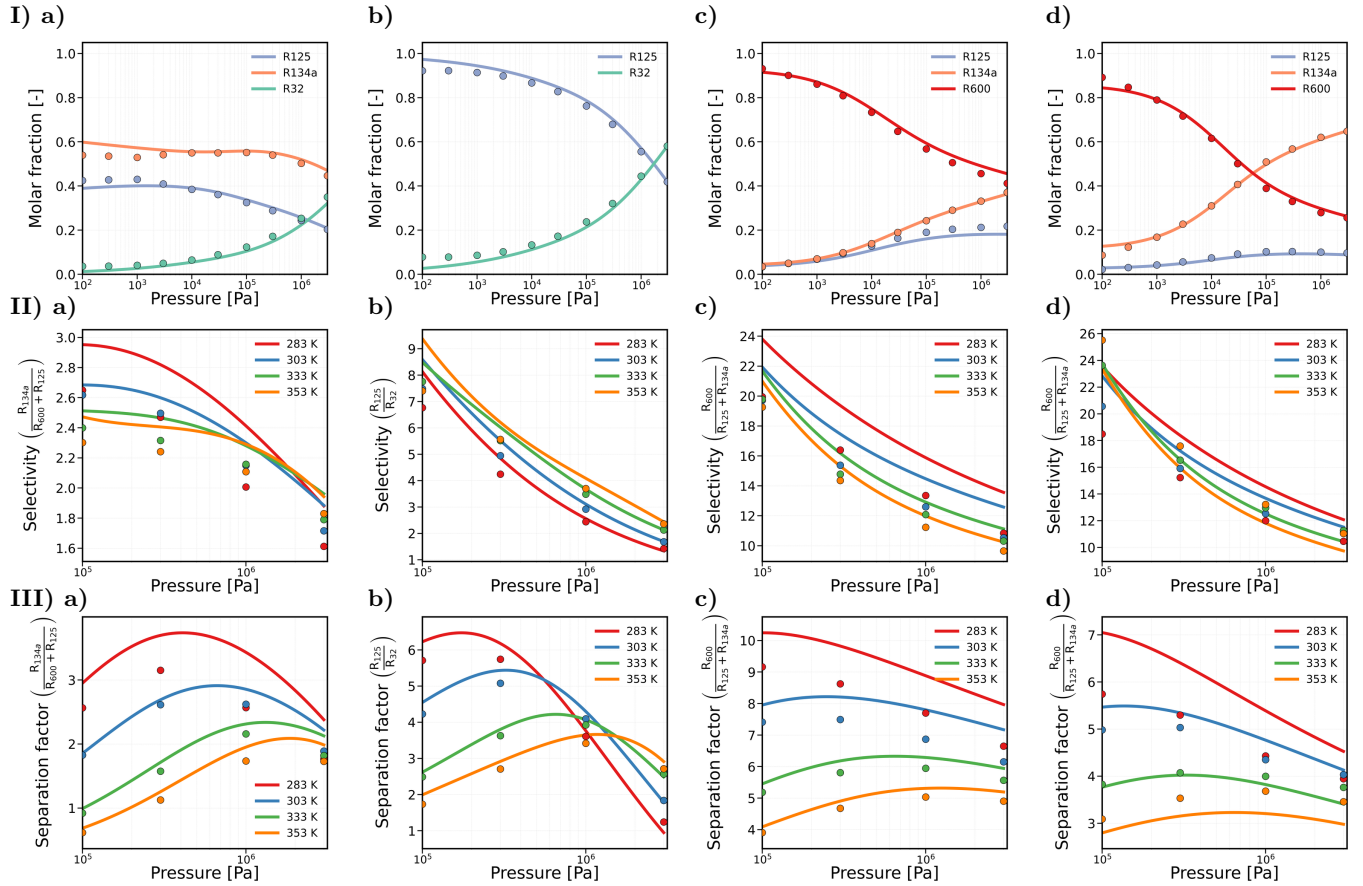

After evaluating the thermochemical energy storage performance, the separation potential of the activated carbons toward the refrigerant blends is investigated. \autoref{fig:IAST_3x4_summary} summarizes the separation performance metrics of the four refrigerant blends in Bhatia-01, adsorbed-phase molar fraction, adsorption selectivity, and separation factor, and compares the IAST predictions with the corresponding GCMC reference data. Similar separation trends are observed for Bhatia-02, Bhatia-03, and CS1000a; see the Supplementary Material \ref{sec: supp seperation}.

For the R417A and R417C mixtures, R600 adsorbs very strongly and dominates the adsorbed phase, leading to pronounced competitive adsorption and deviations from ideal adsorption behavior. As the temperature increases, the adsorption strength of R600 decreases and the system becomes less competitive. Consequently, the IAST predictions move closer to the GCMC reference data, improving the agreement at higher temperatures. In contrast, the opposite trend is observed for R410A and R407F. These mixtures contain R32, which is the most strongly adsorbing component among the non-R600 refrigerants and therefore strongly governs the adsorption behavior of the mixture. At higher temperatures, the adsorption of R32 decreases more rapidly than that of the other components, altering the adsorbed-phase composition and increasing the deviations between IAST and GCMC. These results further demonstrate that the degree of competitive adsorption strongly influences the applicability of ideal adsorption theories for refrigerant blends.

As discussed above, the APT accurately reproduces adsorption isotherms at medium-to-high loading but does not provide sufficiently accurate results for the heat of adsorption calculations due to the amplification of small numerical deviations in derivative-based thermodynamic properties. Nevertheless, because the pressure range investigated for separation ($100$-$3000$ kPa) corresponds predominantly to medium and high loadings, APT can still be combined with IAST to evaluate adsorption-based separation performance. This behavior is confirmed in \autoref{fig:APT_1x2_R407F} in the Supplementary Material, where the selectivities predicted from APT combined with IAST agree closely with those obtained from GCMC combined with IAST for R407F, R410A, R417A, and R417C in Bhatia-01. Therefore, the combined APT/IAST approach can be used as an efficient predictive framework for evaluating separation properties from a reduced number of adsorption isotherms.

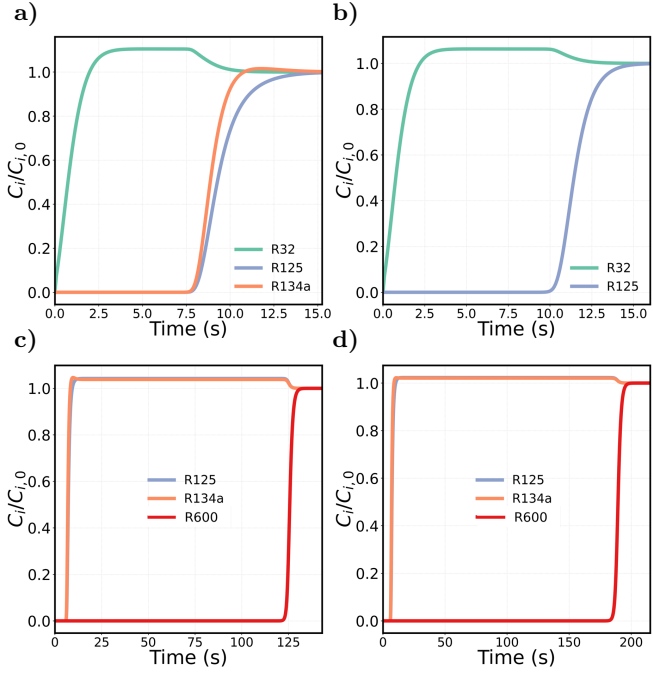
\begin{figure}[t!]
    \centering

    \begin{tikzpicture}
        \node[anchor=south west, inner sep=0] (image) at (0,0)
            {\includegraphics[width=\linewidth]{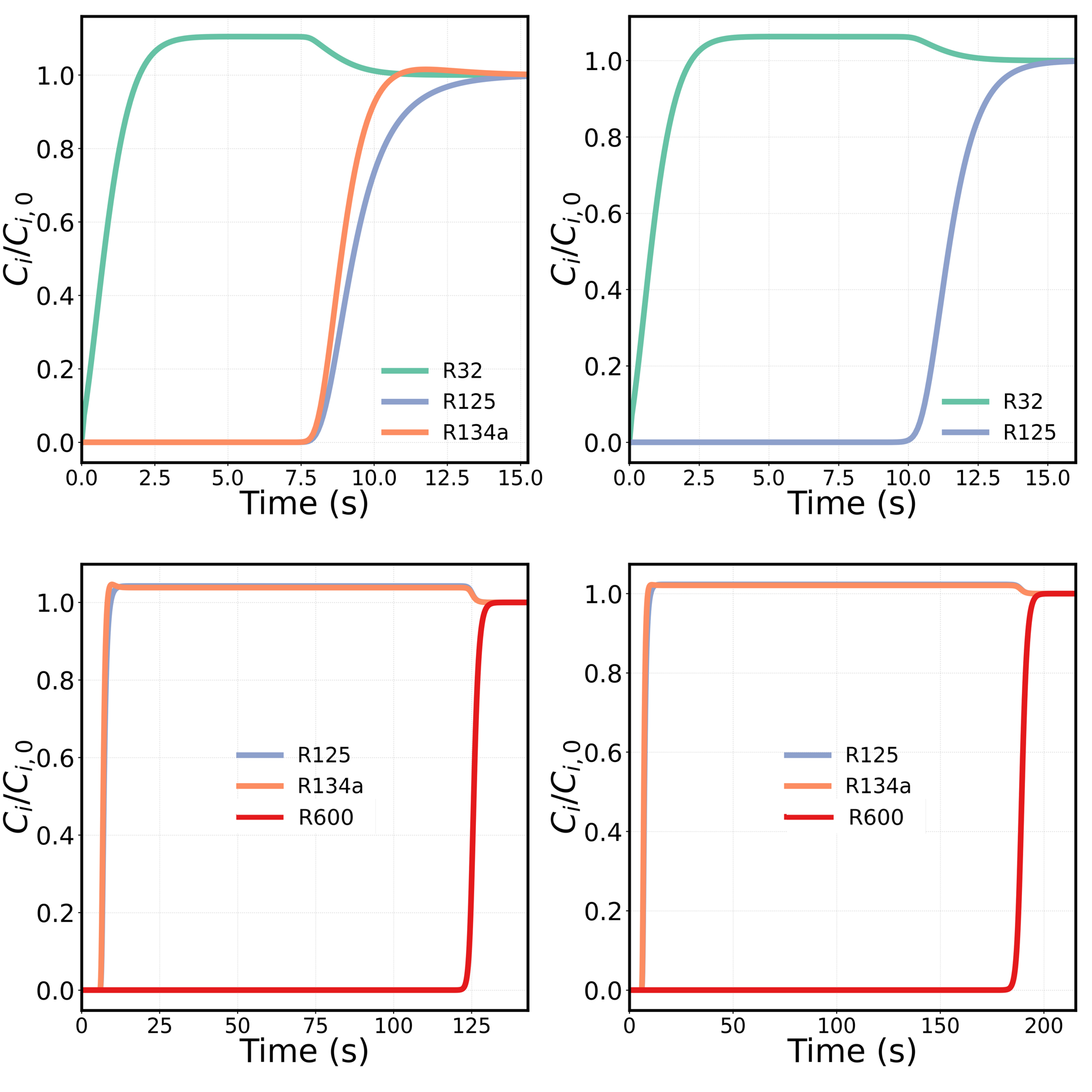}};
            
        \begin{scope}[x={(image.south east)}, y={(image.north west)}]
            \node[font=\bfseries, anchor=west] at (0.00,1.02) {a)};
            \node[font=\bfseries]             at (0.52,1.02) {b)};

            \node[font=\bfseries, anchor=west] at (0.00,0.52) {c)};
            \node[font=\bfseries]             at (0.52,0.52) {d)};
        \end{scope}
    \end{tikzpicture}

    \caption{Breakthrough curves at 303 K for refrigerant blends R407F (a), R410A (b), R417A (c), and R417C (d) in Bhatia-01}.

    \label{fig:breakthrough_2x2}
\end{figure}

The separation dynamics of the refrigerant blends are further evaluated using breakthrough simulations. \autoref{fig:breakthrough_2x2} shows the breakthrough curves for Bhatia-01 at $100$ kPa and $303$ K. For the R32-containing mixtures, R125 and R134a break through first, while R32 remains adsorbed for a longer time. This clear temporal separation enables the efficient recovery of R32 from the mixtures. Increasing temperature slightly enhances the adsorption of R125 relative to R134a, reducing the distinction between these two components. However, the separation between R32 and the remaining refrigerants remains well defined, indicating that the recovery of R32 is only weakly affected by temperature under the investigated conditions.

The behavior of the R600-containing mixtures is markedly different. R600 adsorbs first and strongly dominates the adsorption sites, whereas the breakthrough curves of R125 and R134a almost completely overlap over the investigated temperature range. This indicates that the presence of R600 suppresses the separation between R125 and R134a. The effect arises because R600 occupies the strongest adsorption sites, forcing R125 and R134a to compete for weaker adsorption regions where their adsorption affinities are very similar. In contrast, for mixtures without R600, a measurable separation between R125 and R134a is still observed. These results highlight the strong influence of competitive adsorption on the separation dynamics of refrigerant blends.

The breakthrough behavior of Bhatia-02 and Bhatia-03 is very similar to that of Bhatia-01. In contrast, CS1000a exhibits somewhat different separation characteristics. Although R600 still dominates the adsorption in R417A and R417C, R125 breaks through earlier than R134a, and the overlap between the two curves is reduced. This indicates that CS1000a introduces a measurable separation between R125 and R134a that is absent in the Bhatia materials. For the R32-containing mixtures, R32 remains pure for a shorter time in CS1000a than in the Bhatia materials because R125 begins to appear earlier in the breakthrough curve. Additional breakthrough simulations for the remaining systems are provided in the Supplementary Material \ref{sup sepeation dynamics}.

Overall, the activated carbons show strong potential for adsorption-based refrigerant recovery. The results demonstrate that the separation behavior depends not only on the bulk composition of the refrigerant blend but also on the adsorbed-phase composition and competitive adsorption effects within the pores. In particular, the strong preferential adsorption of R32 or R600 governs both the equilibrium selectivity and the dynamic breakthrough behavior. These findings highlight the importance of combining equilibrium thermodynamics with dynamic separation modeling when evaluating refrigerant blends for circular recovery applications.

\section{Conclusions}
\label{sec:conclusions}

This work introduces a multiscale computational framework for evaluating refrigerant adsorption systems for thermochemical energy storage and circular refrigerant recovery. By combining Monte Carlo simulations, thermodynamic modeling, and breakthrough simulations, the methodology predicts adsorption, energy storage, and separation behavior of refrigerant blends directly from pure-component adsorption data. For the first time, the adsorption thermodynamics and storage performance of refrigerant blends are systematically described in terms of the individual contributions of their pure components within the adsorbed phase. The framework integrates adsorption potential theory (APT), ideal adsorbed solution theory (IAST), and models for the isosteric heat of adsorption, together with an in-house software package for calculating heats of adsorption and energy storage densities of both pure refrigerants and multicomponent mixtures. Since the methodology only requires adsorption isotherms of the pure components as input, it is directly applicable to both computational and experimental studies and enables efficient screening of adsorbent/adsorbate working pairs.

The results demonstrate that refrigerant blends can outperform their pure-component counterparts in thermochemical energy storage due to cooperative adsorption and more efficient molecular packing within the porous structure. R600-containing systems show superior performance at low adsorption pressures, whereas R32-containing systems become more favorable at higher pressures. The results further reveal that the composition of the adsorbed phase can differ substantially from that of the bulk mixture due to competitive adsorption within the pores. This behavior provides new insight into how individual refrigerant components contribute to the overall storage and separation performance and opens the possibility of tailoring blend compositions to maximize performance for specific porous materials and operating conditions. Our result show that optimal refrigerant blends for adsorption systems could not necessarily be those with the best bulk thermodynamic properties, but those whose adsorbed-phase composition under operating conditions maximizes storage density and/or selective adsorption within the porous material. Among the investigated activated carbons, those with broader pore volume and surface area achieve the highest storage densities and enhanced separation performance. In addition, the activated carbons selectively separate key refrigerant components, particularly R32 and R600, highlighting their potential for sustainable refrigerant recovery and circular refrigerant management.

The comparison between theoretical approaches further clarifies the applicability and limitations of current predictive methods for refrigerant mixtures. APT provides reliable predictions of adsorption loadings across a wide range of conditions and is suitable for comparative screening. However, small deviations at low loadings and pressures propagate into large relative errors in derivative-based thermodynamic properties, making APT-derived isotherms less reliable for calculating heats of adsorption. In addition, IAST accurately reproduces overall mixture adsorption but should be applied cautiously for individual component loadings under strong competitive adsorption. These limitations could be further reduced by improving the thermodynamic description of APT in the low-pressure regime and by employing more advanced mixture theories, such as Real Adsorbed Solution Theory (RAST), to better capture competitive adsorption effects. Overall, this work establishes a general framework for the rational design, screening, and optimization of next-generation refrigerant blends and porous materials for adsorption-driven energy storage and separation applications.








\section*{Competing interests}

The authors declare no competing interests.

\onecolumngrid

\bibliography{Refrigerant-blends_ACs}

@article{Thyagarajan2020AMaterials,
    title = {{A Database of Porous Rigid Amorphous Materials}},
    year = {2020},
    journal = {Chemistry of Materials},
    author = {Thyagarajan, R. and Sholl, D.S.},
    number = {18},
    month = {9},
    pages = {8020--8033},
    volume = {32},
    publisher = {American Chemical Society},
    url = {/doi/pdf/10.1021/acs.chemmater.0c03057?ref=article_openPDF},
    doi = {10.1021/ACS.CHEMMATER.0C03057}
}

@misc{Brownlee2021AAlgorithm,
    title = {{A Gentle Introduction to the BFGS Optimization Algorithm }},
    year = {2021},
    author = {Brownlee, J.},
    month = {10},
    url = {https://machinelearningmastery.com/bfgs-optimization-in-python/}
}

@article{Nuhnen2020AMOFs,
    title = {{A practical guide to calculate the isosteric heat/enthalpy of adsorption via adsorption isotherms in metal–organic frameworks, MOFs}},
    year = {2020},
    journal = {Dalton Transactions},
    author = {Nuhnen, A. and Janiak, C.},
    number = {30},
    month = {8},
    pages = {10295--10307},
    volume = {49},
    publisher = {Royal Society of Chemistry},
    url = {https://xlink.rsc.org/?DOI=D0DT01784A},
    doi = {10.1039/D0DT01784A},
    issn = {1477-9226},
    pmid = {32661527}
}

@article{Stavarache2024AdaptedMaterials,
    title = {{Adapted thermodynamical model for the prediction of adsorption in nanoporous materials}},
    year = {2024},
    journal = {Chemical Engineering Journal},
    author = {Stavarache, F. and Luna-Triguero, A. and Calero, S. and Vicent-Luna, J.M.},
    month = {9},
    volume = {496},
    publisher = {Elsevier B.V.},
    doi = {10.1016/j.cej.2024.153480},
    issn = {13858947},
    arxivId = {2310.15885}
}

@article{Vicent-Luna2024AdsorptionFrameworks,
    title = {{Adsorption Characteristics of Refrigerants for Thermochemical Energy Storage in Metal–Organic Frameworks}},
    year = {2024},
    journal = {ACS Applied Engineering Materials},
    author = {Vicent-Luna, J.M. and Luna-Triguero, A.},
    number = {3},
    month = {3},
    pages = {542--552},
    volume = {2},
    publisher = {American Chemical Society (ACS)},
    doi = {10.1021/acsaenm.3c00474},
    issn = {2771-9545},
    arxivId = {2309.01777}
}

@article{El-Sharkawy2016AdsorptionCarbon,
    title = {{Adsorption isotherms and kinetics of a mixture of Pentafluoroethane, 1,1,1,2-Tetrafluoroethane and Difluoromethane (HFC-407C) onto granular activated carbon}},
    year = {2016},
    journal = {Applied Thermal Engineering},
    author = {El-Sharkawy, M. M. and Askalany, A. A. and Harby, K. and Ahmed, M. S.},
    month = {1},
    pages = {988--994},
    volume = {93},
    publisher = {Pergamon},
    url = {https://www.sciencedirect.com/science/article/pii/S1359431115011242},
    doi = {10.1016/J.APPLTHERMALENG.2015.10.077},
    issn = {1359-4311}
}

@article{Askalany2014AdsorptionCarbons,
    title = {{Adsorption isotherms and kinetics of HFC410A onto activated carbons}},
    year = {2014},
    journal = {Applied Thermal Engineering},
    author = {Askalany, Ahmed A. and Saha, Bidyut B. and Ismail, Ibrahim M.},
    number = {2},
    month = {11},
    pages = {237--243},
    volume = {72},
    publisher = {Pergamon},
    url = {https://www.sciencedirect.com/science/article/pii/S1359431114004335},
    doi = {10.1016/J.APPLTHERMALENG.2014.04.075},
    issn = {1359-4311}
}

@article{Yagnamurthy2021AdsorptionCharacterization,
    title = {{Adsorption of difluoromethane onto activated carbon based composites: Thermophysical properties and adsorption characterization}},
    year = {2021},
    journal = {International Journal of Heat and Mass Transfer},
    author = {Yagnamurthy, Sai and Rakshit, Dibakar and Jain, Sanjeev and Rocky, Kaiser Ahmed and Islam, Md Amirul and Saha, Bidyut Baran},
    month = {6},
    pages = {121112},
    volume = {171},
    publisher = {Pergamon},
    url = {https://www.sciencedirect.com/science/article/pii/S0017931021002155},
    doi = {10.1016/J.IJHEATMASSTRANSFER.2021.121112},
    issn = {0017-9310}
}

@article{Sosa2020AdsorptionSeparation,
    title = {{Adsorption of fluorinated greenhouse gases on activated carbons: evaluation of their potential for gas separation}},
    year = {2020},
    journal = {Journal of Chemical Technology and Biotechnology},
    author = {Sosa, J.E. and Malheiro, C. and Ribeiro, R. P.P.L. and Castro, P. J. and Pi{\~{n}}eiro, M. M. and Ara{\'{u}}jo, J. M.M. and Plantier, F. and Mota, J. P.B. and Pereiro, A. B.},
    number = {7},
    month = {7},
    pages = {1892--1905},
    volume = {95},
    publisher = {John Wiley and Sons Ltd},
    doi = {10.1002/jctb.6371},
    issn = {10974660}
}

@article{Madero-Castro2022AdsorptionApplications,
    title = {{Adsorption of Linear Alcohols in Amorphous Activated Carbons: Implications for Energy Storage Applications}},
    year = {2022},
    journal = {ACS Sustainable Chemistry {\&} Engineering},
    author = {Madero-Castro, Rafael M. and Vicent-Luna, José Manuel and Peng, Xuan and Calero, Sofía},
    number = {20},
    month = {5},
    pages = {6509--6520},
    volume = {10},
    publisher = {American Chemical Society},
    url = {/doi/pdf/10.1021/acssuschemeng.1c06315?ref=article_openPDF},
    doi = {10.1021/ACSSUSCHEMENG.1C06315},
    issn = {21680485}
}

@article{Hassan2026AIM:Simulation,
    title = {{AIM: A user-friendly GUI workflow program for isotherm fitting, mixture prediction, isosteric heat of adsorption estimation, and breakthrough simulation}},
    year = {2026},
    journal = {Computer Physics Communications},
    author = {Hassan, Muhammad and Yoon, Sunghyun and Chen, Yu and Kim, Pilseok and Yun, Hongryeol and Kwon, Hyuk Taek and Bae, Youn Sang and Yoo, Chung Yul and Koh, Dong Yeun and Hong, Chang Seop and Lee, Ki Bong and Chung, Yongchul G.},
    month = {2},
    volume = {319},
    publisher = {Elsevier B.V.},
    doi = {10.1016/j.cpc.2025.109944},
    issn = {00104655},
    arxivId = {2504.20713}
}

@article{Madero-Castro2023Alcohol-basedFrameworks,
    title = {{Alcohol-based adsorption heat pumps using hydrophobic metal-organic frameworks}},
    year = {2023},
    journal = {Journal of Materials Chemistry A},
    author = {Madero-Castro, R. M. and Luna-Triguero, A. and Gonz{\'{a}}lez-Gal{\'{a}}n, C. and Vicent-Luna, J.M. and Calero, S.},
    number = {6},
    month = {12},
    pages = {3434--3448},
    volume = {12},
    publisher = {Royal Society of Chemistry},
    doi = {10.1039/d3ta05258c},
    issn = {20507496}
}

@article{Bessa2023AnProducts,
    title = {{An Efficient Strategy for Electroreduction Reactor Outlet Fractioning into Valuable Products}},
    year = {2023},
    journal = {Industrial and Engineering Chemistry Research},
    author = {Bessa, M. C.N. and Luna-Triguero, A. and Vicent-Luna, J. M. and Carmo, P.M.O.C. and Tsampas, M. N. and Ribeiro, A. M. and Rodrigues, A.E. and Calero, S. and Ferreira, A.F.P.},
    number = {22},
    month = {6},
    pages = {8847--8863},
    volume = {62},
    publisher = {American Chemical Society},
    doi = {10.1021/acs.iecr.3c00090},
    issn = {15205045}
}

@article{Heinselman2026AnApproach,
    title = {{An International Laboratory Comparison Study on Approximating the Enthalpy of Adsorption via the Clausius-Clapeyron Approach}},
    year = {2026},
    journal = {ChemPhysChem},
    author = {Heinselman, K.N. and Quine, C. M. and Hurst, K. and Cho, J. and Wenny, M. B. and Mason, J. A. and Verma, G. and Ma, S. and Compton, D. and Stadie, N. P. and Sengupta, D. and Islamoglu, T. and Farha, O. K. and Zlotea, C. and Agafonov, A. and Asgari, M. and Al-Shakhs, A. and Lozano-Castello, D. and Fairen-Jimenez, D. and Furukawa, H. and Yabuuchi, Y. and Broom, D. P. and Benham, M. J. and Villajos, J. A. and Balderas-Xicoht{\'{e}}ncatl, R. and Fackelmann, I. and Hirscher, M. and Hoover, W. J. and Morris, W. and Wang, Timothy C. and Parilla, Philip A. and Gennett, Thomas and Shulda, Sarah},
    number = {7},
    month = {4},
    volume = {27},
    publisher = {John Wiley and Sons Inc},
    doi = {10.1002/cphc.202500332},
    issn = {14397641}
}

@misc{NISTChemistry69,
    title = {{Chemistry WebBook, SRD 69}},
    author = {{NIST}},
    url = {https://webbook.nist.gov/chemistry/},
    doi = {https://doi.org/10.18434/T4D303}
}

@article{Peng2010ComparisonSilicalite-1,
    title = {{Comparison Study on the Adsorption of CFC-115 and HFC-125 on Activated Carbon and Silicalite-1}},
    year = {2010},
    journal = {Industrial {\&} Engineering Chemistry Research},
    author = {Peng, Yong and Zhang, Fumin and Zheng, Xiao and Wang, Huanying and Xu, Chunhui and Xiao, Qiang and Zhong, Yijun and Zhu, Weidong},
    number = {20},
    month = {10},
    pages = {10009--10015},
    volume = {49},
    publisher = { American Chemical Society},
    url = {/doi/pdf/10.1021/ie1010806?ref=article_openPDF},
    doi = {10.1021/IE1010806},
    issn = {0888-5885}
}

@article{Queen2014ComprehensiveZn,
    title = {{Comprehensive study of carbon dioxide adsorption in the metal–organic frameworks M <sub>2</sub> (dobdc) (M = Mg, Mn, Fe, Co, Ni, Cu, Zn)}},
    year = {2014},
    journal = {Chem. Sci.},
    author = {Queen, W.L. and Hudson, M.R. and Bloch, E. D. and Mason, J. A. and Gonzalez, M. I. and Lee, J. S. and Gygi, D. and Howe, J. D. and Lee, K. and Darwish, T. A. and James, M. and Peterson, V. K. and Teat, S. J. and Smit, B. and Neaton, J. B. and Long, J. R. and Brown, C. M.},
    number = {12},
    month = {8},
    pages = {4569--4581},
    volume = {5},
    publisher = {Royal Society of Chemistry},
    url = {https://xlink.rsc.org/?DOI=C4SC02064B},
    doi = {10.1039/C4SC02064B},
    issn = {2041-6520}
}

@misc{TNOEnergyStorage,
    title = {{Energy storage}},
    author = {{TNO}},
    url = {https://www.tno.nl/en/sustainable/energy-built-environment/energy-storage/}
}

@article{Hamid2023EstimationIsotherm,
    title = {{Estimation of isosteric heat of adsorption from generalized Langmuir isotherm}},
    year = {2023},
    journal = {Adsorption},
    author = {Hamid, U. and Vyawahare, P. and Chen, C.C.},
    number = {1},
    month = {1},
    pages = {45--64},
    volume = {29},
    publisher = {Springer},
    doi = {10.1007/s10450-023-00379-x},
    issn = {15728757}
}

@article{Askalany2012ExperimentalPair,
    title = {{Experimental study on adsorption–desorption characteristics of granular activated carbon/R134a pair}},
    year = {2012},
    journal = {International Journal of Refrigeration},
    author = {Askalany, Ahmed A. and Salem, M. and Ismail, I. M. and Ali, Ahmed Hamza H. and Morsy, M. G.},
    number = {3},
    month = {5},
    pages = {494--498},
    volume = {35},
    publisher = {Elsevier},
    url = {https://www.sciencedirect.com/science/article/pii/S0140700711000843?via%3Dihub},
    doi = {10.1016/J.IJREFRIG.2011.04.002},
    issn = {0140-7007}
}

@article{Sosa2024ExploringPotential,
    title = {{Exploring the potential of biomass-derived carbons for the separation of fluorinated gases with high global warming potential}},
    year = {2024},
    journal = {Biomass and Bioenergy},
    author = {Sosa, J.E. and Ribeiro, R. P.P.L. and Matos, I. and Bernardo, M. and Fonseca, I.M. and Mota, J. P.B. and Ara{\'{u}}jo, J.M.M. and Pereiro, A. B.},
    month = {9},
    volume = {188},
    publisher = {Elsevier Ltd},
    doi = {10.1016/j.biombioe.2024.107323},
    issn = {18732909}
}

@article{Sosa2023ExploringPotential,
    title = {{Exploring the Potential of Metal–Organic Frameworks for the Separation of Blends of Fluorinated Gases with High Global Warming Potential}},
    year = {2023},
    journal = {Global Challenges},
    author = {Sosa, J.E. and Malheiro, C. and Castro, P. J. and Ribeiro, R. P.P.L. and Pi{\~{n}}eiro, M.M. and Plantier, F. and Mota, J. P.B. and Ara{\'{u}}jo, J. M.M. and Pereiro, A. B.},
    number = {1},
    month = {1},
    volume = {7},
    publisher = {John Wiley and Sons Inc},
    doi = {10.1002/gch2.202200107},
    issn = {20566646}
}

@article{Elhussien2025ExploringSimulations,
    title = {{Exploring zeolite potential for hydrofluorocarbon capture and recycling: Insights from molecular simulations}},
    year = {2025},
    journal = {Microporous and Mesoporous Materials},
    author = {Elhussien, Abrar A. and Abdulazeez, Ismail and Alasiri, Hassan and Fouad, Wael A.},
    month = {2},
    pages = {113442},
    volume = {384},
    publisher = {Elsevier},
    url = {https://www.sciencedirect.com/science/article/pii/S1387181124004645?via%3Dihub#sec3},
    doi = {10.1016/J.MICROMESO.2024.113442},
    issn = {1387-1811}
}

@misc{TheEngineeringToolboxGasesLimits,
    title = {{Gases - Explosion and Flammability Concentration Limits}},
    author = {{The Engineering Toolbox}},
    url = {https://www.engineeringtoolbox.com/explosive-concentration-limits-d_423.html?utm_source=copilot.com}
}

@article{Islam2024HarnessingSystems,
    title = {{Harnessing waste palm-based activated carbon/difluoromethane pair for sustainable low-emission cooling systems}},
    year = {2024},
    journal = {Journal of Environmental Chemical Engineering},
    author = {Islam, M.A. and Saha, B. B.},
    number = {6},
    month = {12},
    volume = {12},
    publisher = {Elsevier Ltd},
    doi = {10.1016/j.jece.2024.114869},
    issn = {22133437}
}

@article{Critoph1995HeatGases,
    title = {{Heat transfer in granular activated carbon beds in the presence of adsorbable gases}},
    year = {1995},
    journal = {International Journal of Heat and Mass Transfer},
    author = {Critoph, R. E. and Turner, L.},
    number = {9},
    month = {6},
    pages = {1577--1585},
    volume = {38},
    publisher = {Pergamon},
    url = {https://www.sciencedirect.com/science/article/pii/0017931094002762?via%3Dihub},
    doi = {10.1016/0017-9310(94)00276-2},
    issn = {0017-9310}
}

@article{Askalany2016HighlyDifluoromethane,
    title = {{Highly porous activated carbon based adsorption cooling system employing difluoromethane and a mixture of pentafluoroethane and difluoromethane}},
    year = {2016},
    journal = {Heat and Mass Transfer 2016 53:1},
    author = {Askalany, Ahmed A. and Saha, Bidyut B.},
    number = {1},
    month = {3},
    pages = {107--114},
    volume = {53},
    publisher = {Springer},
    url = {https://link.springer.com/article/10.1007/s00231-016-1808-3},
    doi = {10.1007/S00231-016-1808-3},
    issn = {1432-1181}
}

@article{Ibrahim2025HydrocarbonR600a,
    title = {{Hydrocarbon refrigerants as sustainable alternatives to high-GWP refrigerants: a systematic review of R600 and R600a}},
    year = {2025},
    journal = {Journal of Thermal Analysis and Calorimetry 2025 150:21},
    author = {Ibrahim, O. A. A. M. and Kadhim, S. A. and Hammoodi, K. A. and Ashour, A. M. and Bouabidi, A. and Rashid, F. L. and Sathyamurthy, R.},
    number = {21},
    month = {9},
    pages = {17185--17206},
    volume = {150},
    publisher = {Springer},
    url = {https://link.springer.com/article/10.1007/s10973-025-14831-3},
    isbn = {0123456789},
    doi = {10.1007/S10973-025-14831-3},
    issn = {1588-2926}
}

@article{Saha2008IsothermsCarbon,
    title = {{Isotherms and thermodynamics for the adsorption of n-butane on pitch based activated carbon}},
    year = {2008},
    journal = {International Journal of Heat and Mass Transfer},
    author = {Saha, Bidyut Baran and Chakraborty, Anutosh and Koyama, Shigeru and Yoon, Seong Ho and Mochida, Isao and Kumja, M. and Yap, Christopher and Ng, Kim Choon},
    number = {7-8},
    month = {4},
    pages = {1582--1589},
    volume = {51},
    publisher = {Pergamon},
    url = {https://www.sciencedirect.com/science/article/pii/S0017931007005042},
    doi = {10.1016/J.IJHEATMASSTRANSFER.2007.07.031},
    issn = {0017-9310}
}

@article{Yasaka2023Life-CycleDestruction,
    title = {{Life-Cycle Assessment of Refrigerants for Air Conditioners Considering Reclamation and Destruction}},
    year = {2023},
    journal = {Sustainability (Switzerland)},
    author = {Yasaka, Y. and Karkour, S. and Shobatake, K. and Itsubo, N. and Yakushiji, F.},
    number = {1},
    month = {1},
    volume = {15},
    publisher = {MDPI},
    doi = {10.3390/su15010473},
    issn = {20711050}
}

@article{Wanigarathna2020MetalReview,
    title = {{Metal organic frameworks for adsorption-based separation of fluorocompounds: a review}},
    year = {2020},
    journal = {Materials Advances},
    author = {Wanigarathna, Darshika K.J.A. and Gao, Jiajian and Liu, Bin},
    number = {3},
    month = {6},
    pages = {310--320},
    volume = {1},
    publisher = {RSC},
    url = {https://pubs.rsc.org/en/content/articlehtml/2020/ma/d0ma00083c https://pubs.rsc.org/en/content/articlelanding/2020/ma/d0ma00083c},
    doi = {10.1039/d0ma00083c},
    issn = {26335409}
}

@article{Cai2020MolecularNanoparticles,
    title = {{Molecular Simulations of Adsorption and Energy Storage of R1234yf, R1234ze(z), R134a, R32, and their Mixtures in M-MOF-74 (M = Mg, Ni) Nanoparticles}},
    year = {2020},
    journal = {Scientific Reports},
    author = {Cai, Shouyin and Tian, Sen and Lu, Yuyi and Wang, Guangjin and Pu, Yu and Peng, Kang},
    number = {1},
    month = {12},
    pages = {7265-},
    volume = {10},
    publisher = {Nature Research},
    url = {https://www.nature.com/articles/s41598-020-64187-x},
    doi = {10.1038/S41598-020-64187-X;TECHMETA},
    issn = {20452322},
    pmid = {32350321}
}

@article{Hu2025MolecularIRMOF-1,
    title = {{Molecular simulations of increased thermal energy storage in metal–organic heat carriers with R1234ze(E)/R32 mixtures combined with IRMOF-1}},
    year = {2025},
    journal = {Applied Thermal Engineering},
    author = {Hu, Shujing and Cai, Shouyin and Ren, Yunxiu and Huo, Erguang and Song, Jiasheng and Zhang, Lu and Yu, Lin},
    month = {1},
    pages = {124701},
    volume = {258},
    publisher = {Pergamon},
    url = {https://www.sciencedirect.com/science/article/pii/S135943112402369X?via%3Dihub},
    doi = {10.1016/J.APPLTHERMALENG.2024.124701},
    issn = {1359-4311}
}

@article{Dubbeldam2013MONTECodes,
    title = {{MONTE CARLO CODES, TOOLS AND ALGORITHMS: On the inner workings of Monte Carlo codes}},
    year = {2013},
    journal = {Molecular Simulation},
    author = {Dubbeldam, D. and Torres-Knoop, A. and Walton, K.S.},
    number = {14-15},
    month = {12},
    pages = {1253--1292},
    volume = {39},
    publisher = {Taylor and Francis Ltd.},
    doi = {10.1080/08927022.2013.819102},
    issn = {10290435}
}

@article{Beltran-Larrotta2025NewCarbon,
    title = {{New perspectives on the models of porous carbon}},
    year = {2025},
    journal = {Computational and Structural Biotechnology Journal},
    author = {Beltr{\'{a}}n-Larrotta, Jose I. and Moreno-Piraj{\'{a}}n, Juan Carlos and Giraldo, Liliana},
    month = {1},
    pages = {156--165},
    volume = {29},
    publisher = {No longer published by Elsevier},
    url = {https://www.sciencedirect.com/science/article/pii/S200103702500145X?utm_source=copilot.com},
    doi = {10.1016/J.CSBJ.2025.04.024},
    issn = {2001-0370}
}

@article{Madero-Castro2023OnTransfer,
    title = {{On the Use of Water and Methanol with Zeolites for Heat Transfer}},
    year = {2023},
    journal = {ACS Sustainable Chemistry and Engineering},
    author = {Madero-Castro, R.M. and Luna-Triguero, A. and S{\l}awek, A. and Vicent-Luna, J.M. and Calero, S.},
    number = {11},
    month = {3},
    pages = {4317--4328},
    volume = {11},
    publisher = {American Chemical Society},
    doi = {10.1021/acssuschemeng.2c05369},
    issn = {21680485}
}

@misc{MATLABPiecewisePCHIP,
    title = {{Piecewise Cubic Hermite Interpolating Polynomial (PCHIP) }},
    author = {{MATLAB}},
    url = {https://www.mathworks.com/help/matlab/ref/pchip.html}
}

@techreport{EERA2023PolicyIntegration,
    title = {{Policy brief Fossil fuel and GHG emissions reduction through integrating industrial TES Thermal Energy System Integration}},
    year = {2023},
    author = {{EERA}},
    month = {12}
}

@article{Vuppaladadiyam2022ProgressConsequences,
    title = {{Progress in the development and use of refrigerants and unintended environmental consequences}},
    year = {2022},
    journal = {Science of The Total Environment},
    author = {Vuppaladadiyam, A.K. and Antunes, E. and Vuppaladadiyam, S. S. V. and Baig, Z. T. and Subiantoro, A. and Lei, G. and Leu, S.Y. and Sarmah, A. K. and Duan, H.},
    month = {6},
    pages = {153670},
    volume = {823},
    publisher = {Elsevier},
    url = {https://www.sciencedirect.com/science/article/pii/S0048969722007628?via%3Dihub},
    doi = {10.1016/J.SCITOTENV.2022.153670},
    issn = {0048-9697},
    pmid = {35131250}
}

@article{Iacomi2019PyGAPS:Characterisation,
    title = {{pyGAPS: a Python-based framework for adsorption isotherm processing and material characterisation}},
    year = {2019},
    journal = {Adsorption},
    author = {Iacomi, P. and Llewellyn, P. L.},
    number = {8},
    month = {11},
    pages = {1533--1542},
    volume = {25},
    publisher = {Springer New York LLC},
    doi = {10.1007/s10450-019-00168-5},
    issn = {15728757}
}

@techreport{Dubbeldam2020RASPAMaterials,
    title = {{RASPA 2.0: Molecular Software Package for Adsorption and Diffusion in (Flexible) Nanoporous Materials}},
    year = {2020},
    author = {Dubbeldam, D. and Calero, S. and Vlugt, T. J H and Ellis, D. E and Snurr, R. Q}
}

@misc{ApXMachineLearningRegressionMetrics,
    title = {{Regression Model Evaluation Metrics}},
    author = {{ApX Machine Learning}},
    url = {https://apxml.com/courses/probability-statistics-essentials-ml/chapter-6-introduction-regression-analysis/regression-evaluation-metrics}
}

@article{Sharma2023RUPTURA:Models,
    title = {{RUPTURA: simulation code for breakthrough, ideal adsorption solution theory computations, and fitting of isotherm models}},
    year = {2023},
    journal = {Molecular Simulation},
    author = {Sharma, S. and Balestra, S.R.G. and Baur, R. and Agarwal, U. and Zuidema, E. and Rigutto, M. S. and Calero, S. and Vlugt, T. J.H. and Dubbeldam, D.},
    number = {9},
    pages = {893--953},
    volume = {49},
    publisher = {Taylor and Francis Ltd.},
    doi = {10.1080/08927022.2023.2202757},
    issn = {10290435}
}

@article{Xiao2010SimulationStorage,
    title = {{Simulation of heat and mass transfer in activated carbon tank for hydrogen storage}},
    year = {2010},
    journal = {International Journal of Hydrogen Energy},
    author = {Xiao, J and Tong, L and Deng, C and B{\'{e}}nard, P and Chahine, R},
    number = {15},
    month = {8},
    pages = {8106--8116},
    volume = {35},
    publisher = {Pergamon},
    url = {https://www.sciencedirect.com/science/article/pii/S0360319910000625?via%3Dihub},
    doi = {10.1016/J.IJHYDENE.2010.01.021},
    issn = {0360-3199}
}

@article{Sosa2023SupportingPotential,
    title = {{Supporting Information: Exploring the Potential of Metal–Organic Frameworks for the Separation of Blends of Fluorinated Gases with High Global Warming Potential}},
    year = {2023},
    journal = {Global Challenges},
    author = {Sosa, J.E. and Malheiro, C. and Castro, P.J. and Ribeiro, R.P.P.L. and Pi{\~{n}}eiro, M. M. and Plantier, F. and Mota, J.P.B. and Ara{\'{u}}jo, J. M.M. and Pereiro, A. B.},
    number = {1},
    month = {1},
    volume = {7},
    publisher = {John Wiley and Sons Inc},
    doi = {10.1002/gch2.202200107},
    issn = {20566646}
}

@techreport{Gooijer2025SupportingComponents,
    title = {{Supporting Information: TAMOF-1 for Capture and Separation of the main Flue Gas Components}},
    year = {2025},
    author = {Gooijer, S and Capelo-Avil{\'{e}}s, S and Sharma, S and Giancola, S and Gal{\'{a}}n-Mascaros, J R and Vlugt, T J H and Dubbeldam, D and Vicent-Luna, J M and Calero, S}
}

@article{Chen2025SustainableAction,
    title = {{Sustainable Management of Banked Fluorocarbons as a Cost-Effective Climate Action}},
    year = {2025},
    journal = {Environmental Science and Technology},
    author = {Chen, Z. and Purohit, P. and Bai, F. and Gasser, T. and He, Y. and H{\"{o}}glund-Isaksson, L. and Jiang, P. and Wu, J. and Hu, J.},
    number = {31},
    month = {8},
    pages = {16356--16367},
    volume = {59},
    publisher = {American Chemical Society},
    url = {http://iiasa.ac.at/news/aug-2025/sustainable-management-of-refrigerants-could-be-powerful-climate-solution},
    doi = {10.1021/acs.est.5c02575},
    issn = {15205851},
    pmid = {40726035}
}

@article{Garcia2021SystematicRefrigerants,
    title = {{Systematic Search of Suitable Metal–Organic Frameworks for Thermal Energy-Storage Applications with Low Global Warming Potential Refrigerants}},
    year = {2021},
    journal = {ACS Sustainable Chemistry {\&} Engineering},
    author = {Garc{\'{i}}a, Edder J. and Bahamon, Daniel and Vega, Lourdes F.},
    number = {8},
    month = {3},
    pages = {3157--3171},
    volume = {9},
    publisher = {American Chemical Society},
    url = {/doi/pdf/10.1021/acssuschemeng.0c07797?ref=article_openPDF},
    doi = {10.1021/ACSSUSCHEMENG.0C07797},
    issn = {21680485}
}

@article{Gooijer2025TAMOF-1Components,
    title = {{TAMOF-1 for capture and separation of the main flue gas components}},
    year = {2025},
    journal = {Journal of Materials Chemistry A},
    author = {Gooijer, S. and Capelo-Avil{\'{e}}s, S. and Sharma, S. and Giancola, S. and Gal{\'{a}}n-Mascaros, J. R. and Vlugt, T. J.H. and Dubbeldam, D. and Vicent-Luna, J. M. and Calero, S.},
    number = {22},
    month = {4},
    pages = {16879--16892},
    volume = {13},
    publisher = {Royal Society of Chemistry},
    doi = {10.1039/d5ta01362c},
    issn = {20507496}
}

@misc{UnitedNationsTheAgreement,
  author       = {{United Nations}},
  title        = {The Paris Agreement},
  year         = {2015},
  publisher    = {United Nations},
  url          = {https://www.un.org/en/climatechange/paris-agreement}
}

@article{Peng2018UnderstandingCarbons,
    title = {{Understanding the Influence of Pore Heterogeneity on Water Adsorption in Realistic Molecular Models of Activated Carbons}},
    year = {2018},
    journal = {The Journal of Physical Chemistry C},
    author = {Peng, X. and Jain, S.K.},
    number = {50},
    month = {12},
    pages = {28702--28711},
    volume = {122},
    publisher = {American Chemical Society},
    url = {/doi/pdf/10.1021/acs.jpcc.8b09143?ref=article_openPDF},
    doi = {10.1021/ACS.JPCC.8B09143},
    issn = {19327455}
}

@article{Gonzalez-Galan2024UnderstandingMixtures,
    title = {{Understanding the role of open metal sites in MOFs for the efficient separation of benzene/cyclohexane mixtures}},
    year = {2024},
    journal = {Separation and Purification Technology},
    author = {Gonz{\'{a}}lez-Gal{\'{a}}n, C. and Madero-Castro, R.M. and Luna-Triguero, A. and Vicent-Luna, J.M. and Calero, S.},
    month = {11},
    pages = {127606},
    volume = {348},
    publisher = {Elsevier},
    url = {https://www.sciencedirect.com/science/article/pii/S1383586624013455?via%3Dihub},
    doi = {10.1016/J.SEPPUR.2024.127606},
    issn = {1383-5866}
}

@article{Ribeiro2023VacuumBlend,
    title = {{Vacuum swing adsorption for R-32 recovery from R-410A refrigerant blend}},
    year = {2023},
    journal = {International Journal of Refrigeration},
    author = {Ribeiro, Rui P.P.L. and Sosa, Julio E. and Ara{\'{u}}jo, João M.M. and Pereiro, Ana B. and Mota, José P.B.},
    month = {6},
    pages = {253--264},
    volume = {150},
    publisher = {Elsevier},
    url = {https://www.sciencedirect.com/science/article/pii/S0140700723000294},
    doi = {10.1016/J.IJREFRIG.2023.01.020},
    issn = {0140-7007}
}

@article{Peng2020SeparationSimulations,
    title = {{Separation of CF4/N2, C2F6/N2, and SF6/N2 Mixtures in Amorphous Activated Carbons Using Molecular Simulations}},
    year = {2020},
    journal = {ACS Applied Materials {\&} Interfaces},
    author = {Peng, X.. and Vicent-Luna, J.M. and Jin, Q.},
    number = {17},
    month = {4},
    pages = {20044--20055},
    volume = {12},
    publisher = {American Chemical Society},
    url = {https://pubs.acs.org/doi/abs/10.1021/acsami.0c01043},
    doi = {10.1021/ACSAMI.0C01043}
}

@article{Torres-Knoop2017BehaviorConditions,
    title = {{Behavior of the Enthalpy of Adsorption in Nanoporous Materials Close to Saturation Conditions}},
    year = {2017},
    journal = {Journal of Chemical Theory and Computation},
    author = {Torres-Knoop, Ariana and Poursaeidesfahani, Ali and Vlugt, Thijs J.H. and Dubbeldam, David},
    number = {7},
    month = {7},
    pages = {3326--3339},
    volume = {13},
    publisher = {American Chemical Society},
    url = {/doi/pdf/10.1021/acs.jctc.6b01193?ref=article_openPDF},
    doi = {10.1021/ACS.JCTC.6B01193}
}

@article{DeLange2015Adsorption-DrivenFrameworks,
    title = {{Adsorption-Driven Heat Pumps: The Potential of Metal–Organic Frameworks}},
    year = {2015},
    journal = {Chemical Reviews},
    author = {De Lange, M.F. and Verouden, K. J.F.M. and Vlugt, T. J.H. and Gascon, J. and Kapteijn, F.},
    number = {22},
    month = {11},
    pages = {12205--12250},
    volume = {115},
    publisher = {American Chemical Society},
    url = {https://pubs.acs.org/doi/abs/10.1021/acs.chemrev.5b00059},
    doi = {10.1021/ACS.CHEMREV.5B00059}
}

@article{Yang2024ExperimentalSystems,
    title = {{Experimental investigation and thermodynamic modeling of adsorption equilibria of MSC30 with R32 for supercritical adsorption cooling systems}},
    year = {2024},
    journal = {International Journal of Heat and Mass Transfer},
    author = {Yang, Z. and Sultan, M. and Thu, K. and Miyazaki, T.},
    month = {2},
    pages = {124873},
    volume = {219},
    publisher = {Pergamon},
    url = {https://www.sciencedirect.com/science/article/abs/pii/S0017931023010189?via%3Dihub},
    doi = {10.1016/J.IJHEATMASSTRANSFER.2023.124873}
}

@article{Xia2020AdsorptionCOF-5,
    title = {{Adsorption characteristics and cooling/heating performance of COF-5}},
    year = {2020},
    journal = {Applied Thermal Engineering},
    author = {Xia, X. and Liu, Z. and Li, S.},
    month = {7},
    pages = {115442},
    volume = {176},
    publisher = {Pergamon},
    url = {https://www.sciencedirect.com/science/article/abs/pii/S1359431119379815?via%3Dihub},
    doi = {10.1016/J.APPLTHERMALENG.2020.115442}
}

@article{Lehmann2017AssessmentApplications,
    title = {{Assessment of adsorbate density models for numerical simulations of zeolite-based heat storage applications}},
    year = {2017},
    journal = {Applied Energy},
    author = {Lehmann, C. and Beckert, S. and Gl{\"{a}}ser, R. and Kolditz, O. and Nagel, T.},
    month = {1},
    pages = {1965--1970},
    volume = {185},
    publisher = {Elsevier},
    url = {https://www.sciencedirect.com/science/article/abs/pii/S0306261915013665?via%3Dihub},
    doi = {10.1016/J.APENERGY.2015.10.126}
}

@article{Ristic2018ImprovedStorage,
    title = {{Improved performance of binder-free zeolite Y for low-temperature sorption heat storage}},
    year = {2018},
    journal = {Journal of Materials Chemistry A},
    author = {Risti{\'{c}}, A. and Fischer, F. and Hauer, A. and Zabukovec Logar, N.},
    number = {24},
    month = {6},
    pages = {11521--11530},
    volume = {6},
    publisher = {The Royal Society of Chemistry},
    url = {https://pubs.rsc.org/en/content/articlehtml/2018/ta/c8ta00827b https://pubs.rsc.org/en/content/articlelanding/2018/ta/c8ta00827b},
    doi = {10.1039/C8TA00827B}
}

@software{lucassen_2026_20134100,
  author       = {Lucassen, Hilde and
                  Luna-Triguero, Azahara and
                  Vicent-Luna, Jose Manuel},
  title        = {Mixture adsorption thermodynamics},
  month        = may,
  year         = 2026,
  publisher    = {Zenodo},
  version      = {v1.0.0},
  doi          = {10.5281/zenodo.20134100},
  url          = {https://doi.org/10.5281/zenodo.20134100},
}

@article{Hauer2010,
    author = {Hauer, A.},
    title = {Beurteilung fester Adsorbenzien auf der Basis der Adsorptionsgleichgewichte für energetische Anwendungen in offenen Systemen},
    journal = {Chemie Ingenieur Technik},
    volume = {82},
    number = {7},
    pages = {1075-1080},
    doi = {https://doi.org/10.1002/cite.201000012},
    url = {https://onlinelibrary.wiley.com/doi/abs/10.1002/cite.201000012},
    year = {2010}
}

@article{Bathia-Carbon-2016,
title = {Effect of structural anisotropy and pore-network accessibility on fluid transport in nanoporous Ti3SiC2 carbide-derived carbon},
journal = {Carbon},
volume = {103},
pages = {16-27},
year = {2016},
issn = {0008-6223},
doi = {https://doi.org/10.1016/j.carbon.2016.02.093},
url = {https://www.sciencedirect.com/science/article/pii/S0008622316301828},
author = {Amir H. Farmahini and Suresh K. Bhatia}
}

@article{Bathia-JPCC_2017,
author = {Farmahini, Amir H. and Opletal, George and Bhatia, Suresh K.},
title = {Structural Modelling of Silicon Carbide-Derived Nanoporous Carbon by Hybrid Reverse Monte Carlo Simulation},
journal = {The Journal of Physical Chemistry C},
volume = {117},
number = {27},
pages = {14081-14094},
year = {2013},
doi = {10.1021/jp403929r}
}

@article{Bathia-Langmuir-2008,
author = {Nguyen, Thanh X. and Cohaut, Nathalie and Bae, Jun-Seok and Bhatia, Suresh K.},
title = {New Method for Atomistic Modeling of the Microstructure of Activated Carbons Using Hybrid Reverse Monte Carlo Simulation},
journal = {Langmuir},
volume = {24},
number = {15},
pages = {7912-7922},
year = {2008},
doi = {10.1021/la800351d}
}

@article{JAIN20062445,
title = {Molecular modeling and adsorption properties of porous carbons},
journal = {Carbon},
volume = {44},
number = {12},
pages = {2445-2451},
year = {2006},
note = {Carbon for Energy Storage and Environment Protection},
issn = {0008-6223},
doi = {https://doi.org/10.1016/j.carbon.2006.04.034},
url = {https://www.sciencedirect.com/science/article/pii/S000862230600251X},
author = {Surendra K. Jain and Keith E. Gubbins and Roland J.-M. Pellenq and Jorge P. Pikunic}
}

@article{Jain-Langmuir-2006,
author = {Jain, Surendra K. and Pellenq, Roland J.-M. and Pikunic, Jorge P. and Gubbins, Keith E.},
title = {Molecular Modeling of Porous Carbons Using the Hybrid Reverse Monte Carlo Method},
journal = {Langmuir},
volume = {22},
number = {24},
pages = {9942-9948},
year = {2006},
doi = {10.1021/la053402z}
}





\newpage

\renewcommand{\figurename}{Figure}
\renewcommand{\tablename}{Table}
\renewcommand{\thetable}{S\arabic{table}}  
\renewcommand{\thefigure}{S\arabic{figure}} 

\renewcommand{\theHfigure}{S\arabic{figure}}
\renewcommand{\theHtable}{S\arabic{table}}

\setcounter{figure}{0}
\setcounter{table}{0}

\onecolumngrid

\begin{center}
    \textbf{\Huge{Supplementary Material}}

    \vspace{1.0cm}

    for

    \vspace{1.0cm}

    \textbf{\Large{Evaluating Blended Refrigerants for Thermochemical Energy Storage and Circular Refrigerant Recovery using Activated Carbons}}
\end{center}

   \newpage

\clearpage
\setcounter{section}{0}
\renewcommand{\thesection}{S\arabic{section}}

\section*{Supplementary Material}

\section{Methodology}\label{Appendix: Methodology}

\subsection{Python-based software package description}\label{Github repository}
The custom Python-based software package is available through the GitHub repository \nolinkurl{https://github.com/HildeLucassen/mixture-adsorption-thermodynamics.git} and through Zenodo at \nolinkurl{https://doi.org/10.5281/zenodo.20134099}. \autoref{fig:Model workflow overview} presents an overview of the software-package workflow. Detailed instructions for use are provided in the GitHub repository.

\begin{figure}[H]
    \centering
    \includegraphics[width=\linewidth]{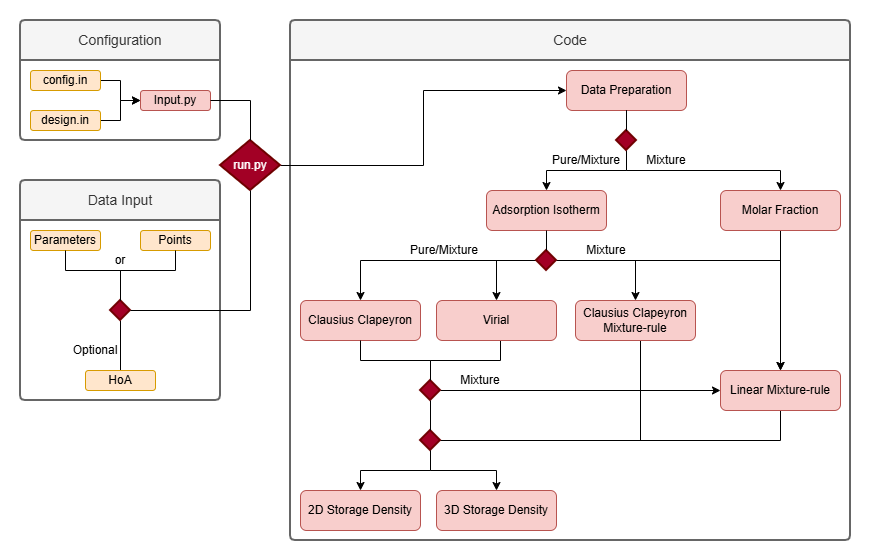}
    \caption{Python-based software package workflow.}
    \label{fig:Model workflow overview}
\end{figure}

 The model requires initial input in the form of equilibrium points obtained experimentally or numerically, or as parameters of multi-site adsorption models, Toth, Langmuir-Freundlich, or Sips. For a mixture, the user must provide adsorption data for each component. In the configuration file, the user specifies the type of input provided and the expected output. The following steps summarize the workflow implemented in the code.

\hspace*{0.5cm} \textbf{1. Data Preparation:} The data-preparation procedure depends on the type of input. When adsorption points are provided, the code creates a new list of points that is used to calculate the heat of adsorption. This new list incorporates the numerical constraint imposed by the Clausius--Clapeyron equation, which requires temperature--pressure points corresponding to the same loading. PCHIP interpolation is applied in regions where the data points are sufficiently dense to support a smooth, shape-preserving curve, while linear interpolation is used in segments where the spacing between points becomes disproportionately large. This prevents artificial curvature in poorly sampled regions.

\hspace*{0.5cm} \textbf{2a. Adsorption Isotherm:} This step plots the adsorption isotherm from the provided parameters or by drawing a line through the original data points.

\hspace*{0.5cm} \textbf{2b. Molar Fraction:} In the case of mixtures, the adsorbed-phase molar fraction is calculated.

\hspace*{0.5cm} \textbf{3a. Clausius--Clapeyron:} \autoref{eq: Clausius Claperyon} is used to calculate the heat of adsorption from the data prepared in step 1.

\hspace*{0.5cm} \textbf{3b. Virial:} \autoref{eq: Virial} is used to calculate the heat of adsorption in combination with a numerical optimizer (BFGS) to determine the optimal polynomial coefficients. This step uses the data prepared in step 1.

\hspace*{0.5cm} \textbf{3c. Provided:} The user can also provide heat-of-adsorption results, for example from experiments, as input for the model.

\hspace*{0.5cm} \textbf{4a. Clausius--Clapeyron Mixing Rule:} \autoref{eq: clausius clapeyron mixture} is used to calculate the heat of adsorption per component and for the overall mixture. This step uses the data prepared in step 1.

\hspace*{0.5cm} \textbf{4b. Linear Mixing Rule:} Depending on the user settings, the linear mixing equation, \autoref{eq: linear mixing rule}, uses the results from step 3 together with the molar fractions to calculate the overall heat of adsorption of the mixture.

\hspace*{0.5cm} \textbf{5a. 2D Storage Density:} This step uses step 3 or step 4 to calculate the storage density with \autoref{eq: Storage density} and plots the results in two dimensions. The 2D storage-density plot can illustrate three operational scenarios: pressure-swing operation, temperature-swing operation, and combined pressure--temperature-swing operation.

\hspace*{0.5cm} \textbf{5b. 3D Storage Density:} This step performs the same calculation as step 5a, but generates three-dimensional storage-density surfaces.

\subsection{Pure Component Adsorption}\label{subsec: Settings Raspa}

For the initial configuration and as a reference for the adsorption potential theory applied to the adsorption isotherm, Monte Carlo simulations were performed using RASPA \cite{Dubbeldam2020RASPAMaterials}. The simulations were conducted for 50,000 production cycles preceded by 10,000 equilibration cycles. In each cycle, trial moves were attempted on randomly selected adsorbed molecules. In \autoref{Rosenbluth Factor and Internal Energy  of refrigerants}, the Rosenbluth factor is reported for each temperature; this factor accounts for the configurational bias associated with molecular displacements. For all activated carbon structures, a $1\times1\times1$ unit cell was used. The refrigerant molecules were considered rigid. The probabilities for translation and rotation moves were each set to 5.0, with a swap probability and fugacity coefficient of 1. The Ewald summation scheme was applied to evaluate electrostatic interactions, using a cutoff of 12~\AA{} \cite{Gooijer2025SupportingComponents}. \autoref{tab:framework_LJ_parameters} reports the Lennard-Jones parameters for both the activated carbons and the refrigerants, together with their corresponding partial charges. Grand Canonical Monte Carlo (GCMC) simulations are additionally employed as a reference for evaluating the proposed methods for calculating the isosteric heat of adsorption. In RASPA, the fluctuation method is implemented; however, the internal energy of each molecule must be subtracted from the RASPA output. The corresponding internal energy values are provided in \autoref{Rosenbluth Factor and Internal Energy  of refrigerants}.

\begin{table}[H]
\centering
\setlength{\tabcolsep}{4pt}
\renewcommand{\arraystretch}{0.95}
\begin{tabular}{c cc cc cc cc}
\hline
\textbf{T (K)}
& \multicolumn{2}{c}{\textbf{R32}}
& \multicolumn{2}{c}{\textbf{R125}}
& \multicolumn{2}{c}{\textbf{R134a}}
& \multicolumn{2}{c}{\textbf{R600}} \\
& \textbf{$U_{R32}$} & \textbf{$W_{R32}$}
& \textbf{$U_{R125}$} & \textbf{$W_{R125}$}
& \textbf{$U_{R134a}$} & \textbf{$W_{R134a}$}
& \textbf{$U_{R600}$} & \textbf{$W_{R600}$} \\
\hline
283 & 5.88 & 1.00 & 12.14 & 0.06 & 11.96 & 0.06 & 8.34 & 0.12 \\
303 & 6.29 & 1.00 & 13.06 & 0.06 & 12.88 & 0.06 & 8.94 & 0.13 \\
333 & 6.92 & 1.00 & 14.42 & 0.06 & 14.28 & 0.06 & 9.84 & 0.15 \\
353 & 7.34 & 1.00 & 15.35 & 0.06 & 15.18 & 0.06 & 10.41 & 0.15 \\
\hline
\end{tabular}
\caption{Internal energy ($U_g$ [kJ/mol]) and Rosenbluth factor ($W_i$ [-]) of refrigerants.}
\label{Rosenbluth Factor and Internal Energy  of refrigerants}
\end{table}

\begin{table}[H]
\centering
\setlength{\tabcolsep}{5pt}
\renewcommand{\arraystretch}{0.95}
\begin{tabular}{l c c c l c c c}
\hline
\multicolumn{8}{c}{\textbf{Refrigerant}} \\
\hline
\textbf{Atom} & \textbf{$\varepsilon/k_B$ [K]} & \textbf{$\sigma$ \AA} & \textbf{$q$ [e$^-$]}
& \textbf{Atom} & \textbf{$\varepsilon/k_B$} & \textbf{$\sigma$} & \textbf{$q$ [e$^-$]} \\
\hline
C\_f1 & 47.00 & 3.60  & 0.002  & CH2    & 56.00  & 3.96  & 0.000 \\
C\_f2 & 47.00 & 3.60  & 0.3085 & CH3    & 108.00 & 3.76  & 0.000 \\
C\_f3 & 47.00 & 3.60  & 0.534  & C\_R32 & 54.60  & 3.15  & 0.214 \\
F     & 24.50 & 2.92  & -0.1915& H\_R32 & 10.00  & 2.17  & 0.097 \\
H\_cf & 10.40 & 2.50  & 0.115  & F\_R32 & 40.00  & 2.975 & -0.2040 \\
\hline
\multicolumn{8}{c}{\textbf{Activated Carbon}} \\
\hline
\textbf{Atom} & \textbf{$\varepsilon/k_B$ {K}} & \textbf{$\sigma$ \AA} & \textbf{$q$[e$^-$]}
& \textbf{Atom} & \textbf{$\varepsilon/k_B$ [K]} & \textbf{$\sigma$ \AA} & \textbf{$q$[e$^-$]} \\
\hline
C & 28.00 & 3.36 & 0.000 & H3 & 15.08 & 2.42 & 0.000 \\
\hline
\end{tabular}
\caption{Lennard--Jones parameters and partial charges.}
\label{tab:framework_LJ_parameters}
\end{table}


\section{Heat of adsorption validation with Reference Literature}\label{appendix: Validation code}

Gooijer et al. (2025) investigated the adsorption of CO\textsubscript{2} and N\textsubscript{2} in TAMOF-1, a recently developed metal--organic framework (MOF) \cite{Gooijer2025TAMOF-1Components}. In their study, GCMC simulations reproduce the experimentally measured adsorption isotherm data points. The resulting isotherms are subsequently correlated using a multi-site Sips adsorption model. For validation, the simulated CO\textsubscript{2} adsorption behavior in TAMOF-1 is compared with experimental measurements. In addition, the study computes the isosteric heat of adsorption via GCMC simulations using the fluctuation method and compares this to the experimental values. The pressure range in the experiments of TAMOF-1 is $10^2$-$10^6$ Pa. Queen et al. (2014) conducted a comprehensive comparative study of the adsorption of CO\textsubscript{2} in the metal-organic framework series M\textsubscript{2}(dobdc) (M = Mg, Mn, Fe, Co, Ni, Cu, Zn) \cite{Queen2014ComprehensiveZn}. Adsorption measurements were carried out at temperatures of $298$, $308$, and $318$ K. The experimental isotherms were correlated using a dual-site Langmuir-Freundlich model. In the case of MOF-74, the experimental pressure range is $10^2$ to $5\cdot 10^6$ Pa. For model validation, only the isotherm data for Co\textsubscript{2}(dobdc) were compared.

First, the experimental and simulation approaches for CO\textsubscript{2}/TAMOF-1 are used to compare the methodology of this paper against the literature. \autoref{supplementary fig: validation code} compares the isosteric heat of adsorption obtained from experimental data points in panel (a) with the corresponding values calculated using the fluctuation method in panel (b). Since the experimental data are not fitted to an adsorption isotherm model, interpolation is used to create intermediate data points needed to calculate the isosteric heat of adsorption. For the Virial method, polynomial degrees of 6 and 5 were selected for polynomials \(a\) and \(b\), respectively. Both the Clausius-Clapeyron and Virial methods show similar agreement, confirming the method. In addition, both equations were validated using the working pair CO\textsubscript{2}/Co\textsubscript{2}(dobdc). For each equation, adsorption-isotherm fitting was employed to determine the isosteric heat of adsorption as a function of loading. The resulting values exhibit good agreement with the experimental data in \autoref{supplementary fig: validation code}c.

\begin{figure}[H]
    \centering
    \begin{tikzpicture}
        \node[anchor=south west, inner sep=0] (image) at (0,0)
            {\includegraphics[width=\linewidth]{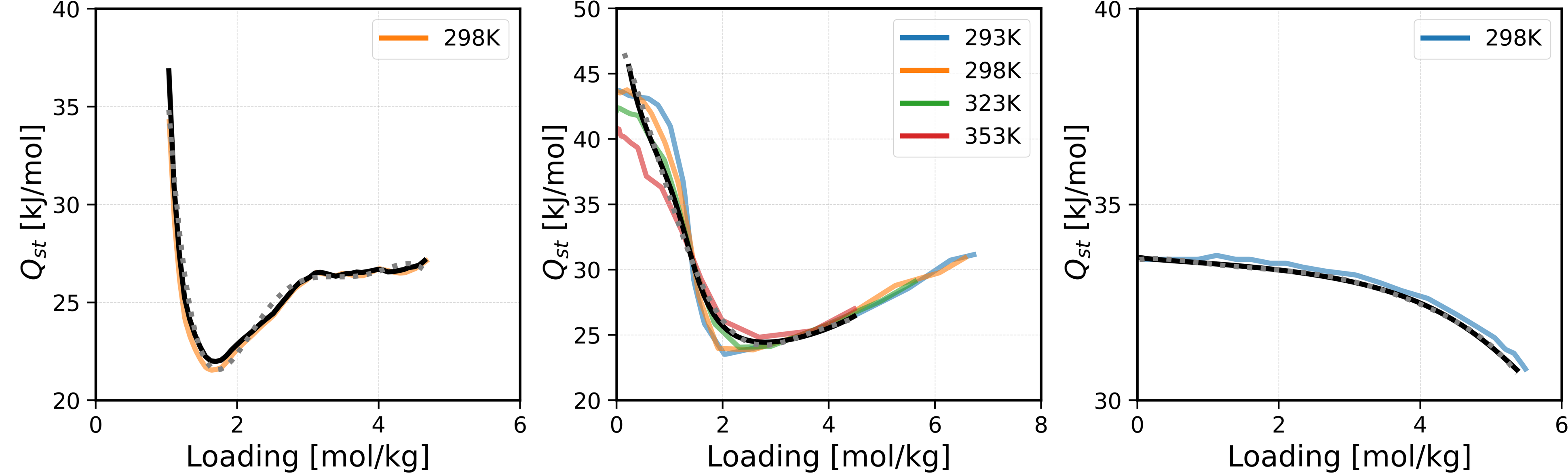}};
        \begin{scope}[x={(image.south east)}, y={(image.north west)}]
            \node[font=\bfseries, anchor=west] at (0.00,1.05) {a)};
            \node[font=\bfseries]             at (0.37,1.05) {b)};
            \node[font=\bfseries]             at (0.70,1.05) {c)};
        \end{scope}
    \end{tikzpicture}
    \caption{Heat of validation for CO\textsubscript{2}/TAMOF-1 using Clausius-Clapeyron (-) and Virial (:) approaches of experimental (a) and simulation data (b) and CO\textsubscript{2}/MOF-74 using Clausius-Clapeyron (-) and Virial (:) with adsorption isotherms of experimental data (c). The reference data are represented by colored solid lines.}
    \label{supplementary fig: validation code}
\end{figure}

\newpage

\section{Results and Discussion}\label{Appendix: Results and Discussion}

\subsection{Pure component adsorption}\label{supplementary: Adsorption isotherm supplementary}
Isotherm fitting was performed using the RUPTURA software package \cite{Sharma2023RUPTURA:Models}. Two adsorption isotherm models were evaluated and compared. Both models yielded comparable fitting quality; however, the dual-site Sips model slightly outperforms the dual-site Langmuir-Freundlich model, as summarized in \autoref{tab: Fit quality metrics for the Dual-site Sips and Dual-site Langmuir-Freundlich models.}.

\begin{table}[H]
\centering
\setlength{\tabcolsep}{4pt}
\renewcommand{\arraystretch}{0.95}
\begin{tabular}{l l r r r r r r}
\hline
& &
\multicolumn{3}{c}{\textbf{Dual-site Sips}} &
\multicolumn{3}{c}{\textbf{Dual-site LF}} \\
\textbf{Material} &
\textbf{Refrigerant} &
\textbf{RSS} &
\textbf{$R^2$} &
\textbf{RMSE} &
\textbf{RSS} &
\textbf{$R^2$} &
\textbf{RMSE} \\
\hline
\multirow{4}{*}{Bhatia-01}
& R32   & 0.0000102 & 1.0000 & 0.000853 & 0.01417 & 0.9999 & 0.03181 \\
& R125  & 0.002855  & 1.0000 & 0.01428  & 0.01462 & 0.9998 & 0.03232 \\
& R134a & 0.01460   & 0.9998 & 0.03229  & 0.01253 & 0.9999 & 0.02992 \\
& R600  & 0.04339   & 0.9994 & 0.05567  & 0.07874 & 0.9989 & 0.07500 \\
\hline
\multirow{4}{*}{CS1000a}
& R32   & 0.0005627 & 1.0000 & 0.00634  & 0.004257 & 1.0000 & 0.01744 \\
& R125  & 0.004849  & 1.0000 & 0.01861  & 0.02708  & 0.9998 & 0.04398 \\
& R134a & 0.09290   & 0.9994 & 0.08146  & 0.10310  & 0.9993 & 0.08581 \\
& R600  & 0.02749   & 0.9998 & 0.04431  & 0.14360  & 0.9990 & 0.10130 \\
\hline
\end{tabular}
\caption{Fit quality metrics for the dual-site Sips and dual-site Langmuir-Freundlich (LF) models.}
\label{tab: Fit quality metrics for the Dual-site Sips and Dual-site Langmuir-Freundlich models.}
\end{table}

\subsubsection{Dual-site Sips parameters}\label{sec: Dual-site Sips parameters}
Based on \autoref{tab: Fit quality metrics for the Dual-site Sips and Dual-site Langmuir-Freundlich models.}, the equilibrium simulation points from GCMC were fitted using the dual-site Sips adsorption model (\autoref{Eq: Sips}). The resulting parameters are reported in \autoref{Sips parameters for Bhatia_01}-\autoref{Sips parameters for CS1000a}. $q_{sat}$ is given in mol/kg, $b$ in Pa$^{-1}$, and $\upsilon$ is dimensionless. The fitted adsorption isotherm models and equilibrium simulation points are visualized in \autoref{fig: Adsorption isotherms pure components Bhatia-1 and CS1000a} and \autoref{supplementary fig: Adsorption isotherms pure components other structures}.

\begin{table}[H]
\centering
\setlength{\tabcolsep}{5pt}
\renewcommand{\arraystretch}{0.95}
\begin{tabular}{c c c c c c c}
\hline
\multirow{2}{*}{\textbf{T(K)}} 
& \multicolumn{3}{c}{\textbf{R32}}
& \multicolumn{3}{c}{\textbf{R125}} \\
& $q_{sat}$ & $b$ & $\upsilon$
& $q_{sat}$ & $b$ & $\upsilon$ \\
\hline
\multirow{2}{*}{283}
& 9.47 & \(3.69 \cdot 10^{-6}\) & 0.9433
& 3.61 & \(1.69 \cdot 10^{-5}\) & 0.9945 \\
& 2.84 & \(2.13 \cdot 10^{-6}\) & 0.3723
& 2.99 & \(8.89 \cdot 10^{-5}\) & 1.556 \\
\hline
\multirow{2}{*}{303}
& 8.40 & \(1.34 \cdot 10^{-6}\) & 0.5943
& 3.07 & \(4.26 \cdot 10^{-5}\) & 1.4248 \\
& 3.00 & \(6.83 \cdot 10^{-6}\) & 0.9504
& 3.20 & \(8.26 \cdot 10^{-6}\) & 0.9044 \\
\hline
\multirow{2}{*}{333}
& 4.73 & \(7.20 \cdot 10^{-7}\) & 0.4323
& 0.5574 & \(1.24 \cdot 10^{-4}\) & 1.1140 \\
& 5.47 & \(1.61 \cdot 10^{-6}\) & 0.9783
& 5.57 & \(4.78 \cdot 10^{-6}\) & 1.1399 \\
\hline
\multirow{2}{*}{353}
& 2.26 & \(2.47 \cdot 10^{-6}\) & 0.6430
& 1.41 & \(2.60 \cdot 10^{-5}\) & 1.1771 \\
& 5.12 $\cdot 10^{2}$ & \(3.15 \cdot 10^{-9}\) & 1.0097
& 4.34 & \(2.44 \cdot 10^{-6}\) & 0.9641 \\
\hline
\multicolumn{1}{c}{}
& \multicolumn{3}{c}{\textbf{R134a}}
& \multicolumn{3}{c}{\textbf{R600}} \\
\multicolumn{1}{c}{}
& $q_{sat}$ & $b$ & $\upsilon$
& $q_{sat}$ & $b$ & $\upsilon$ \\
\hline
\multirow{2}{*}{283}
& 1.0012 & \(6.11 \cdot 10^{-4}\) & 1.359
& 1.8544 & \(3.69 \cdot 10^{-4}\) & 0.6929 \\
& 6.09 & \(2.88 \cdot 10^{-5}\) & 0.9879
& 4.5116 & \(1.04 \cdot 10^{-3}\) & 2.431 \\
\hline
\multirow{2}{*}{303}
& 4.21 & \(1.19 \cdot 10^{-5}\) & 0.9515
& 5.2116 & \(1.64 \cdot 10^{-4}\) & 1.3365 \\
& 2.74 & \(3.39 \cdot 10^{-5}\) & 1.4958
& 0.8355 & \(4.65 \cdot 10^{-2}\) & 1.4318 \\
\hline
\multirow{2}{*}{333}
& 1.97 & \(2.49 \cdot 10^{-5}\) & 1.2929
& 1.9720 & \(6.42 \cdot 10^{-4}\) & 1.9677 \\
& 4.52 & \(4.47 \cdot 10^{-6}\) & 0.9085
& 3.86 & \(3.94 \cdot 10^{-5}\) & 1.1628 \\
\hline
\multirow{2}{*}{353}
& 5.41 & \(3.68 \cdot 10^{-6}\) & 1.0801
& 2.1628 & \(2.03 \cdot 10^{-4}\) & 1.9144 \\
& 1.46 & \(1.24 \cdot 10^{-6}\) & 2.0065
& 3.56 & \(2.05 \cdot 10^{-5}\) & 1.1215 \\
\hline
\end{tabular}
\caption{Sips parameters for Bhatia-01.}
\label{Sips parameters for Bhatia_01}
\end{table}

\begin{table}[H]
\centering
\setlength{\tabcolsep}{5pt}
\renewcommand{\arraystretch}{0.95}
\begin{tabular}{c c c c c c c}
\hline
\multirow{2}{*}{\textbf{T(K)}} 
& \multicolumn{3}{c}{\textbf{R32}}
& \multicolumn{3}{c}{\textbf{R125}} \\
& $q_{sat}$ & $b$ & $\upsilon$
& $q_{sat}$ & $b$ & $\upsilon$ \\
\hline
\multirow{2}{*}{283}
& 3.15 & \(1.69 \cdot 10^{-6}\) & 1.554
& 3.02 & \(2.67 \cdot 10^{-5}\) & 1.932 \\
& 8.86 & \(4.03 \cdot 10^{-6}\) & 0.5979
& 3.46 & \(5.74 \cdot 10^{-5}\) & 0.7742 \\
\hline
\multirow{2}{*}{303}
& 8.97 & \(1.91 \cdot 10^{-6}\) & 0.5602
& 4.53 & \(1.91 \cdot 10^{-5}\) & 1.480 \\
& 1.61 & \(9.93 \cdot 10^{-6}\) & 0.9761
& 1.65 & \(2.45 \cdot 10^{-5}\) & 0.5370 \\
\hline
\multirow{2}{*}{333}
& 5.58 & \(9.29 \cdot 10^{-7}\) & 0.4962
& 3.61 & \(8.03 \cdot 10^{-6}\) & 1.435 \\
& 4.37 & \(1.50 \cdot 10^{-6}\) & 1.008
& 2.22 & \(8.23 \cdot 10^{-6}\) & 0.7408 \\
\hline
\multirow{2}{*}{353}
& 3.94 & \(6.63 \cdot 10^{-7}\) & 0.4297
& 0.747 & \(1.04 \cdot 10^{-5}\) & 1.732 \\
& 5.41 & \(8.91 \cdot 10^{-7}\) & 0.9721
& 4.79 & \(4.72 \cdot 10^{-6}\) & 1.015 \\
\hline
\multicolumn{1}{c}{}
& \multicolumn{3}{c}{\textbf{R134a}}
& \multicolumn{3}{c}{\textbf{R600}} \\
\multicolumn{1}{c}{}
& $q_{sat}$ & $b$ & $\upsilon$
& $q_{sat}$ & $b$ & $\upsilon$ \\
\hline
\multirow{2}{*}{283}
& 3.30 & \(3.14 \cdot 10^{-5}\) & 1.835
& 1.219 & \(9.81 \cdot 10^{-4}\) & 0.3639 \\
& 3.71 & \(6.43 \cdot 10^{-5}\) & 0.5725
& 4.592 & \(1.54 \cdot 10^{-3}\) & 1.781 \\
\hline
\multirow{2}{*}{303}
& 5.50 & \(2.27 \cdot 10^{-5}\) & 1.236
& 3.818 & \(4.54 \cdot 10^{-4}\) & 1.939 \\
& 1.13 & \(2.76 \cdot 10^{-5}\) & 0.1864
& 1.927 & \(4.30 \cdot 10^{-4}\) & 0.6426 \\
\hline
\multirow{2}{*}{333}
& 3.36 & \(8.86 \cdot 10^{-6}\) & 0.7079
& 2.238 & \(1.32 \cdot 10^{-4}\) & 0.7426 \\
& 2.91 & \(8.58 \cdot 10^{-6}\) & 1.485
& 3.39 & \(9.85 \cdot 10^{-5}\) & 1.963 \\
\hline
\multirow{2}{*}{353}
& 4.97 & \(5.13 \cdot 10^{-6}\) & 1.168
& 1.31 & \(6.11 \cdot 10^{-5}\) & 0.5840 \\
& 1.09 & \(4.96 \cdot 10^{-6}\) & 0.4524
& 4.16 & \(5.33 \cdot 10^{-5}\) & 1.652 \\
\hline
\end{tabular}
\caption{Sips parameters for Bhatia-02.}
\end{table}

\begin{table}[H]
\centering
\setlength{\tabcolsep}{5pt}
\renewcommand{\arraystretch}{0.95}
\begin{tabular}{c c c c c c c}
\hline
\multirow{2}{*}{\textbf{T(K)}} 
& \multicolumn{3}{c}{\textbf{R32}}
& \multicolumn{3}{c}{\textbf{R125}} \\
& $q_{sat}$ & $b$ & $\upsilon$
& $q_{sat}$ & $b$ & $\upsilon$ \\
\hline
\multirow{2}{*}{283}
& 7.20 & \(1.94 \cdot 10^{-6}\) & 0.5077
& 3.84 & \(1.27 \cdot 10^{-5}\) & 0.8341 \\
& 1.68 & \(2.47 \cdot 10^{-5}\) & 0.6019
& 1.0975 & \(7.34 \cdot 10^{-4}\) & 0.9800 \\
\hline
\multirow{2}{*}{303}
& 5.44 & \(9.48 \cdot 10^{-7}\) & 0.5551
& 3.96 & \(1.39 \cdot 10^{-5}\) & 1.606 \\
& 3.31 & \(4.07 \cdot 10^{-6}\) & 0.9666
& 1.08 & \(5.54 \cdot 10^{-6}\) & 0.2761 \\
\hline
\multirow{2}{*}{333}
& 2.34 & \(2.19 \cdot 10^{-6}\) & 1.001
& 3.49 & \(2.59 \cdot 10^{-6}\) & 0.7910 \\
& 8.52 & \(3.21 \cdot 10^{-7}\) & 0.7620
& 1.08 & \(8.65 \cdot 10^{-5}\) & 1.006 \\
\hline
\multirow{2}{*}{353}
& 1.23 & \(4.57 \cdot 10^{-6}\) & 0.7981
& 3.49 & \(1.84 \cdot 10^{-6}\) & 0.8093 \\
& 5.31 & \(5.37 \cdot 10^{-7}\) & 0.5320
& 0.883 & \(5.88 \cdot 10^{-5}\) & 0.9034 \\
\hline
\multicolumn{1}{c}{}
& \multicolumn{3}{c}{\textbf{R134a}}
& \multicolumn{3}{c}{\textbf{R600}} \\
\multicolumn{1}{c}{}
& $q_{sat}$ & $b$ & $\upsilon$
& $q_{sat}$ & $b$ & $\upsilon$ \\
\hline
\multirow{2}{*}{283}
& 4.28 & \(3.19 \cdot 10^{-5}\) & 1.598
& 2.849 & \(1.68 \cdot 10^{-3}\) & 2.499 \\
& 1.34 & \(1.90 \cdot 10^{-5}\) & 0.2268
& 1.892 & \(1.96 \cdot 10^{-4}\) & 0.4369 \\
\hline
\multirow{2}{*}{303}
& 3.42 & \(2.46 \cdot 10^{-5}\) & 1.447
& 2.559 & \(8.77 \cdot 10^{-4}\) & 2.182 \\
& 1.80 & \(6.53 \cdot 10^{-6}\) & 0.3542
& 1.99 & \(6.58 \cdot 10^{-5}\) & 0.4871 \\
\hline
\multirow{2}{*}{333}
& 1.29 & \(4.20 \cdot 10^{-5}\) & 1.391
& 3.45 & \(8.23 \cdot 10^{-5}\) & 1.988 \\
& 3.78 & \(2.93 \cdot 10^{-6}\) & 0.8314
& 1.13 & \(1.98 \cdot 10^{-5}\) & 0.3057 \\
\hline
\multirow{2}{*}{353}
& 1.63 & \(2.02 \cdot 10^{-5}\) & 1.109
& 2.56 & \(1.09 \cdot 10^{-5}\) & 0.7686 \\
& 3.00 & \(1.80 \cdot 10^{-6}\) & 0.6033
& 1.694 & \(2.93 \cdot 10^{-4}\) & 1.584 \\
\hline
\end{tabular}
\caption{Sips parameters for Bhatia-03.}
\end{table}

\begin{table}[H]
\centering
\setlength{\tabcolsep}{5pt}
\renewcommand{\arraystretch}{0.95}
\begin{tabular}{c c c c c c c}
\hline
\multirow{2}{*}{\textbf{T(K)}} 
& \multicolumn{3}{c}{\textbf{R32}}
& \multicolumn{3}{c}{\textbf{R125}} \\
& $q_{sat}$ & $b$ & $\upsilon$
& $q_{sat}$ & $b$ & $\upsilon$ \\
\hline
\multirow{2}{*}{283}
& 2.21 & \(3.84 \cdot 10^{-6}\) & 1.001
& 1.01 & \(2.56 \cdot 10^{-5}\) & 1.779 \\
& 0.501 & \(8.53 \cdot 10^{-7}\) & 0.2702
& 0.535 & \(4.37 \cdot 10^{-8}\) & 1.506 \\
\hline
\multirow{2}{*}{303}
& 1.01 & \(5.35 \cdot 10^{-6}\) & 0.8936
& 0.630 & \(3.89 \cdot 10^{-5}\) & 1.343 \\
& 1.39 & \(8.68 \cdot 10^{-7}\) & 0.6339
& 0.401 & \(1.22 \cdot 10^{-6}\) & 1.037 \\
\hline
\multirow{2}{*}{333}
& 2.00 & \(8.41 \cdot 10^{-7}\) & 1.053
& 0.165 & \(4.38 \cdot 10^{-5}\) & 0.8643 \\
& 0.830 & \(1.26 \cdot 10^{-7}\) & 0.7072
& 0.871 & \(1.46 \cdot 10^{-6}\) & 1.571 \\
\hline
\multirow{2}{*}{353}
& 0.850 & \(8.12 \cdot 10^{-7}\) & 1.079
& 0.760 & \(1.17 \cdot 10^{-6}\) & 1.357 \\
& 1.41 & \(3.19 \cdot 10^{-7}\) & 0.9571
& 0.145 & \(3.16 \cdot 10^{-5}\) & 0.8881 \\
\hline
\multicolumn{1}{c}{}
& \multicolumn{3}{c}{\textbf{R134a}}
& \multicolumn{3}{c}{\textbf{R600}} \\
\multicolumn{1}{c}{}
& $q_{sat}$ & $b$ & $\upsilon$
& $q_{sat}$ & $b$ & $\upsilon$ \\
\hline
\multirow{2}{*}{283}
& 1.53 & \(1.76 \cdot 10^{-8}\) & 1.291
& 0.490 & \(2.45 \cdot 10^{-6}\) & 1.926 \\
& 1.08 & \(3.68 \cdot 10^{-5}\) & 1.475
& 0.7995 & \(3.52 \cdot 10^{-3}\) & 1.552 \\
\hline
\multirow{2}{*}{303}
& 0.776 & \(8.65 \cdot 10^{-7}\) & 1.727
& 0.572 & \(9.27 \cdot 10^{-7}\) & 2.369 \\
& 0.602 & \(3.39 \cdot 10^{-5}\) & 1.234
& 0.737 & \(1.02 \cdot 10^{-3}\) & 1.464 \\
\hline
\multirow{2}{*}{333}
& 0.500 & \(7.33 \cdot 10^{-7}\) & 0.9332
& 0.7041 & \(1.99 \cdot 10^{-4}\) & 1.391 \\
& 0.598 & \(1.46 \cdot 10^{-5}\) & 1.164
& 0.796 & \(8.45 \cdot 10^{-8}\) & 2.461 \\
\hline
\multirow{2}{*}{353}
& 0.241 & \(1.79 \cdot 10^{-5}\) & 1.035
& 0.759 & \(7.59 \cdot 10^{-5}\) & 1.410 \\
& 0.783 & \(1.13 \cdot 10^{-6}\) & 1.220
& 0.253 & \(9.47 \cdot 10^{-7}\) & 1.046 \\
\hline
\end{tabular}
\caption{Sips parameters for CS400.}
\end{table}

\begin{table}[H]
\centering
\setlength{\tabcolsep}{5pt}
\renewcommand{\arraystretch}{0.95}
\begin{tabular}{c c c c c c c}
\hline
\multirow{2}{*}{\textbf{T(K)}} 
& \multicolumn{3}{c}{\textbf{R32}}
& \multicolumn{3}{c}{\textbf{R125}} \\
& $q_{sat}$ & $b$ & $\upsilon$
& $q_{sat}$ & $b$ & $\upsilon$ \\
\hline
\multirow{2}{*}{283}
& 1.09 & \(7.89 \cdot 10^{-8}\) & 1.631
& 0.361 & \(7.84 \cdot 10^{-6}\) & 2.251 \\
& 1.25 & \(1.05 \cdot 10^{-5}\) & 1.159
& 0.2251 & \(2.11 \cdot 10^{-3}\) & 1.321 \\
\hline
\multirow{2}{*}{303}
& 0.461 & \(5.68 \cdot 10^{-7}\) & 0.8257
& 0.284 & \(5.92 \cdot 10^{-7}\) & 1.544 \\
& 1.10 & \(6.69 \cdot 10^{-6}\) & 1.102
& 0.3274 & \(3.75 \cdot 10^{-4}\) & 1.485 \\
\hline
\multirow{2}{*}{333}
& 0.228 & \(4.73 \cdot 10^{-6}\) & 0.9174
& 0.397 & \(5.57 \cdot 10^{-8}\) & 1.748 \\
& 1.22 & \(1.03 \cdot 10^{-6}\) & 1.233
& 0.325 & \(7.80 \cdot 10^{-5}\) & 1.489 \\
\hline
\multirow{2}{*}{353}
& 0.140 & \(9.68 \cdot 10^{-7}\) & 0.02810
& 0.120 & \(8.51 \cdot 10^{-7}\) & 0.9799 \\
& 0.962 & \(1.73 \cdot 10^{-6}\) & 1.074
& 0.312 & \(3.91 \cdot 10^{-5}\) & 1.438 \\
\hline
\multicolumn{1}{c}{}
& \multicolumn{3}{c}{\textbf{R134a}}
& \multicolumn{3}{c}{\textbf{R600}} \\
\multicolumn{1}{c}{}
& $q_{sat}$ & $b$ & $\upsilon$
& $q_{sat}$ & $b$ & $\upsilon$ \\
\hline
\multirow{2}{*}{283}
& 0.2740 & \(1.47 \cdot 10^{-3}\) & 1.289
& 0.266 & \(4.14 \cdot 10^{-6}\) & 2.317 \\
& 0.382 & \(7.70 \cdot 10^{-6}\) & 2.086
& 0.4088 & \(6.02 \cdot 10^{-2}\) & 2.155 \\
\hline
\multirow{2}{*}{303}
& 0.313 & \(2.24 \cdot 10^{-6}\) & 1.705
& 0.4649 & \(6.86 \cdot 10^{-3}\) & 2.285 \\
& 0.3080 & \(4.46 \cdot 10^{-4}\) & 1.334
& 0.257 & \(2.46 \cdot 10^{-7}\) & 1.764 \\
\hline
\multirow{2}{*}{333}
& 0.187 & \(6.69 \cdot 10^{-7}\) & 1.414
& 0.368 & \(2.90 \cdot 10^{-5}\) & 2.615 \\
& 0.383 & \(5.48 \cdot 10^{-5}\) & 1.551
& 0.2057 & \(5.94 \cdot 10^{-3}\) & 1.466 \\
\hline
\multirow{2}{*}{353}
& 0.297 & \(4.91 \cdot 10^{-5}\) & 1.285
& 0.363 & \(5.09 \cdot 10^{-6}\) & 2.413 \\
& 0.187 & \(1.49 \cdot 10^{-6}\) & 1.068
& 0.2323 & \(1.91 \cdot 10^{-6}\) & 1.517 \\
\hline
\end{tabular}
\caption{Sips parameters for CS1000.}
\end{table}

\begin{table}[H]
\centering
\setlength{\tabcolsep}{5pt}
\renewcommand{\arraystretch}{0.95}
\begin{tabular}{c c c c c c c}
\hline
\multirow{2}{*}{\textbf{T(K)}} 
& \multicolumn{3}{c}{\textbf{R32}}
& \multicolumn{3}{c}{\textbf{R125}} \\
& $q_{sat}$ & $b$ & $\upsilon$
& $q_{sat}$ & $b$ & $\upsilon$ \\
\hline
\multirow{2}{*}{283}
& 1.31 & \(1.65 \cdot 10^{-5}\) & 0.8148
& 3.10 & \(1.31 \cdot 10^{-5}\) & 1.848 \\
& 14.2 & \(1.77 \cdot 10^{-6}\) & 0.4634
& 6.00 & \(1.17 \cdot 10^{-5}\) & 0.7308 \\
\hline
\multirow{2}{*}{303}
& 0.673 & \(1.81 \cdot 10^{-5}\) & 0.8331
& 1.57 & \(6.84 \cdot 10^{-6}\) & 0.2754 \\
& 14.4 & \(1.04 \cdot 10^{-6}\) & 0.5694
& 7.32 & \(5.56 \cdot 10^{-6}\) & 1.277 \\
\hline
\multirow{2}{*}{333}
& 4.62 & \(5.75 \cdot 10^{-7}\) & 0.5826
& 2.74 & \(2.09 \cdot 10^{-6}\) & 0.5417 \\
& 84.3 & \(2.12 \cdot 10^{-8}\) & 1.129
& 5.14 & \(4.00 \cdot 10^{-6}\) & 1.192 \\
\hline
\multirow{2}{*}{353}
& 5.63 & \(5.02 \cdot 10^{-7}\) & 0.3706
& 6.47 & \(1.54 \cdot 10^{-6}\) & 0.7977 \\
& 5.58 & \(5.37 \cdot 10^{-7}\) & 0.9987
& 0.981 & \(2.16 \cdot 10^{-5}\) & 1.087 \\
\hline
\multicolumn{1}{c}{}
& \multicolumn{3}{c}{\textbf{R134a}}
& \multicolumn{3}{c}{\textbf{R600}} \\
\multicolumn{1}{c}{}
& $q_{sat}$ & $b$ & $\upsilon$
& $q_{sat}$ & $b$ & $\upsilon$ \\
\hline
\multirow{2}{*}{283}
& 6.35 & \(1.18 \cdot 10^{-5}\) & 1.382
& 6.635 & \(2.05 \cdot 10^{-4}\) & 1.487 \\
& 3.73 & \(1.70 \cdot 10^{-5}\) & 0.3560
& 1.847 & \(1.89 \cdot 10^{-4}\) & 0.2514 \\
\hline
\multirow{2}{*}{303}
& 8.44 & \(6.75 \cdot 10^{-6}\) & 0.7003
& 5.01 & \(6.59 \cdot 10^{-5}\) & 1.861 \\
& 0.6509 & \(2.54 \cdot 10^{-4}\) & 0.7047
& 3.41 & \(7.91 \cdot 10^{-5}\) & 0.4538 \\
\hline
\multirow{2}{*}{333}
& 5.06 & \(2.91 \cdot 10^{-6}\) & 1.303
& 0.2419 & \(9.94 \cdot 10^{-3}\) & 0.9898 \\
& 3.95 & \(2.85 \cdot 10^{-6}\) & 0.5510
& 7.49 & \(2.43 \cdot 10^{-5}\) & 0.9822 \\
\hline
\multirow{2}{*}{353}
& 0.904 & \(2.69 \cdot 10^{-5}\) & 0.9571
& 2.51 & \(1.26 \cdot 10^{-5}\) & 0.5559 \\
& 7.12 & \(1.72 \cdot 10^{-6}\) & 0.6873
& 5.15 & \(1.41 \cdot 10^{-5}\) & 1.429 \\
\hline
\end{tabular}
\caption{Sips parameters for CS1000a.}
\label{Sips parameters for CS1000a}
\end{table}

\begin{figure}[H]
    \centering

    \begin{tikzpicture}
        \node[anchor=south west, inner sep=0] (image) at (0,0)
            {\includegraphics[width=\linewidth]{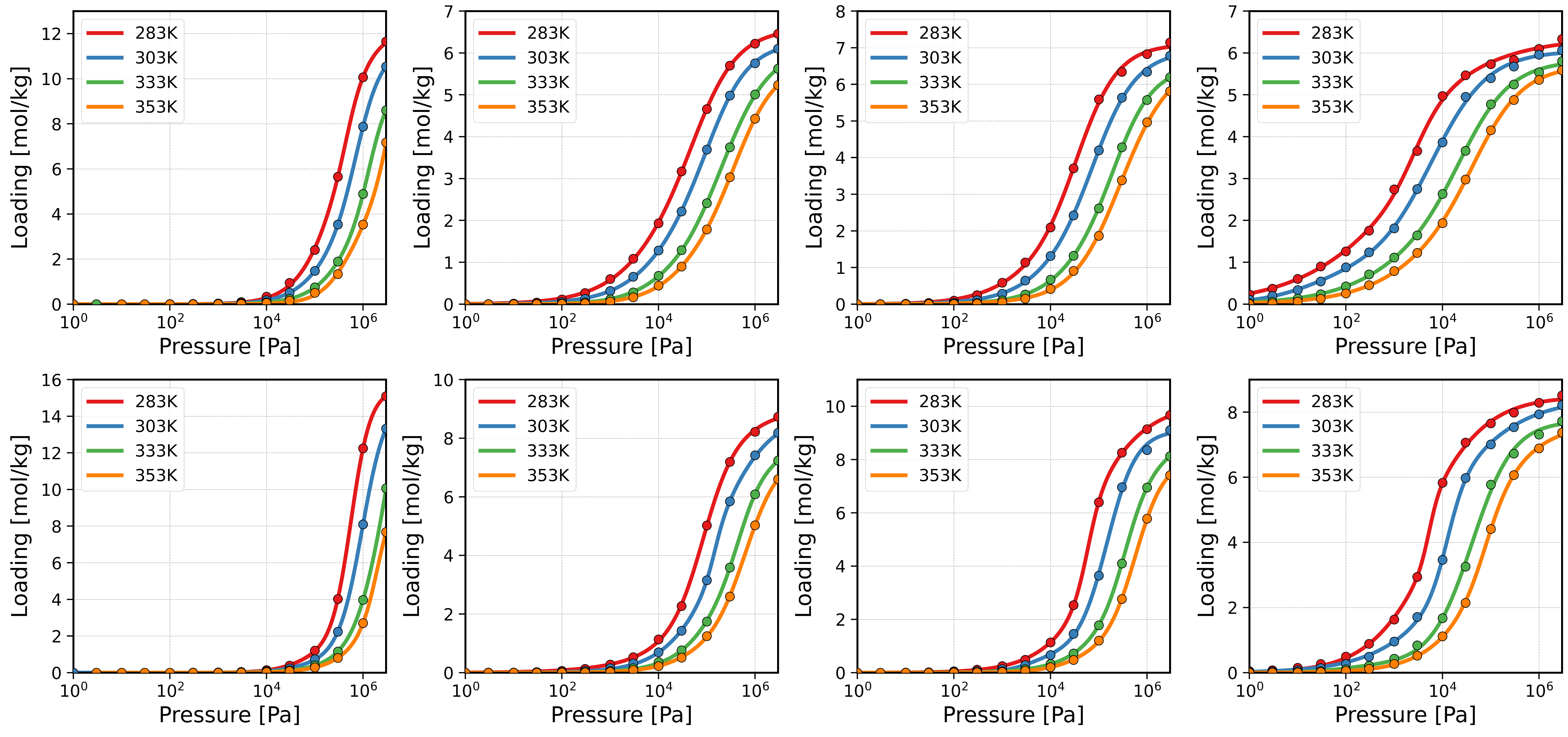}};
            
        \begin{scope}[x={(image.south east)}, y={(image.north west)}]
            \node[font=\bfseries, anchor=west] at (0.00,1.02) {I) a)};
            \node[font=\bfseries]             at (0.27,1.02) {b)};
            \node[font=\bfseries]             at (0.52,1.02) {c)};
            \node[font=\bfseries]             at (0.77,1.02) {d)};

            \node[font=\bfseries, anchor=west] at (0.00,0.52) {II) a)};
            \node[font=\bfseries]             at (0.27,0.52) {b)};
            \node[font=\bfseries]             at (0.52,0.52) {c)};
            \node[font=\bfseries]             at (0.77,0.52) {d)};
        \end{scope}
    \end{tikzpicture}

    \caption{Adsorption isotherms of Bhatia-01 (I) and CS1000a (II) for R32 (a), R125 (b), R134a (c), and R600 (d) with GCMC ($\bullet$), and dual-site Sips adsorption isotherm (—).}

    \label{fig: Adsorption isotherms pure components Bhatia-1 and CS1000a}
\end{figure}

\begin{figure}[H]
    \centering
    \begin{tikzpicture}
        \node[anchor=south west, inner sep=0] (image) at (0,0)
            {\includegraphics[width=\linewidth]{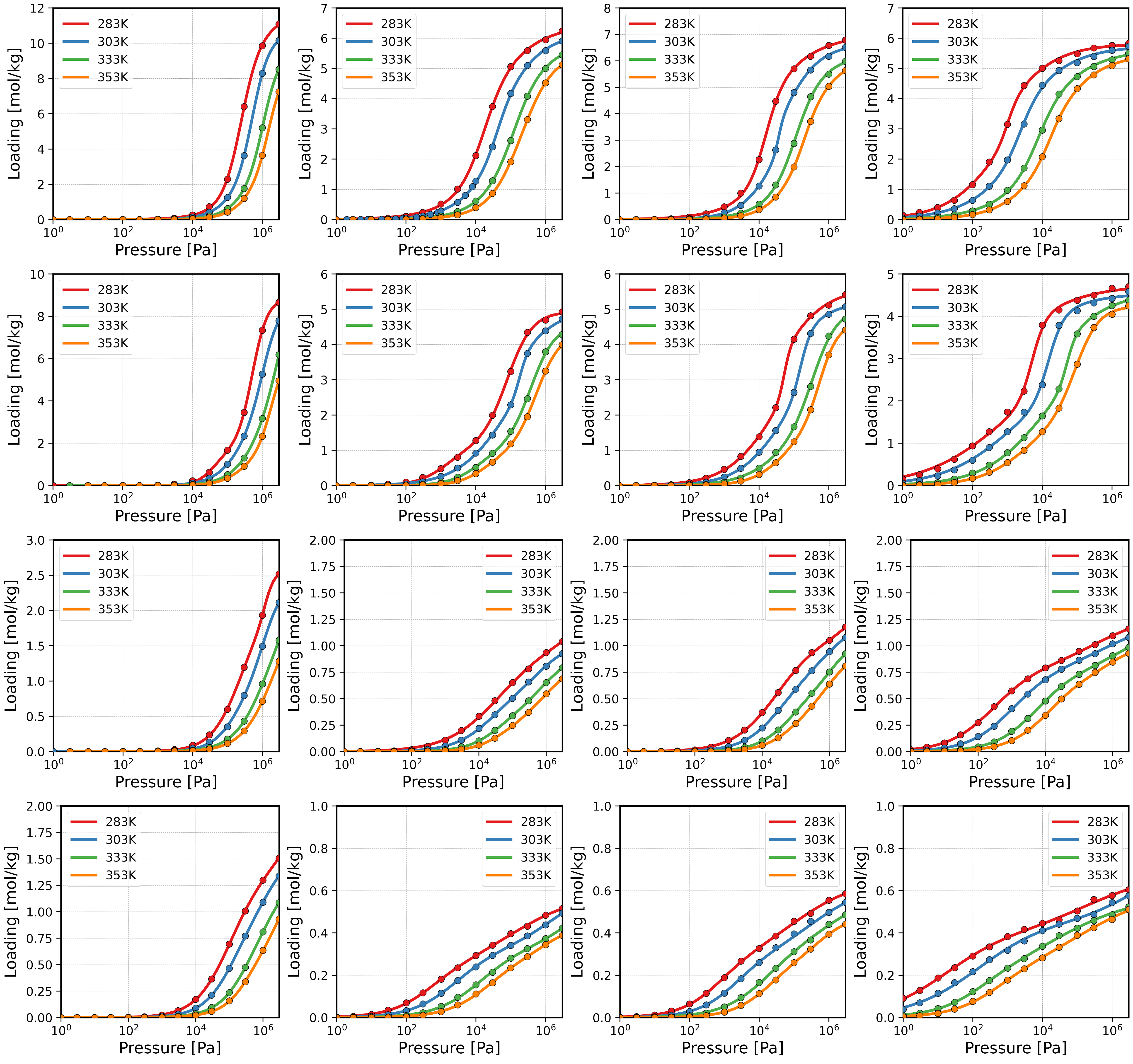}};
        \begin{scope}[x={(image.south east)}, y={(image.north west)}]
            \node[font=\bfseries, anchor=west] at (0.00,1.02) {I) a)};
            \node[font=\bfseries]             at (0.27,1.02) {b)};
            \node[font=\bfseries]             at (0.52,1.02) {c)};
            \node[font=\bfseries]             at (0.77,1.02) {d)};

            \node[font=\bfseries, anchor=west] at (0.00,0.76) {II) a)};
            \node[font=\bfseries]             at (0.27,0.76) {b)};
            \node[font=\bfseries]             at (0.52,0.76) {c)};
            \node[font=\bfseries]             at (0.77,0.76) {d)};

            \node[font=\bfseries, anchor=west] at (0.00,0.51) {III) a)};
            \node[font=\bfseries]             at (0.27,0.51) {b)};
            \node[font=\bfseries]             at (0.52,0.51) {c)};
            \node[font=\bfseries]             at (0.77,0.51) {d)};

            \node[font=\bfseries, anchor=west] at (0.00,0.26) {IV) a)};
            \node[font=\bfseries]             at (0.27,0.26) {b)};
            \node[font=\bfseries]             at (0.52,0.26) {c)};
            \node[font=\bfseries]             at (0.77,0.26) {d)};
        \end{scope}
    \end{tikzpicture}
    \caption{Adsorption isotherms of Bhatia-02 (I), Bhatia-03 (II), CS400 (III) and CS1000 (IV) for R32 (a), R125 (b), R134a (c) and R600 (d) with GCMC ($\bullet$), and dual-site Sips adsorption isotherm (---).}
    \label{supplementary fig: Adsorption isotherms pure components other structures}
\end{figure}

\newpage

\subsubsection{Adsorption Potential Theory}\label{Supplementary: APT Isotherms}
For the adsorption potential theory (APT), the adsorption model at $303$ K is used as reference. In \autoref{fig:supplementaryIsotherm_APT}, the GCMC adsorption isotherms are compared with APT predictions for all activated carbons at $283$ K and $333$ K, including the relative error of the loadings.

\begin{figure}[H]
    \centering
    \begin{tikzpicture}
        \node[anchor=south west, inner sep=0] (image) at (0,0)
            {\includegraphics[width=\linewidth]{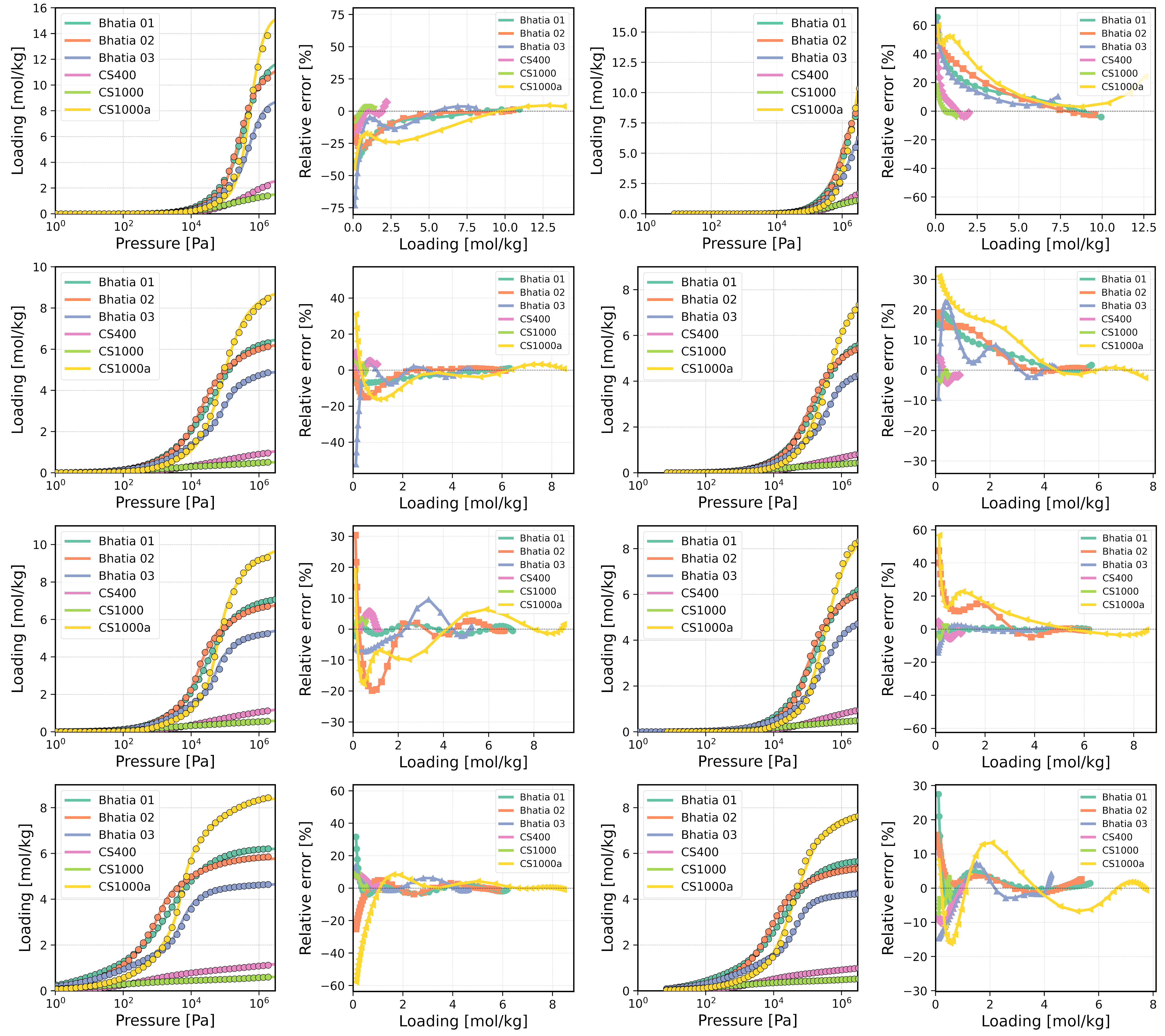}};
        \begin{scope}[x={(image.south east)}, y={(image.north west)}]
            \node[font=\bfseries, anchor=west] at (0.00,1.02) {I) a)};
            \node[font=\bfseries]             at (0.27,1.02) {b)};
            \node[font=\bfseries]             at (0.52,1.02) {c)};
            \node[font=\bfseries]             at (0.77,1.02) {d)};

            \node[font=\bfseries, anchor=west] at (0.00,0.76) {II) a)};
            \node[font=\bfseries]             at (0.27,0.76) {b)};
            \node[font=\bfseries]             at (0.52,0.76) {c)};
            \node[font=\bfseries]             at (0.77,0.76) {d)};

            \node[font=\bfseries, anchor=west] at (0.00,0.51) {III) a)};
            \node[font=\bfseries]             at (0.27,0.51) {b)};
            \node[font=\bfseries]             at (0.52,0.51) {c)};
            \node[font=\bfseries]             at (0.77,0.51) {d)};

            \node[font=\bfseries, anchor=west] at (0.00,0.26) {IV) a)};
            \node[font=\bfseries]             at (0.27,0.26) {b)};
            \node[font=\bfseries]             at (0.52,0.26) {c)};
            \node[font=\bfseries]             at (0.77,0.26) {d)};
        \end{scope}
    \end{tikzpicture}
    \caption{Adsorption isotherm at $283\,\mathrm{K}$ (a,b) and $333\,\mathrm{K}$ (c,d) for R32 (I), R125 (II), R134a (III) and R600 (IV) with GCMC ($\bullet$), dual-site Sips adsorption isotherm ($-$), and APT ($\bullet$).}
    \label{fig:supplementaryIsotherm_APT}
\end{figure}

\newpage

\subsubsection{Isosteric heat of adsorption}
In \autoref{subsection: Sensitivity Temperature Selection simulations}, the sensitivities of the Clausius-Clapeyron and Virial methods are tested by varying the type of adsorption model, temperature range, and number of temperatures. In conclusion, the number of adsorption isotherms measured at distinct temperatures exerts a negligible influence, while the temperature range has a significant impact.

\begin{figure}[H]
    \centering
    \begin{tikzpicture}
        \node[anchor=south west, inner sep=0] (image) at (0,0)
            {\includegraphics[width=0.95\linewidth]{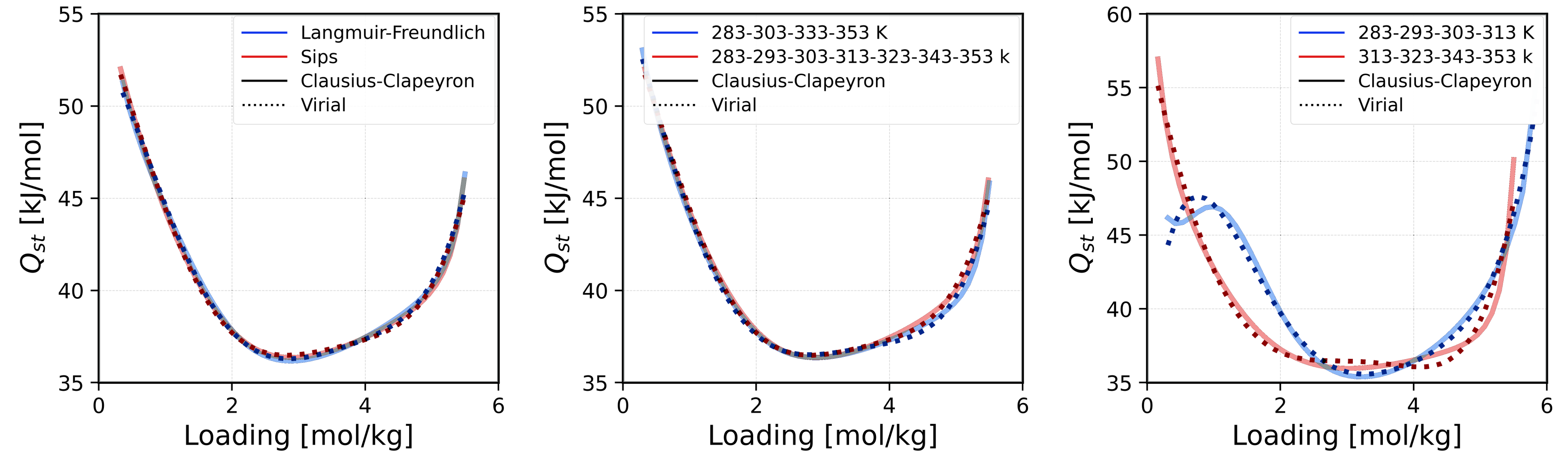}};
        \begin{scope}[x={(image.south east)}, y={(image.north west)}]
            \node[font=\bfseries, anchor=west] at (0.00,1.02) {a)};
            \node[font=\bfseries]             at (0.36,1.02) {b)};
            \node[font=\bfseries]             at (0.69,1.02) {c)};
        \end{scope}
    \end{tikzpicture}
    \caption{Comparison of the Clausius-Clapeyron and Virial results based on different data selections for Bhatia-01/R600.}
    \label{subsection: Sensitivity Temperature Selection simulations}
\end{figure}

\newpage

\subsubsection{Storage Density}\label{sec:Pure component storage density}
In \autoref{supplementary_all_3D_SD}, the three-dimensional representations of the energy storage density associated with the pressure-temperature swing process are presented for an adsorption pressure of \(P_{\mathrm{ads}} =10^{5}\,\mathrm{Pa}\).

\begin{figure}[H]
    \centering
    \begin{tikzpicture}
        \node[anchor=south west, inner sep=0] (image) at (0,0)
            {\includegraphics[width=\linewidth]{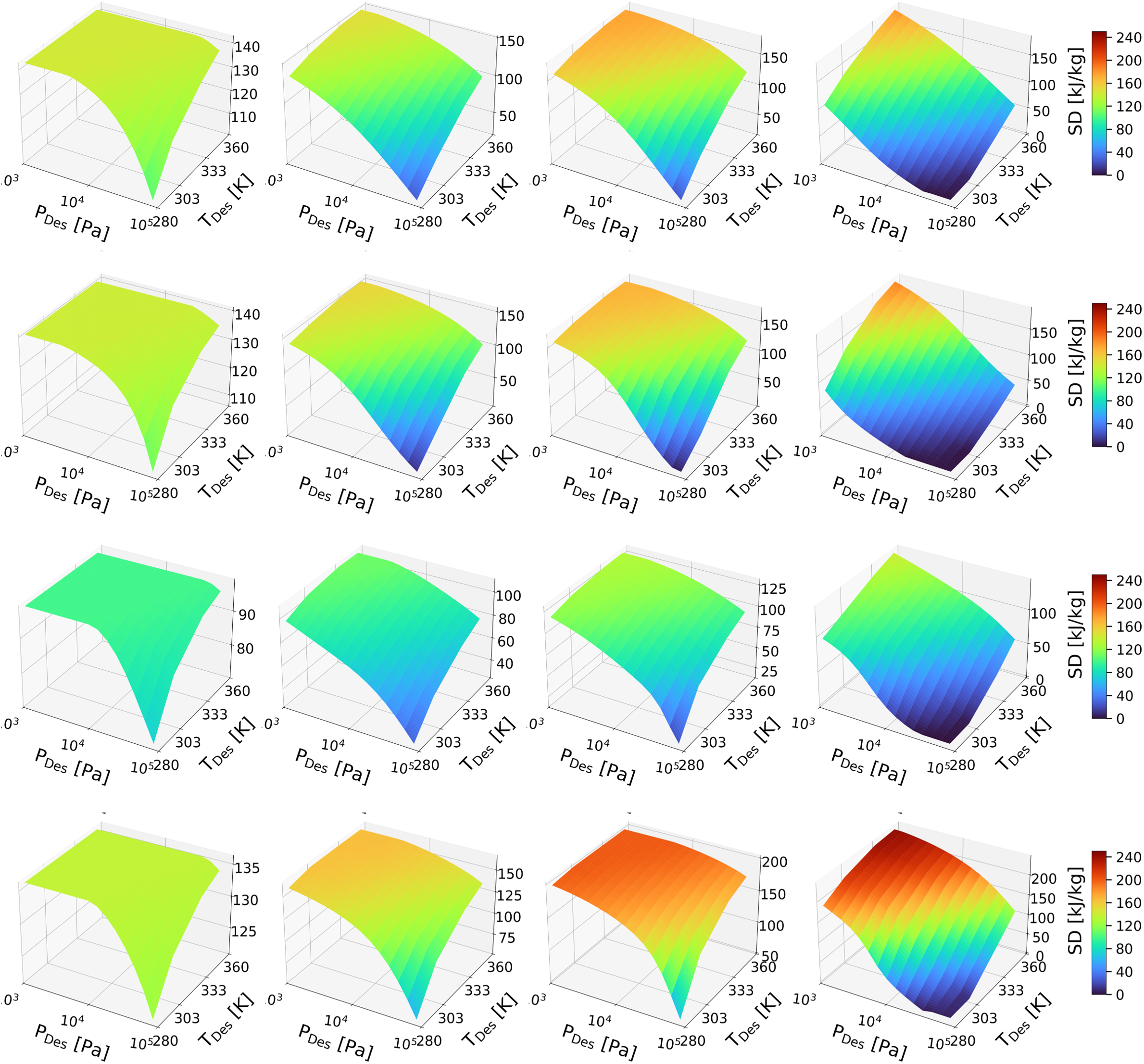}};
        \begin{scope}[x={(image.south east)}, y={(image.north west)}]
            \node[font=\bfseries, anchor=west] at (0.00,1.02) {I) a)};
            \node[font=\bfseries]             at (0.27,1.02) {b)};
            \node[font=\bfseries]             at (0.52,1.02) {c)};
            \node[font=\bfseries]             at (0.77,1.02) {d)};

            \node[font=\bfseries, anchor=west] at (0.00,0.76) {II) a)};
            \node[font=\bfseries]             at (0.27,0.76) {b)};
            \node[font=\bfseries]             at (0.52,0.76) {c)};
            \node[font=\bfseries]             at (0.77,0.76) {d)};

            \node[font=\bfseries, anchor=west] at (0.00,0.51) {III) a)};
            \node[font=\bfseries]             at (0.27,0.51) {b)};
            \node[font=\bfseries]             at (0.52,0.51) {c)};
            \node[font=\bfseries]             at (0.77,0.51) {d)};

            \node[font=\bfseries, anchor=west] at (0.00,0.26) {IV) a)};
            \node[font=\bfseries]             at (0.27,0.26) {b)};
            \node[font=\bfseries]             at (0.52,0.26) {c)};
            \node[font=\bfseries]             at (0.77,0.26) {d)};
        \end{scope}
    \end{tikzpicture}
    \caption{3D storage density of Bhatia-01 (I), Bhatia-02 (II), Bhatia-03 (III) and CS1000a (IV) for R32 (a), R125 (b), R134a (c) and R600 (d) at $P_{\mathrm{ads}} = 10^{5}$ Pa.}
    \label{supplementary_all_3D_SD}
\end{figure}

\newpage

\subsubsection{Maximum Storage Density}\label{Storage density maximum}
A limitation of the storage density is the loading range of the isosteric heat of adsorption. \autoref{tab: max pressure sd pure comp} reports the maximum achievable loading and the corresponding maximum pressure. Consequently, this pressure defines the upper bound of the pressure range over which the storage density can be reliably predicted.

\begin{table}[H]
\centering
\setlength{\tabcolsep}{6pt}
\renewcommand{\arraystretch}{0.95}
\begin{tabular}{l c c c}
\hline
\textbf{Activated-carbon} & \textbf{Refrigerant} & \textbf{$L_{max}$ (mol/kg)} & \textbf{$P_{max}$ (kPa)} \\
\hline
Bhatia-01 & \multirow{6}{*}{R32}   & 7    & 418 \\
Bhatia-02 &                         & 7.12 & 363 \\
Bhatia-03 &                         & 4.91 & 468 \\
CS1000    &                         & 0.93 & 223 \\
CS1000a   &                         & 7.59 & 514 \\
CS400     &                         & 1.25 & 337 \\
\hline
Bhatia-01 & \multirow{6}{*}{R600}  & 5.48 & 35 \\
Bhatia-02 &                         & 5.22 & 20 \\
Bhatia-03 &                         & 4.19 & 27 \\
CS1000    &                         & 0.51 & 88 \\
CS1000a   &                         & 7.15 & 38 \\
CS400     &                         & 0.92 & 78 \\
\hline
Bhatia-01 & \multirow{6}{*}{R125}  & 5.16 & 160 \\
Bhatia-02 &                         & 5.1  & 112 \\
Bhatia-03 &                         & 3.96 & 196 \\
CS1000    &                         & 0.39 & 89 \\
CS1000a   &                         & 6.46 & 194 \\
CS400     &                         & 0.68 & 121 \\
\hline
Bhatia-01 & \multirow{6}{*}{R134a} & 5.79 & 130 \\
Bhatia-02 &                         & 5.56 & 81 \\
Bhatia-03 &                         & 4.37 & 136 \\
CS1000    &                         & 0.43 & 78 \\
CS1000a   &                         & 7.28 & 149 \\
CS400     &                         & 0.79 & 112 \\
\hline
\end{tabular}
\caption{Limitation of the storage density by the loading range of the isosteric heat of adsorption for the maximum achievable loading and corresponding maximum pressure.}
\label{tab: max pressure sd pure comp}
\end{table}

\newpage

\subsection{Multicomponent adsorption}\label{sup overall isotherms mixtures}
\autoref{total_adsorption_isotherm_mixtures} shows the total adsorption isotherms of the mixtures R407F, R410A, R417A, and R417C.

\begin{figure}[H]
    \centering
    \begin{tikzpicture}
        \node[anchor=south west, inner sep=0] (image) at (0,0)
            {\includegraphics[width=\linewidth]{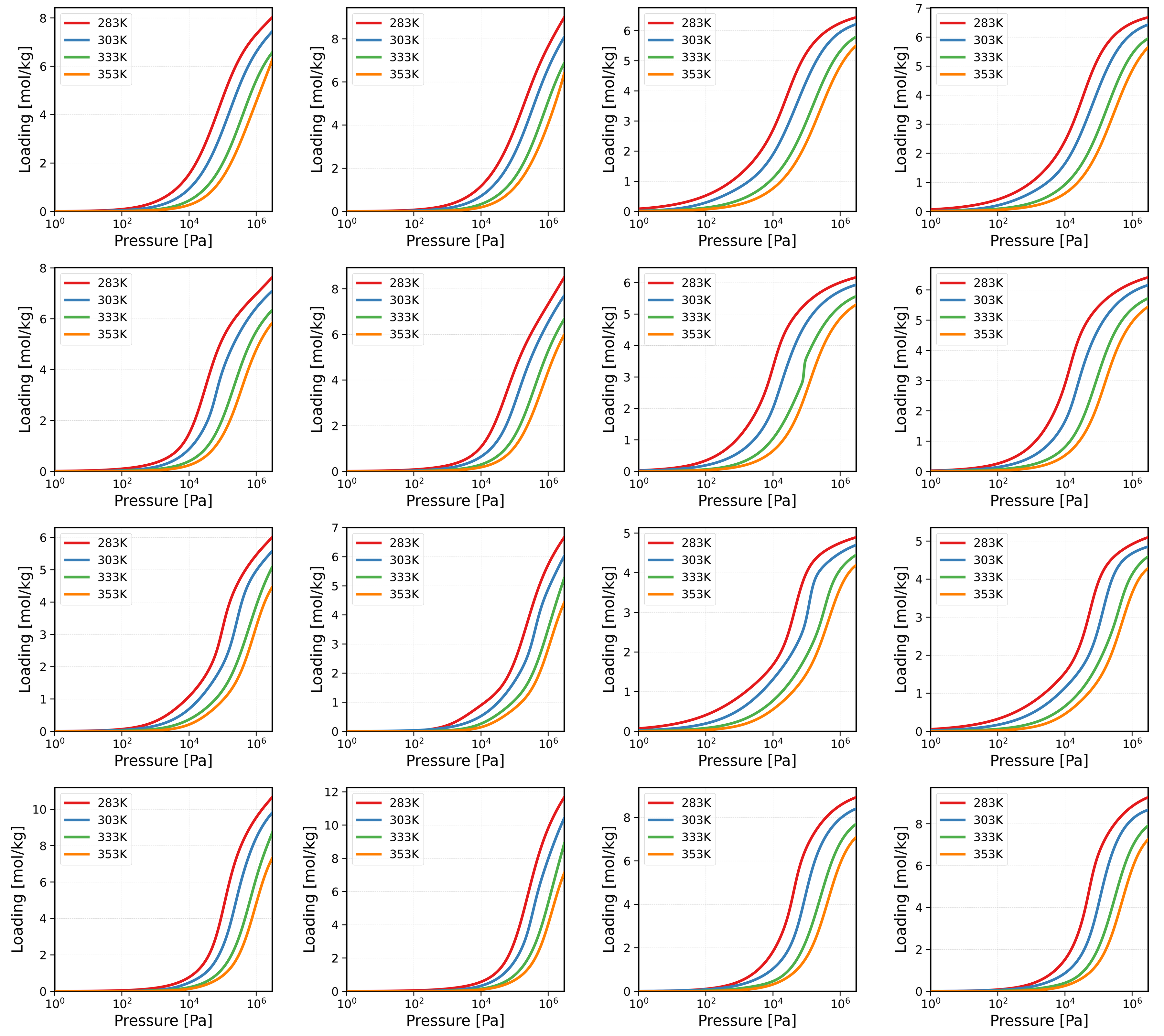}};
        \begin{scope}[x={(image.south east)}, y={(image.north west)}]
            \node[font=\bfseries, anchor=west] at (0.00,1.02) {I) a)};
            \node[font=\bfseries]             at (0.27,1.02) {b)};
            \node[font=\bfseries]             at (0.52,1.02) {c)};
            \node[font=\bfseries]             at (0.77,1.02) {d)};

            \node[font=\bfseries, anchor=west] at (0.00,0.76) {II) a)};
            \node[font=\bfseries]             at (0.27,0.76) {b)};
            \node[font=\bfseries]             at (0.52,0.76) {c)};
            \node[font=\bfseries]             at (0.77,0.76) {d)};

            \node[font=\bfseries, anchor=west] at (0.00,0.51) {III) a)};
            \node[font=\bfseries]             at (0.27,0.51) {b)};
            \node[font=\bfseries]             at (0.52,0.51) {c)};
            \node[font=\bfseries]             at (0.77,0.51) {d)};

            \node[font=\bfseries, anchor=west] at (0.00,0.26) {IV) a)};
            \node[font=\bfseries]             at (0.27,0.26) {b)};
            \node[font=\bfseries]             at (0.52,0.26) {c)};
            \node[font=\bfseries]             at (0.77,0.26) {d)};
        \end{scope}
    \end{tikzpicture}
    \caption{Adsorption isotherms of Bhatia-01 (I), Bhatia-02 (II), Bhatia-03 (III) and CS1000a (IV) for R407F (a), R410A (b), R417A (c) and R417C (d).}
    \label{total_adsorption_isotherm_mixtures}
\end{figure}

\newpage

In \autoref{sup: seperation R410A R417C}, the remaining mixtures for Bhatia-01 are presented. 

\begin{figure}[H]
    \centering
    \begin{tikzpicture}
        \node[anchor=south west, inner sep=0] (image) at (0,0)
            {\includegraphics[width=\linewidth]{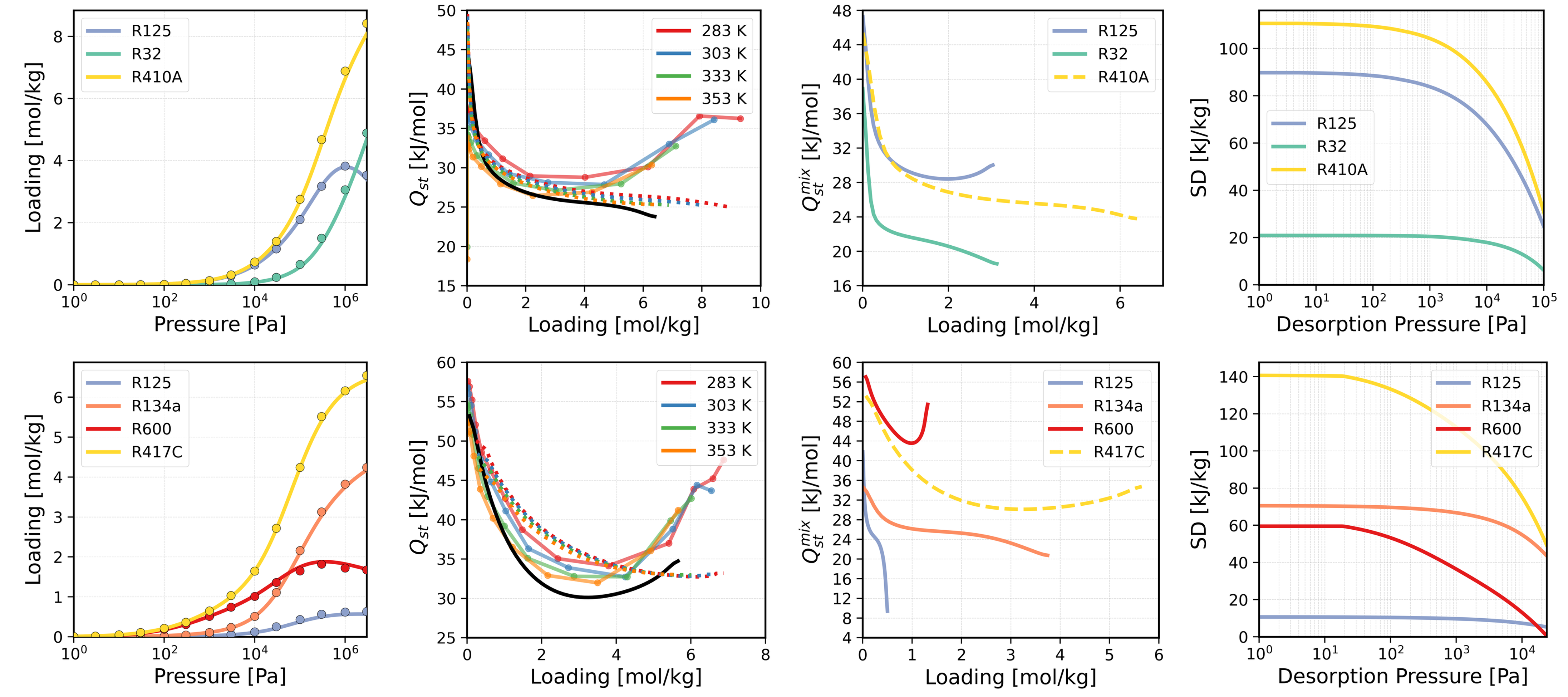}};
        \begin{scope}[x={(image.south east)}, y={(image.north west)}]
            \node[font=\bfseries, anchor=west] at (0,1.02) {I) a)};
            \node[font=\bfseries]             at (0.27,1.02) {b)};
            \node[font=\bfseries]             at (0.52,1.02) {c)};
            \node[font=\bfseries]             at (0.77,1.02) {d)};

            \node[font=\bfseries, anchor=west] at (0,0.52) {II) a)};
            \node[font=\bfseries]             at (0.27,0.52) {b)};
            \node[font=\bfseries]             at (0.52,0.52) {c)};
            \node[font=\bfseries]             at (0.77,0.52) {d)};
        \end{scope}
    \end{tikzpicture}
    \caption{Adsorption isotherm (a), isosteric heat of adsorption methods (b), isosteric heat of adsorption per component (c), and storage density per component (d) for R410A (I) and R417C (II), in Bathia-01 using GCMC ($\bullet$), IAST ($-$), Clausius-Clapeyron total (black $-$), linear mixing equation ($- -$), and fluctuation method ($-\bullet$).}
    \label{sup: seperation R410A R417C}
\end{figure}

\newpage

\subsubsection{Isosteric heat of adsorption}\label{sup: HoA linear and CC}

\begin{figure}[H]
    \centering
        \begin{tikzpicture}[inner sep=0, outer sep=0]
        \node[anchor=south west] (image) at (0,0)
        {\includegraphics[width=0.55\linewidth]{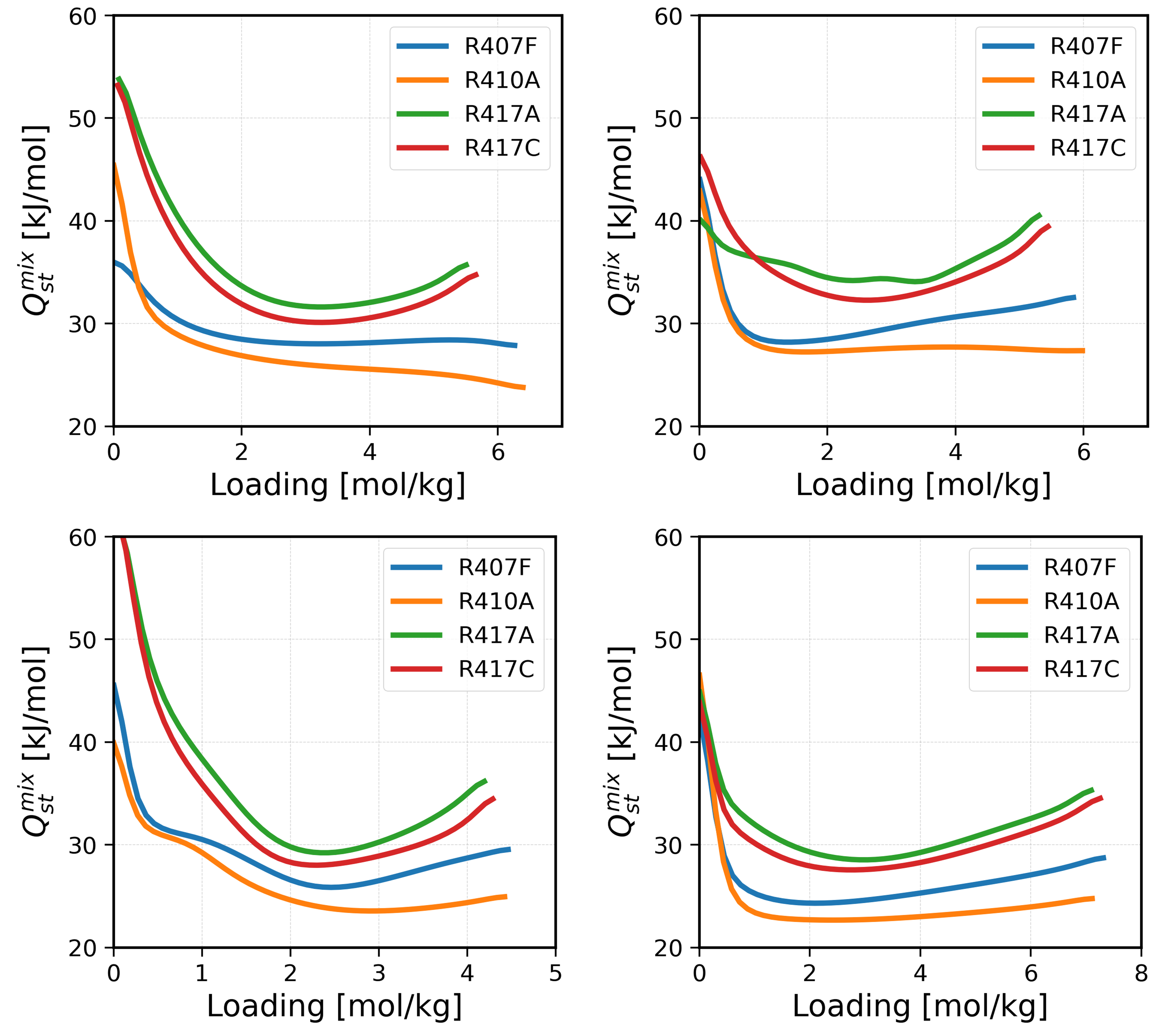}};
        \begin{scope}[x={(image.south east)}, y={(image.north west)}]
        \node[font=\bfseries] at (0.02, 1.02) {a)};
        \node[font=\bfseries] at (0.52, 1.02) {b)};
        \node[font=\bfseries] at (0.02, 0.52) {c)};
        \node[font=\bfseries] at (0.52, 0.52) {d)};
        \end{scope}
        \end{tikzpicture}
        \label{fig:two_subfigs_METHODSleft}
        \caption{Isosteric heat of adsorption for the mixtures in Bhatia-01 (a), Bhatia-02 (b), Bhatia-03 (c) and CS1000a (d).  Clausius-Clapeyron for mixtures (\autoref{eq: clausius clapeyron mixture})}
\end{figure}

\begin{figure}[H]
        \centering
        \begin{tikzpicture}[inner sep=0, outer sep=0]
            \node[anchor=south west] (image) at (0,0)
            {\includegraphics[width=0.55\linewidth]{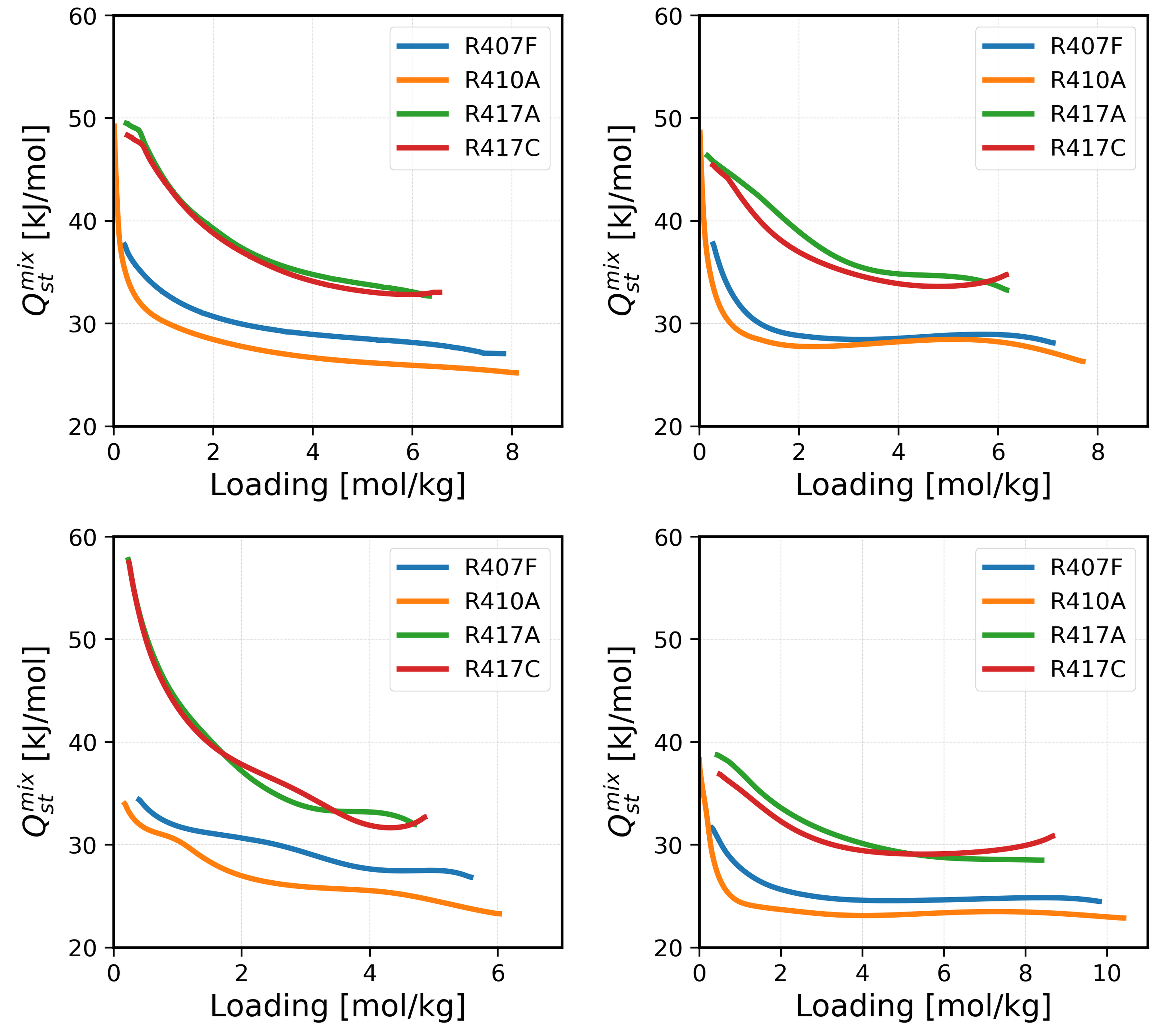}};
            \begin{scope}[x={(image.south east)}, y={(image.north west)}]
        \node[font=\bfseries] at (0.02, 1.02) {a)};
        \node[font=\bfseries] at (0.52, 1.02) {b)};
        \node[font=\bfseries] at (0.02, 0.52) {c)};
        \node[font=\bfseries] at (0.52, 0.52) {d)};
            \end{scope}
        \end{tikzpicture}
        \label{fig:two_subfigs_METHODSright}
    \caption{Isosteric heat of adsorption for the mixtures in Bhatia-01 (a), Bhatia-02 (b), Bhatia-03 (c) and CS1000a (d). Linear mixing (\autoref{eq: linear mixing rule}) with pure-component heats calculated by Clausius-Clapeyron at 303 K}
\end{figure}

\newpage

\subsubsection{Validation Storage density}\label{sec: control sd mixture}
In \autoref{fig:storage_density_mixtures_valdity}, the Clausius-Clapeyron mixing rule, the linear mixing rule, and the fluctuation-based method are compared based on storage density, demonstrating good agreement among the approaches.

\begin{figure}[H]
    \centering
    \begin{tikzpicture}
        \node[anchor=south west, inner sep=0] (image) at (0,0)
            {\includegraphics[width=0.7\linewidth]{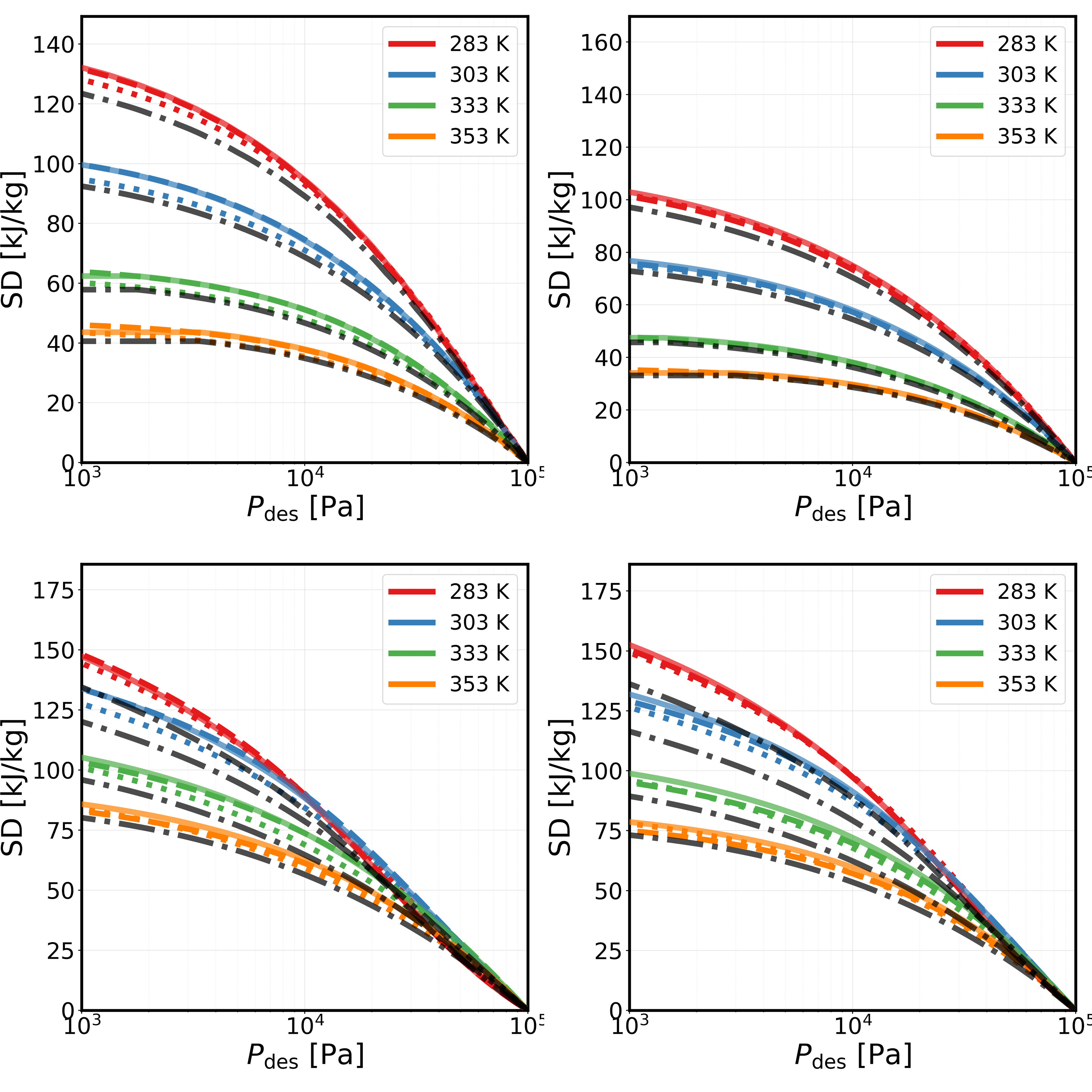}};
        \begin{scope}[x={(image.south east)}, y={(image.north west)}]
        \node[font=\bfseries] at (0.02, 1.02) {a)};
        \node[font=\bfseries] at (0.52, 1.02) {b)};
        \node[font=\bfseries] at (0.02, 0.52) {c)};
        \node[font=\bfseries] at (0.52, 0.52) {d)};
        \end{scope}
    \end{tikzpicture}
    \caption{Storage density validation for refrigerant mixtures R407F, R410A, R417A, and R417C at $283$-$353\,\mathrm{K}$ using the fluctuation approach (:), mixture approach Clausius-Clapeyron ($-$), mixture approach Virial ($- -$), and Clausius-Clapeyron ($\cdot-$).}
    \label{fig:storage_density_mixtures_valdity}
\end{figure}

\newpage

\subsubsection{Storage density 3D Plots}\label{3D plots mix}

In \autoref{supplementary all 3D SD}, the three-dimensional representations of the storage density associated with the pressure-temperature swing process are presented for an adsorption pressure of \(P_{\mathrm{ads}} = 10^{5}\,\mathrm{Pa}\).

\begin{figure}[H]
    \centering
    \begin{tikzpicture}
        \node[anchor=south west, inner sep=0] (image) at (0,0)
            {\includegraphics[width=\linewidth]{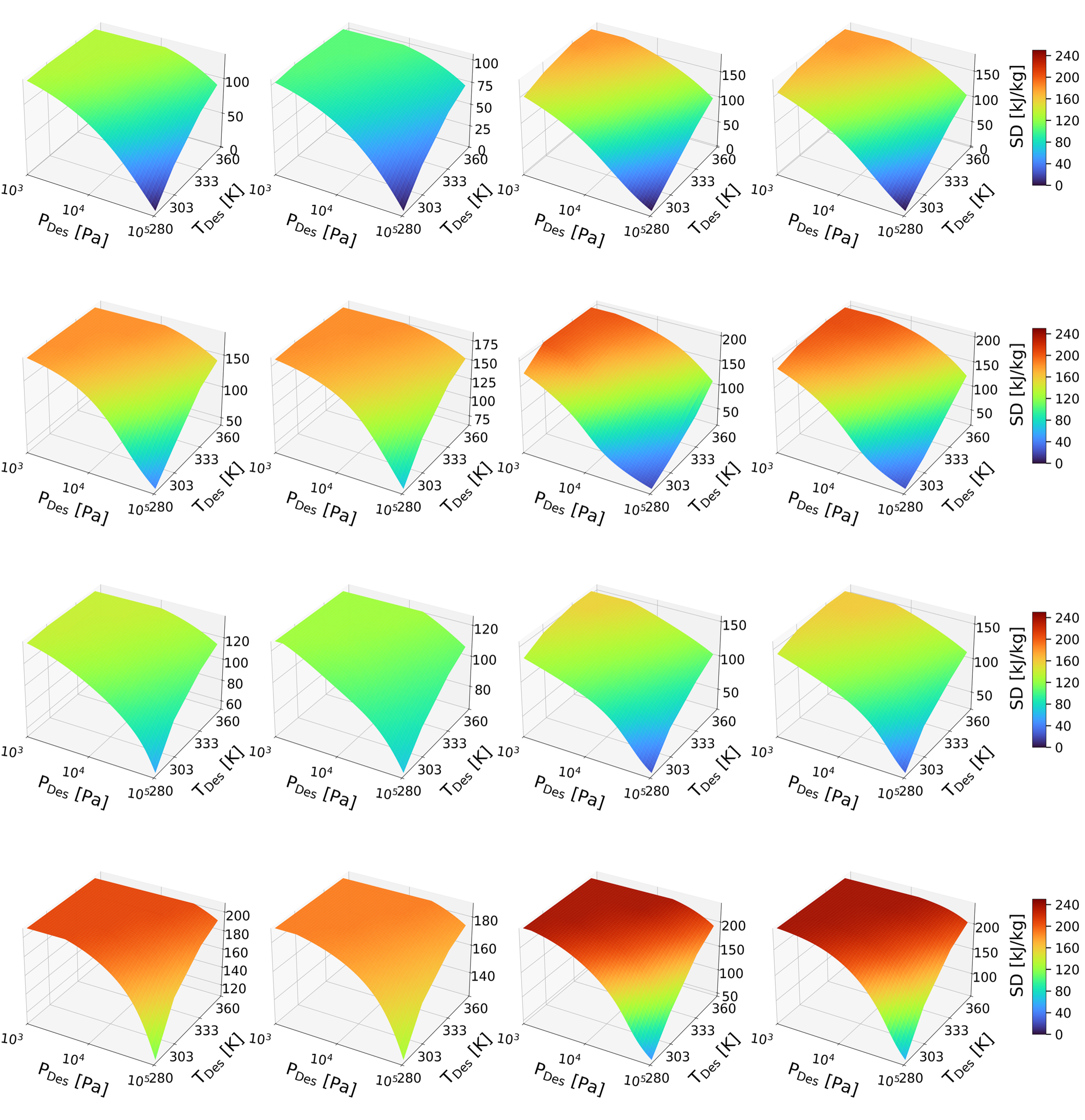}};
        \begin{scope}[x={(image.south east)}, y={(image.north west)}]
            \node[font=\bfseries, anchor=west] at (0.00,1.02) {I) a)};
            \node[font=\bfseries]             at (0.27,1.02) {b)};
            \node[font=\bfseries]             at (0.52,1.02) {c)};
            \node[font=\bfseries]             at (0.77,1.02) {d)};

            \node[font=\bfseries, anchor=west] at (0.00,0.76) {II) a)};
            \node[font=\bfseries]             at (0.27,0.76) {b)};
            \node[font=\bfseries]             at (0.52,0.76) {c)};
            \node[font=\bfseries]             at (0.77,0.76) {d)};

            \node[font=\bfseries, anchor=west] at (0.00,0.51) {III) a)};
            \node[font=\bfseries]             at (0.27,0.51) {b)};
            \node[font=\bfseries]             at (0.52,0.51) {c)};
            \node[font=\bfseries]             at (0.77,0.51) {d)};

            \node[font=\bfseries, anchor=west] at (0.00,0.26) {IV) a)};
            \node[font=\bfseries]             at (0.27,0.26) {b)};
            \node[font=\bfseries]             at (0.52,0.26) {c)};
            \node[font=\bfseries]             at (0.77,0.26) {d)};
        \end{scope}
    \end{tikzpicture}
    \caption{3D storage density of Bhatia-01 (I), Bhatia-02 (II), Bhatia-03 (III) and CS1000a (IV) for R407F (a), R410A (b), R417A (c) and R417C (d) at $P_{ads}=10^5$ Pa.}
    \label{supplementary all 3D SD}
\end{figure}

\newpage

\subsubsection{Maximum Storage Density for mixtures}\label{Maximum Storage Density}
A limitation of the storage density is the loading range of the isosteric heat of adsorption. \autoref{mix max sd loading mixture} reports the maximum achievable loading and the corresponding maximum pressure. Consequently, this pressure defines the upper bound of the pressure range over which the storage density can be reliably predicted.

\begin{table}[H]
\centering
\setlength{\tabcolsep}{4pt}
\renewcommand{\arraystretch}{0.95}
\begin{tabular}{l c c c c c c c c}
\hline
\textbf{Activated-carbon} &
\multicolumn{2}{c}{\textbf{R407F}} &
\multicolumn{2}{c}{\textbf{R410A}} &
\multicolumn{2}{c}{\textbf{R417A}} &
\multicolumn{2}{c}{\textbf{R417C}} \\
& \textbf{$L_{max}$} & \textbf{$P_{max}$}
& \textbf{$L_{max}$} & \textbf{$P_{max}$}
& \textbf{$L_{max}$} & \textbf{$P_{max}$}
& \textbf{$L_{max}$} & \textbf{$P_{max}$} \\
\hline
Bhatia-01 & 6.26 & 303 & 6.39 & 425 & 5.50 & 141 & 5.65 & 153 \\
Bhatia-02 & 5.84 & 186 & 5.98 & 295 & 5.30 & 89  & 5.44 & 98  \\
Bhatia-03 & 4.46 & 248 & 4.42 & 364 & 4.19 & 132 & 4.29 & 153 \\
CS1000a   & 7.31 & 255 & 7.11 & 365 & 7.09 & 143 & 7.26 & 155 \\
\hline
\end{tabular}
\caption{Maximum loading ($L_{max}$ [mol/kg]) and corresponding pressure ($P_{max}$ [kPa]) for all refrigerant blends.}
\label{mix max sd loading mixture}
\end{table}

In \autoref{maximum storage density 50kpa}-\autoref{maximum storage density 100kpa}, the values corresponding to \autoref{fig:Summary_3D_maximum_storage_densities} are given. 

\begin{table}[H]
\centering
\setlength{\tabcolsep}{4pt}
\renewcommand{\arraystretch}{0.95}
\begin{tabular}{l c c c c c c c c}
\hline
\textbf{Material} & \textbf{R125} & \textbf{R134a} & \textbf{R32} & \textbf{R407F} & \textbf{R410A} & \textbf{R417A} & \textbf{R417C} & \textbf{R600} \\
\hline
Bhatia-01 & 111.232 & 136.793 & 26.8033 & 105.781 & 81.6818 & 163.818 & 152.403 & 204.146 \\
Bhatia-02 & 131.120 & 157.767 & 19.7424 & 126.554 & 92.3936 & 177.710 & 174.959 & 208.297 \\
Bhatia-03 & 70.4142 & 89.4338 & 19.3664 & 66.2006 & 51.1154 & 113.381 & 102.561 & 156.900 \\
CS1000a   & 83.7624 & 104.151 & 9.6401 & 65.8638 & 46.6493 & 159.052 & 137.746 & 243.337 \\
\hline
\end{tabular}
\caption{Maximum storage density $SD_{\max}$ [kJ/kg] at $P_{ads} = 50$ kPa.}
\label{maximum storage density 50kpa}
\end{table}

\begin{table}[H]
\centering
\setlength{\tabcolsep}{4pt}
\renewcommand{\arraystretch}{0.95}
\begin{tabular}{l c c c c c c c c}
\hline
\textbf{Material} & \textbf{R125} & \textbf{R134a} & \textbf{R32} & \textbf{R407F} & \textbf{R410A} & \textbf{R417A} & \textbf{R417C} & \textbf{R600} \\
\hline
Bhatia-01 & 136.462 & 168.149 & 50.8036 & 137.859 & 109.230 & 185.561 & 178.486 & 204.146 \\
Bhatia-02 & 153.362 & 169.446 & 46.0163 & 158.255 & 126.383 & 188.196 & 187.464 & 208.297 \\
Bhatia-03 & 92.1472 & 124.458 & 33.0594 & 93.5863 & 68.5782 & 138.934 & 130.853 & 156.900 \\
CS1000a   & 127.874 & 173.694 & 19.1755 & 114.581 & 75.8840 & 206.310 & 196.168 & 243.337 \\
\hline
\end{tabular}
\caption{Maximum storage density $SD_{\max}$ [kJ/kg] at $P_{ads} = 100$ kPa.}
\end{table}

\begin{table}[H]
\centering
\setlength{\tabcolsep}{4pt}
\renewcommand{\arraystretch}{0.95}
\begin{tabular}{l c c c c c c c c}
\hline
\textbf{Material} & \textbf{R125} & \textbf{R134a} & \textbf{R32} & \textbf{R407F} & \textbf{R410A} & \textbf{R417A} & \textbf{R417C} & \textbf{R600} \\
\hline
Bhatia-01 & 149.335 & 175.818 & 142.120 & 177.091 & 168.442 & 190.410 & 189.780 & 204.146 \\
Bhatia-02 & 153.446 & 169.446 & 140.392 & 176.813 & 164.472 & 188.196 & 187.464 & 208.297 \\
Bhatia-03 & 109.397 & 129.699 & 98.5806 & 125.189 & 110.765 & 143.702 & 141.382 & 156.900 \\
CS1000a   & 164.361 & 201.102 & 136.029 & 180.468 & 159.597 & 215.462 & 216.733 & 243.337 \\
\hline
\end{tabular}
\caption{Maximum storage density $SD_{\max}$ [kJ/kg] at $P_{ads} = 500$ kPa.}
\label{maximum storage density 100kpa}
\end{table}

\newpage

\subsection{Refrigerant blends separation: Adsorption equilibrium}\label{sec: supp seperation}

In \autoref{supp:separation_D8}, the separation performance of the mixtures in CS1000a is presented. In \autoref{fig:APT_1x2_R407F}, the validation of the APT to predict adsorption selectivities is shown. In  \autoref{supp:separation_summary_D14}, the selectivity and separation factor of the mixtures per activated carbon are given. For R407F, both selectivity and separation factors are given for R125 and R134a, since both score above 1. The selectivity of R125 is slightly higher than that of R134a. Bhatia-01, Bhatia-02, and Bhatia-03 show a similar ordering as before, where Bhatia-01 has slightly higher values than Bhatia-02, and both outperform Bhatia-03. For the separation factor, CS1000a outperforms the rest in mixtures R407F and R410A at high pressures, but is outperformed in mixtures R417A and R417C by Bhatia-01 and Bhatia-02.

\begin{figure}[H]
    \centering
    \begin{tikzpicture}
        \node[anchor=south west, inner sep=0] (image) at (0,0)
            {\includegraphics[width=\linewidth]{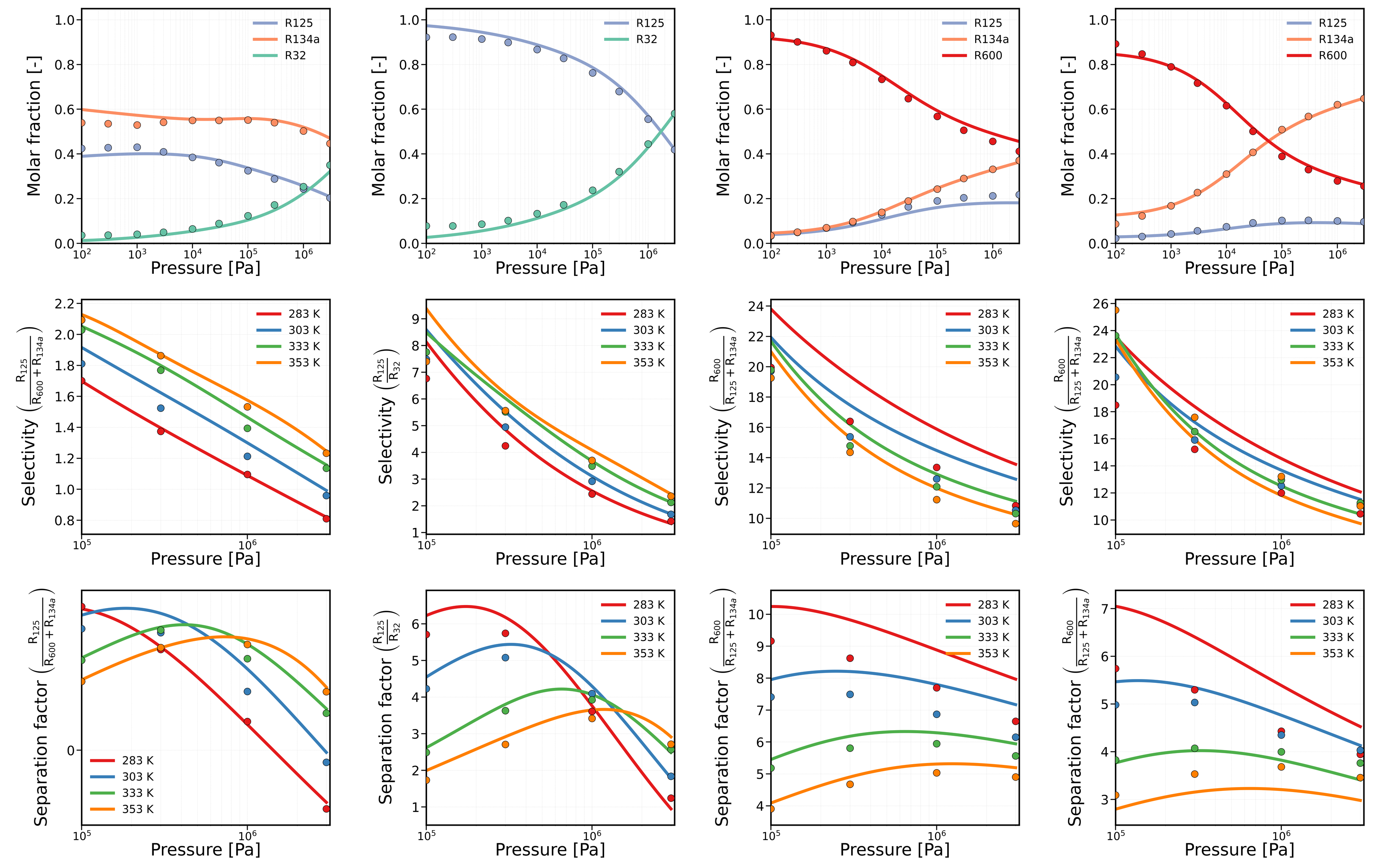}};
        \begin{scope}[x={(image.south east)}, y={(image.north west)}]
            \node[font=\bfseries, anchor=west] at (0.00,1.02) {I) a)};
            \node[font=\bfseries]             at (0.28,1.02) {b)};
            \node[font=\bfseries]             at (0.52,1.02) {c)};
            \node[font=\bfseries]             at (0.78,1.02) {d)};

            \node[font=\bfseries, anchor=west] at (0.00,0.68) {II) a)};
            \node[font=\bfseries]             at (0.28,0.68) {b)};
            \node[font=\bfseries]             at (0.52,0.68) {c)};
            \node[font=\bfseries]             at (0.78,0.68) {d)};

            \node[font=\bfseries, anchor=west] at (0.00,0.35) {III) a)};
            \node[font=\bfseries]             at (0.28,0.35) {b)};
            \node[font=\bfseries]             at (0.52,0.35) {c)};
            \node[font=\bfseries]             at (0.78,0.35) {d)};
        \end{scope}
    \end{tikzpicture}
    \caption{IAST mixture mol fractions at 303 K (I), selectivity (II), and separation factors (III) for refrigerant blends R407F (a), R410A (b), R417A (c), and R417C (d) of CS1000a.}
    \label{supp:separation_D8}
\end{figure}

\begin{figure}[H]
    \centering

    \begin{tikzpicture}
        \node[anchor=south west, inner sep=0] (image) at (0,0)
            {\includegraphics[width=0.7\linewidth]{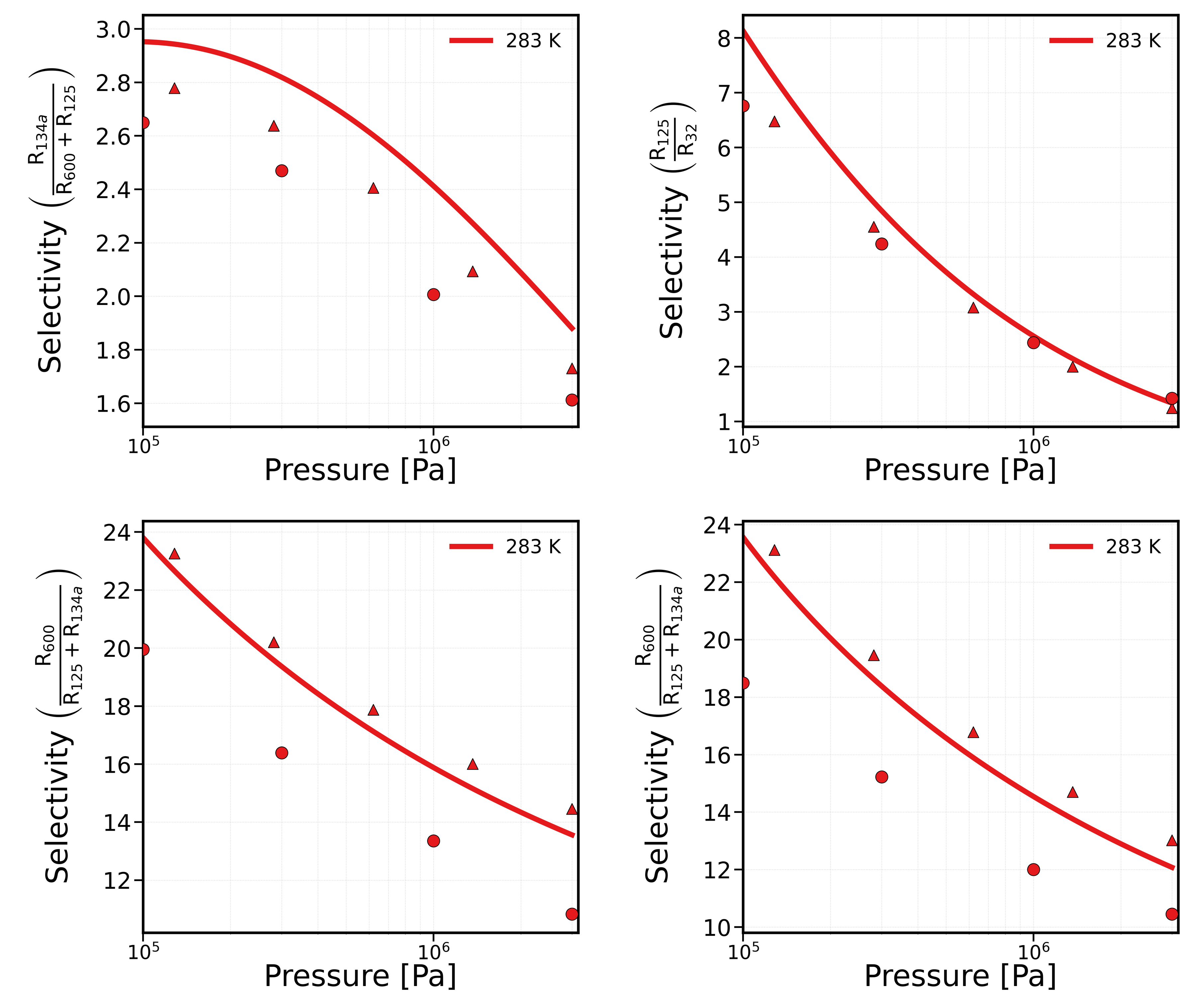}};
            
        \begin{scope}[x={(image.south east)}, y={(image.north west)}]
            \node[font=\bfseries, anchor=west] at (0.00,1.02) {a)};
            \node[font=\bfseries]             at (0.55,1.02) {b)};

            \node[font=\bfseries, anchor=west] at (0.00,0.55) {c)};
            \node[font=\bfseries]             at (0.55,0.55) {d)};
        \end{scope}
    \end{tikzpicture}

    \caption{APT selectivity for R407F (a), R410A (b), R417A (c) and R417C (d) with GCMC ($\bullet$), APT ($\triangle$) and IAST ($-$) in Bhatia-01.}

    \label{fig:APT_1x2_R407F}
\end{figure}

\begin{figure}[H]
    \centering
    \begin{tikzpicture}
        \node[anchor=south west, inner sep=0] (image) at (0,0)
            {\includegraphics[width=\linewidth]{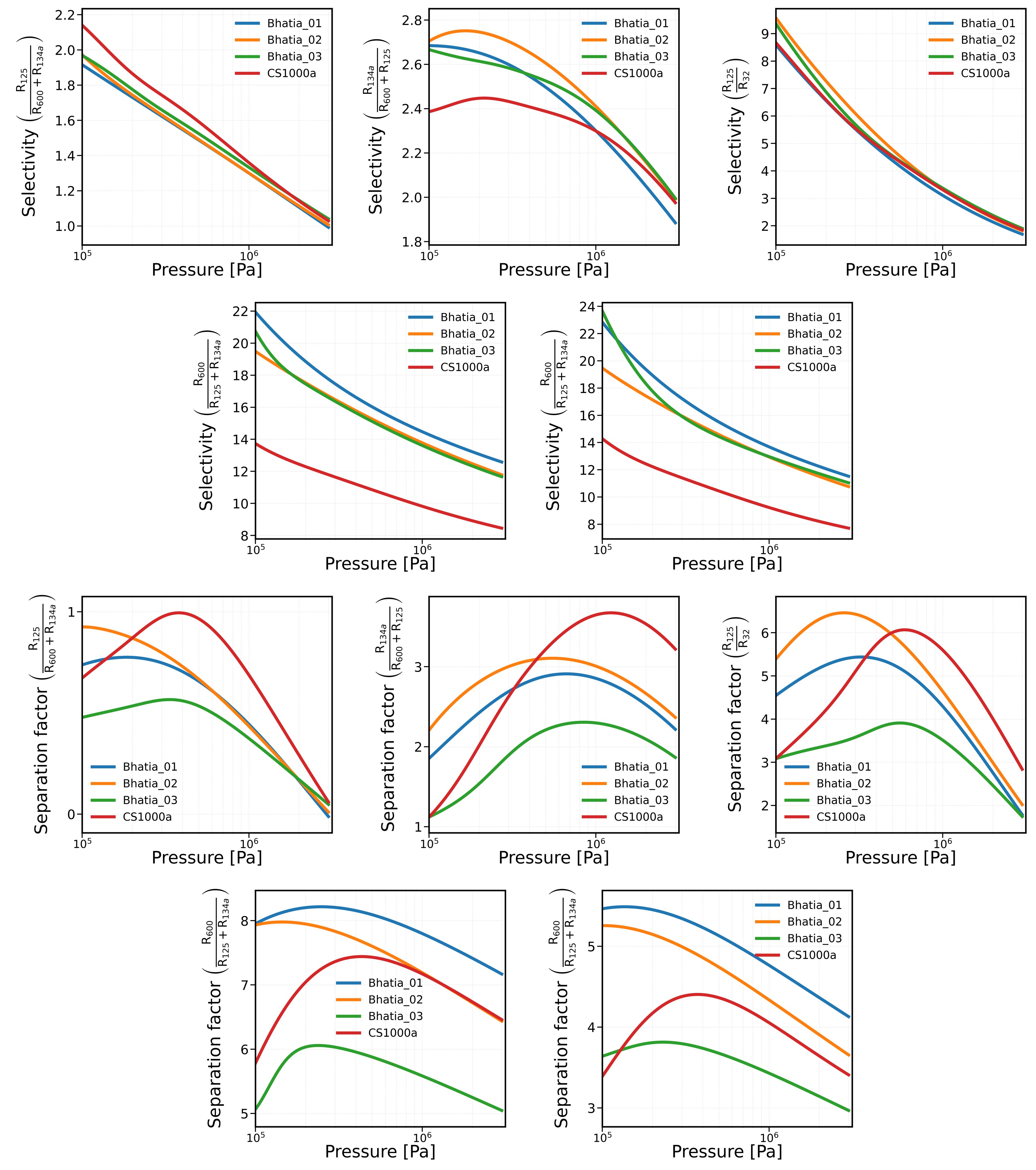}};
        \begin{scope}[x={(image.south east)}, y={(image.north west)}]
            \node[font=\bfseries, anchor=west] at (0.00,1.02) {I)};
            \node[font=\bfseries]             at (0.03,0.78) {a)};
            \node[font=\bfseries]             at (0.37,0.78) {b)};
            \node[font=\bfseries]             at (0.72,0.78) {c)};
            \node[font=\bfseries]             at (0.22,0.52) {d)};
            \node[font=\bfseries]             at (0.55,0.52) {e)};

            \node[font=\bfseries, anchor=west] at (0.00,0.52) {II)};
            \node[font=\bfseries]             at (0.03,0.27) {a)};
            \node[font=\bfseries]             at (0.37,0.27) {b)};
            \node[font=\bfseries]             at (0.72,0.27) {c)};
            \node[font=\bfseries]             at (0.22,0.00) {d)};
            \node[font=\bfseries]             at (0.55,0.00) {e)};
        \end{scope}
    \end{tikzpicture}
    \caption{Selectivity (I) and separation factor (II) of R407F (a,b), R410A (c), R417A (d) and R417C (e) at 303 K.}
    \label{supp:separation_summary_D14}
\end{figure}

\newpage

\subsection{Refrigerant blends separation: Adsorption dynamics}
In \autoref{supp:separation_D9}, the breakthrough curves of all activated carbons for the respective mixtures at 303 K are reported.

\label{sup sepeation dynamics}
\begin{figure}[H]
    \centering
    \begin{tikzpicture}
        \node[anchor=south west, inner sep=0] (image) at (0,0)
            {\includegraphics[width=\linewidth]{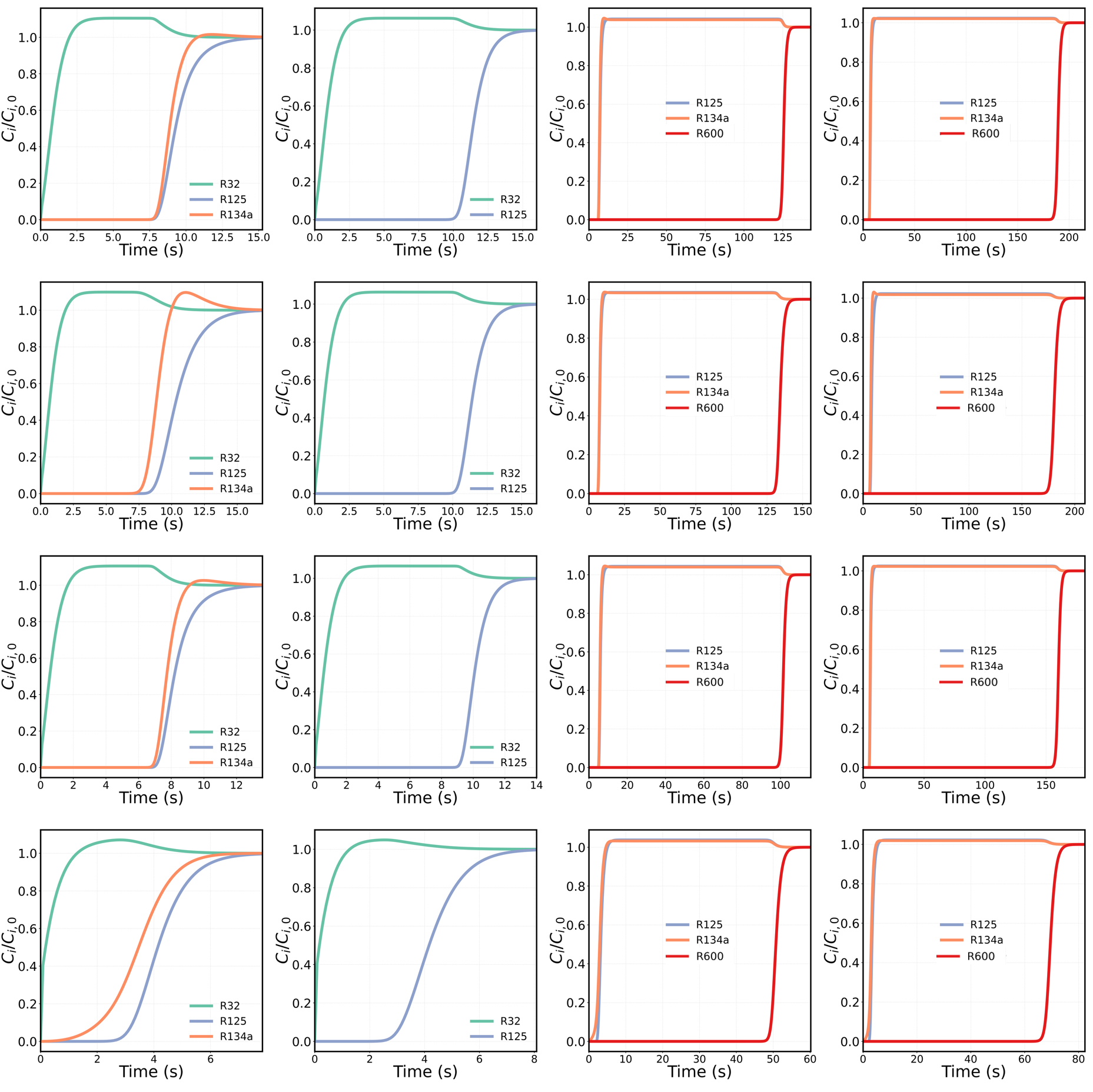}};
        \begin{scope}[x={(image.south east)}, y={(image.north west)}]
            \node[font=\bfseries, anchor=west] at (0.00,1.02) {I) a)};
            \node[font=\bfseries]             at (0.27,1.02) {b)};
            \node[font=\bfseries]             at (0.52,1.02) {c)};
            \node[font=\bfseries]             at (0.77,1.02) {d)};

            \node[font=\bfseries, anchor=west] at (0.00,0.76) {II) a)};
            \node[font=\bfseries]             at (0.27,0.76) {b)};
            \node[font=\bfseries]             at (0.52,0.76) {c)};
            \node[font=\bfseries]             at (0.77,0.76) {d)};

            \node[font=\bfseries, anchor=west] at (0.00,0.51) {III) a)};
            \node[font=\bfseries]             at (0.27,0.51) {b)};
            \node[font=\bfseries]             at (0.52,0.51) {c)};
            \node[font=\bfseries]             at (0.77,0.51) {d)};

            \node[font=\bfseries, anchor=west] at (0.00,0.26) {IV) a)};
            \node[font=\bfseries]             at (0.27,0.26) {b)};
            \node[font=\bfseries]             at (0.52,0.26) {c)};
            \node[font=\bfseries]             at (0.77,0.26) {d)};
        \end{scope}
    \end{tikzpicture}
    \caption{Breakthrough curves at $303$ K of Bhatia-01 (I), Bhatia-02 (II), Bhatia-03 (III) and CS1000a (IV) for refrigerant blends R407F (a), R410A (b), R417A (c), and R417C (d).}
    \label{supp:separation_D9}
\end{figure}

\end{document}